\documentclass[preprint,prd,aps,amsmath,nofootinbib,tightenlines,superscriptaddress]{revtex4}
\usepackage[hypertex]{hyperref}
\usepackage{graphicx}
\usepackage{bm}
\usepackage{epsfig}
\usepackage{graphics}
\usepackage{xspace}

\def\Dslash{D\!\!\!\!\slash}

\def\nslash{n\!\!\!\slash}
\def\bnslash{\bar n\!\!\!\slash}

\def\OMIT#1{}

\newcommand{\CH}[2]{\chi_{#1,#2}}

\newcommand{\bCH}[2]{\overline\chi_{#1,#2}}

\newcommand{\nn}{\nonumber} 

\newcommand{\bn}{{\bar n}}
\newcommand{\bea}{\begin{eqnarray}}
\newcommand{\eea}{\end{eqnarray}}

\newcommand{\bnP}{\bar {\cal P}}

\newcommand{\cP}{{\cal P}}

\newcommand{\mcdot}{\!\cdot\!}

\newcommand{\bt}{\bar{t}}

\newcommand{\SCETa}{\ensuremath{{\rm SCET}_{\rm I}}\xspace}
\newcommand{\SCETb}{\ensuremath{{\rm SCET}_{\rm II}}\xspace}

\newcommand{\plus}{\ensuremath{\! + \!}}
\newcommand{\minus}{\ensuremath{\! - \!}}

\newcommand{\shat}{\hat{s}}

\begin{document}
\setlength\baselineskip{17pt}


\preprint{ \vbox{ \hbox{MIT-CTP 3791} \hbox{CALT-68-2624} \hbox{MPP-2007-9}  
 \hbox{hep-ph/0703207} 
} 
}

\title{\boldmath 
Jets from Massive Unstable Particles:\\ Top-Mass Determination
} 

\vspace*{1cm}

\author{Sean Fleming} 
  \affiliation{Department of Physics, University of Arizona, Tucson,
  AZ 85721 \footnote{Electronic address: fleming@physics.arizona.edu}}  

\author{Andre H. Hoang}
  \affiliation{Max-Planck-Institut f\"ur Physik, 
  F\"ohringer Ring 6, M\"unchen, Germany, 80805 \footnote{Electronic address:
    ahoang@mppmu.mpg.de}}

\author{Sonny Mantry}
  \affiliation{California Institute of Technology, Pasadena, CA 91125
  \footnote{Electronic address:  mantry@theory.caltech.edu}} 

\author{Iain W.~Stewart\vspace{0.4cm}}
  \affiliation{Department of Physics, Massachusetts Institute of Technology, 
Boston, MA 02139 \footnote{Electronic address: iains@mit.edu}\vspace*{0.5cm}}
        


\begin{abstract}
 \vspace*{0.3cm}
  
 We construct jet observables for energetic top quarks that can be used to
 determine a short distance top quark mass from reconstruction in $e^+e^-$
 collisions with accuracy better than $\Lambda_{\rm QCD}$. Using a sequence of
 effective field theories we connect the production energy, mass, and top width
 scales, $Q\gg m\gg\Gamma$, for the top jet cross section, and derive a QCD
 factorization theorem for the top invariant mass spectrum. Our analysis
 accounts for: $\alpha_s$ corrections from the production and mass scales,
 corrections due to constraints in defining invariant masses, non-perturbative
 corrections from the cross-talk between the jets, and $\alpha_s$ corrections to
 the Breit-Wigner line-shape.  This paper mainly focuses on deriving the
 factorization theorem for hemisphere invariant mass distributions and other
 event shapes in $e^+e^-$ collisions applicable at a future Linear Collider. We
 show that the invariant mass distribution is not a simple Breit-Wigner
 involving the top width. Even at leading order it is shifted and broadened by
 non-perturbative soft QCD effects. We predict that the invariant mass peak
 position increases linearly with $Q/m$ due to these non-perturbative effects.
 They are encoded in terms of a universal soft function that also describes soft
 effects for massless dijet events. In a future paper we compute $\alpha_s$
 corrections to the jet invariant mass spectrum, including a summation of large
 logarithms between the scales $Q$, $m$ and $\Gamma$.

\end{abstract}

\maketitle

\newpage

\tableofcontents

\newpage
\section{Introduction}
\label{sectionintroduction}

Precise determinations of the top quark mass $m$ are among the most important
Standard Model measurements being carried out at the Tevatron, and being planned
at the Large Hadron Collider (LHC) and a future International Linear Collider
(ILC). A precise top mass determination is important for precision electroweak
constraints, as well as extensions to the standard model like minimal
supersymmetry~\cite{Heinemeyer:2003ud}. The present combined measurement from
the Tevatron is $m=171.4\pm 2.1$~GeV~\cite{Brubaker:2006xn,Heinson:2006yq} and
mainly relies on methods where a number of top-mass-dependent kinematical
quantities and observables are used in a global fit to determine the most likely
top quark mass.  For these fitting methods~\cite{Abe:1994st,Abazov:2004cs} the
observable most sentitive to the top quark mass is the top invariant
distribution.  It is obtained from reconstructing the total invariant mass of
the top decay products. At the Tevatron the invariant mass distribution is being
used in connection with other top mass dependent observables due to the limited
statistics.

In principle the reconstruction of the top invariant mass distribution
provides the most natural way to measure the top quark mass since the
peaked structure at the resonance is most closely related to the
notion of the mass of a propagating massive and unstable degree of
freedom. This method can be applied at the LHC and ILC where larger
statistics are available. Experimental studies have concluded that at
the LHC top mass measurements with uncertainties at the level of
1~GeV~\cite{Borjanovic:2004ce,Etienvre:2006ph} can be achieved, while
at the ILC even smaller uncertainties can be
expected~\cite{Chekanov:2002sa,Chekanov:2003cp}.  However, since the
top quark is a parton carrying non-vanishing color charge, its mass is
a priori not directly observable. In fact the top mass should be
considered as a renormalization scheme-dependent coupling in the QCD
Lagrangian rather than a physical object, just like the strong
coupling $\alpha_s$. As such, the top mass obtained from
reconstruction also depends on the method and prescription that is
used to define the top invariant mass since the latter is not a unique
physical quantity. In fact the notion of a physical particle whose
squared four-momentum is the mass does not apply to the top quark if
one asks for a precision in the mass value that is comparable to the
hadronic scale.  This is also reflected in a number of conceptual and
experimental issues for top quark mass determinations that are
associated with gluon radiation, underlying events, and the jet energy
scales -- effects that can never be fully eliminated for measurements
of the top quark mass from reconstruction.  Moreover certain top quark
mass renormalization schemes are more suitable for precision
measurements than others since the choice of scheme can affect the
higher order behavior of the perturbative corrections as well as the
organization of power corrections.  Suitable quark mass schemes are
compatible with the power counting and also lead to an optimal
behavior of the perturbative expansion. Such schemes can be identified
and defined unambiguously if the precise relation of the observable to
a given Lagrangian top quark mass scheme can be established.

For all jet based methods of top quark mass determination, and for
reconstruction in particular, these issues have been intrinsically
difficult to address in the past. Previous work has not provided a
coherent analytic framework in which perturbative and non-perturbative
effects could be described in a systematic manner. Considering the
expected precision for top quark mass measurements in the upcoming
experiments such a framework is imperative.
  
A top mass determination method where a systematic analytic framework
exists and where the relation between the Lagrangian top mass
parameter $m$ and the measured top mass can be established to high
precision is the threshold scan of the line-shape of the total
hadronic cross section in the top-antitop threshold region, $Q\approx
2m$, at a future Linear
Collider~\cite{Fadin:1987wz,Strassler:1990nw,Fadin:1991zw,Jezabek:1992np,Sumino:1992ai},
where $Q$ is the c.m.\,energy. In this case the system of interest is
a top-antitop quark pair in a color singlet state and the observable
is related to a comparatively simple counting measurement. The
line-shape of the cross section rises near a center of mass energy
that is related to a toponium-like top-antitop bound state with a mass
that can be computed perturbatively to very high
precision~\cite{Hoang:2000yr,Hoang:2000ib,Hoang:2001mm,Pineda:2006ri,Hoang:2004tg}
using non-relativistic QCD (NRQCD)~\cite{Bodwin:1994jh,Luke:1999kz} an
effective field theory (EFT) for nonrelativistic heavy quark
pairs. The short lifetime of the top quark, $\tau=1/\Gamma\approx
(1.5\,\mbox{GeV})^{-1}$, provides an infrared cutoff for all kinematic
scales governing the top-antitop dynamics and leads to a strong power
suppression of non-perturbative QCD effects. Experimental studies
concluded that theoretical as well as experimental systematic
uncertainties for this method are at a level of only
$100$~MeV~\cite{Peralta:etal,Martinez:2002st}. The most suitable top
quark mass schemes are the so-called threshold
masses~\cite{Hoang:2000yr}, which can be related accurately to other
short-distance mass schemes such as the running $\overline{\rm MS}$
mass.  Unfortunately, the threshold scan method cannot be used at the
LHC because the top-antitop invariant mass can only be determined with
a relative uncertainty of around 5\%~\cite{Beneke:2000hk}, which is
not sufficient to resolve the top-antitop threshold region.

\begin{figure}
  \centerline{ 
   \hspace{2cm}\includegraphics[width=12cm]{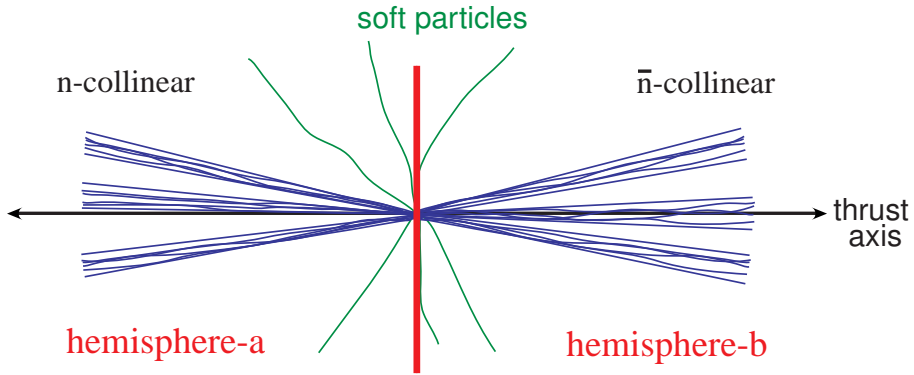}  
  } 
\caption{ Six jet event initiated by a top quark pair, $t\bar t\to bW \bar b
  W\to b qq' \bar b qq'$.  The plane separating the two hemispheres is
  perpendicular to the thrust axis and intersects the thrust axis at
  the interaction point. The total invariant mass inside each
  hemisphere is measured. Our analysis applies equally well to the
  lepton+jets and the dilepton channels (not shown).}
\label{fig:6topjet}
\end{figure}

In this work we use EFT's to provide, for the first time, an analytic framework
that can be applied to systematically describe the perturbative and
nonperturbative aspects of top quark invariant mass distributions obtained from
reconstruction. As a first step towards developing a detailed framework for the
LHC, we focus in this work on jets in a $e^+e^-$ Linear Collider environment at
c.m.~energies far above threshold $Q\sim 0.5-1$~TeV. For $e^+e^-$ collisions
strong interaction effects arising from the initial state can be neglected and
there is no need to identify or remove any `beam remnant' or underlying events.
Also, in the $e^+e^-$ framework it is easier to formulate shape variables like
thrust that control the jet-likeness and the soft dynamics of an event. We
consider the double differential top and antitop invariant mass distribution,
where each of the invariant masses, $M_t^2$ and $M_{\bar t}^2$, are defined from
all particles in each of the two hemispheres that are determined by the event's
thrust axis. In Fig.~\ref{fig:6topjet} we show a sketch of such an event. Other
invariant mass definitions, e.g.~based on $k_T$ algorithms and criteria to
identify jets from top and antitop decay can be employed as well. Our approach
also works for all-jet and lepton plus jet final states.
Our focus is to study the double differential invariant mass distribution in
the peak region close to the top mass, so that $M_t^2-m^2\sim m\Gamma$ and
$M_{\bar t}^2-m^2 \sim m\Gamma$. It is convenient to introduce
the 
shifted variables
\begin{eqnarray}
\label{massshell} 
 \hat s_{t, \bar{t}}\equiv \frac{s_{t, \bar{t}}}{m}
\equiv \frac{M_{{t,\bar{t}}}^2-m^2}{m} 
\sim \Gamma \ll m\,, 
\end{eqnarray}
because it is only the invariant mass distribution close to the peak
that we wish to predict.  Here the top width $\Gamma$ is setting a
lower bound on the width of the invariant mass distribution and the
shifted variable $\hat s_{t,\bar t}$ can also be larger than $\Gamma$
as long as $\hat s_{t,\bar t}\ll m$.  However, for simplicity we will often
write $\hat s_{t,\bar t}\sim \Gamma$ as we did in Eq.~(\ref{massshell}).

There are three relevant disparate scales governing the dynamics of the system,
\begin{eqnarray} \label{threescales} 
 Q \gg m \gg \Gamma > \Lambda_{\rm  QCD} \,.
\end{eqnarray}
This kinematic situation is characterized by energy deposits contained
predominantly in two back-to-back regions of the detector with opening angles of
order $m/Q$ associated to the energetic jets coming from the top quark decay and
collinear radiation. Frequently in this work we refer to the jets coming from
the top and antitop quark collectively as top and antitop jet, respectively, but
we stress that we do not require the jets from the top and antitop decay
products to be unresolved as pictured in Fig.~\ref{fig:6topjet} (for example one
can still identify a $W$ and do $b$-tagging). The region between the top jets is
predominantly populated by soft particles with energies of order of the hadronic
scale.

\begin{figure}
  \centerline{ 
   \includegraphics[width=12cm]{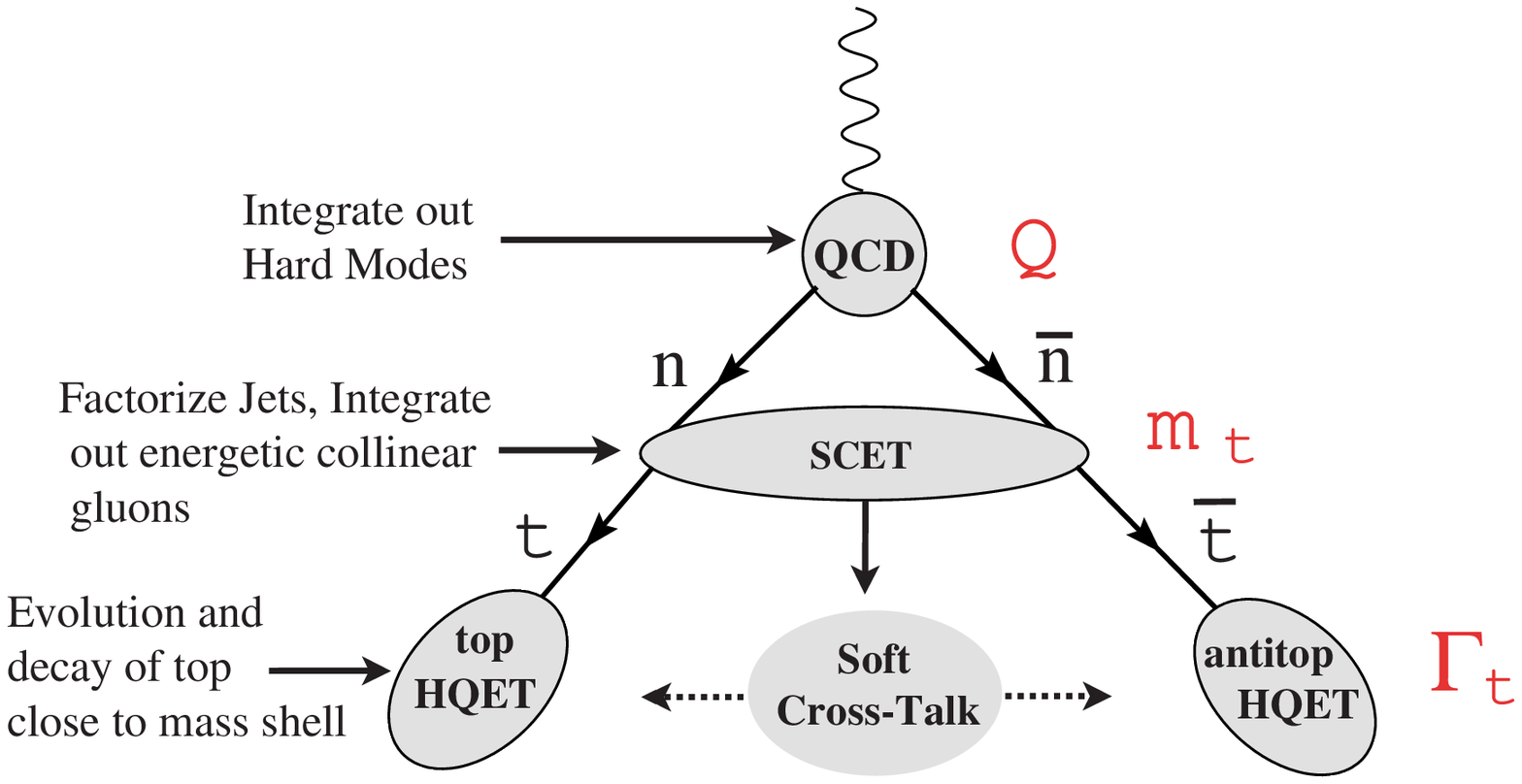}  
  } 
\caption{Sequence of effective field theories used to compute the
top/antitop invariant mass distribution in the peak region. }
\label{fig:efts}
\end{figure}

The EFT setup used to describe the dynamics in this kinematic
situation is illustrated in Fig.~\ref{fig:efts} and represents a
sequence of different EFT's.  The use of different EFT's is mandatory
to separate the various relevant physical fluctuations.  The high
energy dynamics for the top quarks at the scale $Q\gg m$ can be
described by quark and gluon degrees of freedom that are collinear to
the top and antitop jet axes, and by soft degrees of freedom that can
freely propagate between the jets. The appropriate EFT for this
situation is the Soft-Collinear Effective Theory
(SCET)~\cite{Bauer:2000ew,Bauer:2000yr,Bauer:2001ct,Bauer:2001yt} with
a nonzero top quark mass term~\cite{Leibovich:2003jd}, which
represents an expansion in $\lambda \sim m/Q\sim 0.2-0.3$.  The
leading order soft-collinear decoupling~\cite{Bauer:2001ct} properties
of SCET allows a factorization of the process into three sectors: top
jet dynamics, antitop jet dynamics, and dynamics of the soft cross
talk between the top and antitop jets, which corresponds quite
intuitively to the situation pictured in Fig.~\ref{fig:6topjet}. In
SCET the typical fluctuation of the jet invariant masses around the
top mass are still of order $m$, $\hat s_{t, \bar{t}}\sim m$.  Thus to
describe invariant masses in the peak region $\hat s_{t, \bar{t}}\sim
\Gamma$ the top and antitop jets are finally computed in Heavy-Quark
Effective Theory (HQET)~\cite{Manohar:2000dt} which represents an
expansion $\hat s/m$ and $\Gamma/m\sim 0.01$. We have in fact two
copies of HQET, one for the top and one for the antitop, plus soft
interactions between them. In these EFT's the top decay can be treated
as inclusive and is therefore described by the total top width term
$\Gamma$ that acts as an imaginary residual mass
term~\cite{Fadin:1987wz,Beneke:2004km}. Since HQET is usually
understood as being formulated close to the rest frame of the heavy
quark without the soft cross-talk interactions, we refer to these two
EFT's as boosted HQET's (bHQET's).\footnote{We adopt the acronym bHQET
in cases where we wish to remind the reader that the residual momentum
components of the heavy quark in the $e^+e^-$ c.m. frame are not
homogeneous, and that additional gluon interactions occur which are
not simply the soft gluons of standard HQET. }

At leading order in the expansion in $m/Q$ and $\Gamma/m$ we show that
the double differential invariant hemisphere mass distribution can be
factorized in the form
\begin{align} 
\label{FactThm}
  \bigg(\frac{d\sigma}{ dM_t^2\, dM_{\bar t}^2}\bigg)_{\rm hemi} &=
  \sigma_0 \: H_Q(Q,\mu_Q,\mu_m) H_m\Big(m,\frac{Q}{m},\mu_m,\mu\Big)\!
  \\
 &\times \int\! d\ell^+ d\ell^- B_+\Big(\hat s_t- \frac{Q\ell^+}{m},\Gamma,\mu\Big)\:
   B_-\Big(\hat s_{\bar t}-\frac{Q\ell^-}{m},\Gamma,\mu\Big)  
   S_{\rm hemi}(\ell^+,\ell^-,\mu)\,,\,\,\,\,\,\mbox{} \nn
\end{align}
where $\hat s_t$ and $\hat s_{\bar t}$ are defined in terms of
$M_{t,\bar t}^2$ in Eq.~(\ref{massshell}). The term $\sigma_0$ is a
normalization factor, and the factors $H_Q$ and $H_m$ are matching
corrections that are derived from matching and running in SCET and the
bHQET's, respectively. $H_Q$ and $H_m$ are independent of $\hat s_t$
and $\hat s_{\bar t}$ and do not affect the form of the invariant mass
distributions. The jet functions $B_\pm$ describe the QCD dynamics of
collinear radiation in the top/antitop direction, and the decay of the
top and antitop quarks near mass shell within the top/antitop jets.
They can be computed perturbatively at the scale $\mu\gtrsim\Gamma$ since
the top width $\Gamma$ provides an infrared cutoff from
hadronization. At tree level they are Breit-Wigner functions
\begin{align}
  B_\pm(\hat s,\Gamma) \, 
  &=  \,\frac{1}{\pi m} \: 
\frac{\Gamma}{\hat s^2 + \Gamma^2} \,+\,\ldots \,,
\end{align}
where the ellipses indicate QCD corrections that distort the
Breit-Wigners.  For the computation of the $B_{\pm}$ it is mandatory
to employ properly defined short-distance top mass schemes, to achieve
a well-behaved perturbative expansion. Finally, the soft function
$S_{\rm hemi}(\ell^+,\ell^- )$ describes the physics of the soft
nonperturbative gluons through which the top and antitop jets can
communicate. The low energy fluctuations of these soft gluons are not
cut off by the large top quark width. This can be intuitively
understood due to the lifetime dilation of the boosted top quarks. As
explained in Sec.~\ref{sectionefts}, using soft-collinear
factorization, we can show that the soft function is universal, namely
that the same function governs the low energy dynamics for massless
jets in the dijet
limit~\cite{Korchemsky:1998ev,Korchemsky:1999kt,Bauer:2002ie,Bauer:2002aj,Lee:2006nr}.
So, information on the form of $S_{\rm hemi}(\ell^+,\ell^-)$ can be
gained in a model-independent way from experimental data on massless
dijet events. The form of the factorization theorem in
Eq.~(\ref{FactThm}) is based on the same principles as the
factorization formula for massless dijet event
shapes~\cite{Korchemsky:1998ev,Korchemsky:1999kt,Bauer:2002ie,Lee:2006nr},
but it differs due to the need to treat massive quark jets and effects
related to the large top quark width.  We also use our results to
derive a factorization theorem for thrust and the heavy-jet mass event
shape for $t\bar t$ production in the peak region. These distributions
can also be used to measure the top-quark mass.

The convolution in Eq.~(\ref{FactThm}) shows that the observed hemisphere mass
distributions are inevitably distorted by the nonperturbative soft momentum
distribution, and that top and antitop jets can only interact indirectly through
exchange of different light-cone momentum components that are governed by the
soft function.  We can also show that for invariant masses $M_{t,\bar t}$ that
are defined through the identification of the jets from top and antitop decay,
which are determined from a $k_T$ jet algorithm, the same factorization formula
as in Eq.~(\ref{FactThm}) can be derived up to a different soft function.

We believe that the factorization approach proposed in this work, and the
factorization formula in Eq.~(\ref{FactThm}), represent advancements concerning
the following points:
\begin{itemize}
\item We give a well defined relation between a jet observable
sensitive to the top mass and the Lagrangian mass.  This allows the
definition of a short-distance top mass which we call the
``jet-mass''. Theoretically the jet-mass can be determined with a
precision better than $\Lambda_{\rm QCD}$, once the soft function
governing nonperturbative effects is known by other means. We expect
that the jet-mass will be useful for a broad range of observables
involving jets and parton showering from massive quarks.
\item The soft function appearing in the massive-jet factorization formula is
universal, and appears in massless dijet event shapes. This universality can
reduce the dependence of uncertainties in the top mass from
reconstruction on parton shower monte carlos and hadronization models.
\item The factorization approach opens up the possibility to systematically
construct top mass observables where nonperturbative effects are
suppressed.
\end{itemize}
While the focus of this paper is on $t\bt$ production at a $e^+e^-$ Linear
Collider, the main ideas and tools developed are general, and will also play an
important role for the environment of the LHC where a substantial number of top
events with large $p_T$ will be available. Other applications of our approach to
factorization of jets from massive particles may include processes such as
single top production~\cite{Cortese:1991fw,Willenbrock:1986cr}, W pair
production, or processes involving new colored unstable
particles~\cite{Butterworth:2002tt,Skiba:2007fw}. We briefly comment on these
applications in the summary.

The outline of the paper is as follows. In Sec.~\ref{sectionefts} we describe
the relevant EFT formalism for our computation. In Sec.~\ref{section3} we derive
the factorization theorem in SCET, introduce the hemisphere jet invariant
masses, and perform the factorization of mass effects in boosted HQET. The
result of the analysis in this section is the complete factorization theorem for
the double invariant mass distribution, and the extension to thrust and the
heavy-jet mass event shapes. In this section we also define the short-distance
jet-mass scheme.  In Sec.~\ref{section4} we study the factorization theorem
numerically at leading order, and discuss implications for top-mass
measurements.  We also display numerical results for the shape of the peak
region. In Sec.~\ref{sectionotheralgo} we discuss the relation between the
factorization theorem for the hemisphere invariant masses used in our work, to a
factorization theorem for the reconstruction method based on $k_T$~jet
algorithms employed in Refs.~\cite{Chekanov:2002sa,Chekanov:2003cp}.  Finally we
summarize and conclude in Sec.~\ref{section6}.  

This paper concentrates on the derivation of the factorization theorem, on field
theoretic issues, and on the basic phenomenological implications of our result.
Readers only interested in the final result may skip over the analysis in
sections~\ref{sectionefts} and \ref{section3}, and go directly to
section~\ref{section4}. In a future paper we present the computation of
$\alpha_s$ corrections to the jet invariant mass cross-section, and the
summation of large logarithms between the scales $Q, m,\Gamma$.

\section{The Effective Field Theories}
\label{sectionefts}

In this section we discuss the EFT's required to compute the double
differential invariant mass distribution $d^2\sigma/dM^2_t dM^2_{\bar
t}$ in the peak region.  The relevant energy scales are:
\begin{eqnarray}
  Q \gg m \gg \hat s_t\sim \hat s_{\bar t} \sim \Gamma \,,
\end{eqnarray} 
where the hatted $s$-variables were defined in terms of $M_t^2$ and $M_{\bar
  t}^2$ in Eq.~(\ref{massshell}).  Once radiative corrections are included,
large logarithms arise through ratios of the above energy scales, some of which
are double logs, and thus can be quite large.  For example $\Gamma/m\approx
1/120$, so $\ln^2(\Gamma/m)\approx 25$. It is obviously important to understand
the appearance of all large logs as accurately as possible, and to sum them
systematically. This summation is accomplished by matching onto a sequence of
EFTs and using renormalization group equations (RGE's).

Starting from QCD we first switch to the Soft Collinear Effective
Theory (SCET)~\cite{Bauer:2000yr,Bauer:2002nz,Bauer:2001yt,Bauer:2001ct}
for massive quarks and then to Heavy Quark Effective
Theory (HQET)~\cite{Eichten:1989zv,Isgur:1989vq,Isgur:1989ed,Grinstein:1990mj,Georgi:1990um}
combined with the unstable particle EFT 
method~\cite{Beneke:2003xh,Beneke:2004km,Hoang:2006pd,Beenakker:1999hi}.
This scheme includes systematically effects related to the large top
quark width, as well as interactions related to the soft cross-talk:
\begin{eqnarray}
\text{QCD} \longrightarrow \text{SCET} \longrightarrow \text{boosted-HQET with
  unstable heavy quarks}.
\end{eqnarray}
An intuitive picture which displays why this sequence of EFTs is relevant is
shown in Fig.~\ref{fig:efts}.  We are interested in events where the top quarks
are produced close to their mass shell as characterized by the condition in
Eq.~(\ref{massshell}). At the production scale $Q$, the invariant mass of the top
and antitop quarks can still fluctuate with $\hat s_{t, \bar{t}} \sim Q$ due to
it's interactions with hard gluons of characteristic momentum $p_h\sim Q$. In
the first step, when switching to SCET, these hard modes are integrated out and
we expand in $m/Q\ll 1$.  SCET makes it simple to
separate the physics associated with i) the top-jet, ii) the antitop jet, and
iii) the soft-cross talk between the jets. After the implementation of this
factorization theorem, each top jet and the soft-cross talk can be studied
independently in the field theory. The factorization theorem tells us how to tie
them together. Now in SCET the invariant mass of the top quark fluctuates with
$\hat s_{t,\bar{t}} \sim m$, so we still have to remove these large momentum
fluctuations to describe the desired kinematic region where $\Gamma
\sim\hat s_{t,\bar{t}}  \ll m$. Such invariant mass fluctuations are analogous to those
encountered in HQET for a bottom quark inside a $B$-meson
\begin{eqnarray}
  (mv +k)^2 -m^2 = 2 m v\cdot k + k^2 \sim 2 m \Lambda_{\rm QCD}  \,,
\end{eqnarray}
with the difference that for the unstable top quark $v\cdot k\to v\cdot k +
i\Gamma/2$.  Since top-quarks decay before they have a chance to hadronize, the
top-width $\Gamma$ adopts the role $\Lambda_{\rm QCD}$ plays for the $B$ meson.
Keeping in mind that the tops are highly boosted and unstable, we actually match
onto two boosted versions of HQET, one for the top and one for the antitop.  A
discussion of the necessary SCET and HQET theoretical ingredients is given in
the following subsections.

\subsection{SCET with Masses}\label{mass-scet}

SCET is an effective theory describing the interactions of soft and collinear
particles, which are characterized by the scaling of their momenta. In
this framework it is convenient to introduce the four-vectors
\begin{eqnarray}
n^\mu = (1,\vec n), \qquad\quad  \bn^\mu=(1,-\vec n), 
\end{eqnarray}
where $\vec n$ can be thought of as the direction of the top jet and
$-\vec n$ as the direction of the antitop jet ($\vec n^2 =1$, $n^2=0$, $\bn^2=0$) .
Any momentum can then be decomposed as
\begin{eqnarray}
p^\mu = n\cdot p \>\frac{\bn^\mu}{2} + \bn \cdot p\> \frac{n^\mu}{2} +
p_\perp^\mu \,,
\end{eqnarray}
and we denote momentum components in this light cone basis as
$(p^+,p^-,p_\perp)=(n\cdot p,\bn \cdot p, p_\perp)$. The square of the
momentum vector $p^\mu$ then reads $p^2=p^+ p^-+p_\perp^2$.
It is also convenient to denote the momentum of collinear particles in
the $\vec{n}$ and $-\vec{n}$ directions by the subscripts $n$ and $\bn$
respectively, which corresponds to the large energy modes in the
corresponding jets. Thus we have collinear labels
\begin{align}
  & n\ \ \text{for the top-jet,}\  &\bn& \ \ \text{for the antitop-jet} \,.
\end{align}
The momentum of soft particles that communicate between the jets will be denoted
by a subscript $s$.  We also have mass-modes that are required in order to
describe certain top-quark vacuum polarization loops. The momenta of the
collinear, mass, and soft modes\footnote{ In some factorization theorems it is
  necessary to distinguish between soft and ultrasoft particles, and between two
  versions of SCET: called \SCETa and \SCETb.  In this paper we only deal with
  \SCETa with ultrasoft gluons. For simplicity we will therefore simply use the
  term soft modes. For modes with momenta $p^\mu\sim (m,m,m)$ that are specific
  to the massive SCET theory, we use the term ``mass-modes''.}  have the typical
scalings shown in table~\ref{table_fields} in the SCET column, where here
$\lambda$ is the small expansion parameter. 
\begin{table}[t!]
\begin{center}
\begin{tabular}{|crl|rl|}
\hline
  & 
  \multicolumn{2}{c}{SCET [$\lambda\sim m/Q\ll 1$] }  
  \vline  &   \multicolumn{2}{c}{bHQET [$\Gamma/m\ll 1$] } \vline
   \\  \hline 
  & $n$-collinear ($\xi_n$, $A_n^\mu$)
    & \hspace{0.2cm}  $p_n^\mu \!\sim\! Q(\lambda^2,1,\lambda)$ 
  & $n$-ucollinear  ($h_{v_+}$, $A_{+}^\mu$)
    & \hspace{0.2cm}  $k^\mu\!\sim\! \Gamma(\lambda,\lambda^{-1},1)$  \\
  & $\bn$-collinear ($\xi_\bn$, $A_\bn^\mu$)
      & \hspace{0.2cm}  $p_\bn^\mu\!\sim\! Q(1,\lambda^2,\lambda)$ 
  & $\bn$-ucollinear ($h_{v_-}$, $A_{-}^\mu$) 
      & \hspace{0.2cm}  $k^\mu\!\sim\! \Gamma(\lambda^{-1},\lambda,1)$ \\
 & mass-modes ($q_m$, $A_m^\mu$) 
     & \hspace{0.2cm}  $p_m^\mu\!\sim\! Q(\lambda,\lambda,\lambda)$ & & 
 \\ 
  Crosstalk: & soft
  ($q_{s}$, $A^\mu_{s}$)
   & \hspace{0.2cm}  $p_s^\mu\!\sim\! Q (\lambda^2,\lambda^2,\lambda^2)$ 
 & same soft ($q_{s}$, $A^\mu_{s}$) 
   & \hspace{0.2cm}  $p_s^\mu\!\sim\! (\Delta,\Delta,\Delta)$
     \\
\hline
\end{tabular}
\end{center}
\vskip-0.4cm
\caption{ \label{table_fields} Summary of the  fields required in SCET and
   bHQET. The first field in each bracket is a quark, and the second
   is a gluon. The scaling of momentum components is given for
   $(p^+,p^-,p^\perp)$. After
   factorization, the soft fields on the last line generate a cross-talk theory 
   that communicates with collinear
   fields in both SCET and bHQET through two kinematic variables. $\Delta$ is
   the scale for the soft modes.}
\end{table}
A particle with components scaling as $(\lambda^2,1,\lambda)$ has a small
$\perp$-momentum relative to its energy, and is said to be collinear to the
$n^\mu$ direction etc. Both $\lambda$ and the hard scale $Q$ have a size that
depends on the particular process under study.  For example, in $B\to X_s
\gamma$ the hard scale is the $b$-quark mass $m_b$, and the expansion parameter
is $\sqrt{{\Lambda_{QCD}}/{m_b}}$. For pair production of top jets, the hard
scale $Q$ is the center of mass energy, and the SCET expansion parameter is
\begin{eqnarray}
\lambda \sim \frac{m}{Q} \,.
\end{eqnarray}
It follows that the typical virtuality  of the collinear, mass, and soft modes in SCET satisfy
\begin{eqnarray}
p_n^2 \sim p_{\bn}^2 \sim  m^2, \qquad p_m^2\sim m^2, 
 \qquad \text{and}\quad p_s^2 \sim \frac{m^4}{Q^2}.
\end{eqnarray}
Since $m^4/Q^2 \gg \Lambda _{QCD}^2$, the soft modes in this theory still
contain perturbative components as well as the underlying non-perturbative
dynamics at smaller scales.  Using $m =171\, \text{GeV}$ this is true for
$Q\lesssim 40\,\text{TeV}$ i.e.\,for any conceivable c.m.\,energy of a future
Linear Collider.  The soft particles correspond to modes with wavelengths that
allow cross talk between the two jets.  In addition, at two-loop order the soft
gluons in SCET interact with virtual top-quarks which are described by the
mass-modes indicated in table~\ref{table_fields}. These mass-modes do not
interact directly with the collinear fields and only appear as virtual effects
for our observable, because we only consider cases where $s_{t,\bar t}\ll Q m$.
(As discussed below Eq.~(\ref{state}).)  In addition we have virtual collinear
top-quarks that can interact with the collinear particles through the collinear
Lagrangian.  The $n$-collinear, $\bn$-collinear, mass modes, and soft modes are
described by separate quark and gluon fields which are also listed in
table~\ref{table_fields}. Hard modes involving momenta $p^\mu\sim Q$ have
already been integrated out when QCD is matched onto SCET.

At leading order the SCET Lagrangian for collinear particles in
different directions can be written as a soft Lagrangian plus a sum of
collinear terms~\cite{Bauer:2002nz}, ${\cal L}^{(0)}={\cal
L}_s+\sum_{n_i} {\cal L}^{(0)}_{n_i} $. The sum satisfies the
constraint $n_i\cdot n_j\gg \lambda^2$ for $i\ne j$, with the choice
of $\lambda$ determining what is meant by distinct collinear
directions. The collinear particles in different sectors only interact
via soft gluon exchange or interactions in external operators.
When the $\perp$-momentum of the collinear particles is of the same
size as the quark mass the result for the leading order collinear
Lagrangian~\cite{Bauer:2000yr,Bauer:2001ct} must include the quark
mass terms derived in Ref.~\cite{Leibovich:2003jd} (see also
Ref.~\cite{Rothstein:2003wh}).  The collinear quark Lagrangian for the
direction $n$ is therefore given by
\begin{eqnarray} 
\label{Lscet}
{\cal L}^{(0)}_{qn} = \bar{\xi}_n\Big [ i n\cdot D_s + gn \cdot
A_n + (i \Dslash _c^\perp \!-\! m) W_n\frac{1}{\bn\mcdot {\cal P}}W_n^\dagger
(i\Dslash _c^\perp \!+\! m) \Big ] \frac{\bnslash}{2} \xi_n  \,,
\end{eqnarray}
with $D_c^\perp \sim m \gg D_s^\perp$. There is also an $n$-collinear Lagrangian
for gluons~\cite{Bauer:2001yt}.  Here the soft and collinear covariant
derivatives are
\begin{eqnarray}
 i D_s^\mu = i \partial ^\mu + g A_s^\mu ,  \qquad
 i D_c^\mu = {\cal P}^\mu + g A_n^\mu \,,
\end{eqnarray}
where ${\cal P}^\mu$ is a label operator picking out the large
collinear momentum  of order $Q$ and $Q\lambda$ of a collinear
field~\cite{Bauer:2001ct}, while the partial derivative 
acts on the residual momentum components $\partial ^\mu\sim\lambda^2$.
The term $W_n$ is the momentum space Wilson line built out of collinear
gluon fields
\begin{eqnarray}
 W_n (x) = \sum _{\text{perms}} \text{exp} \>\big ( - \frac{g}{\bar{{\cal P}}}
   \bn \cdot A_{n}(x) \> \big ) \,.
\end{eqnarray}
We also note that Eq.~(\ref{Lscet}) is the bare Lagrangian. In particular, any
mass definition can be chosen for $m$ through an appropriate renormalization
condition without breaking the power-counting. At ${\cal O}(\alpha_s)$ these
mass-schemes are the same as those in QCD~\cite{Chay:2005ck}, because the
self-energy graphs are directly related.
\begin{figure}
  \centerline{ 
   \hspace{2cm}\includegraphics[width=10cm]{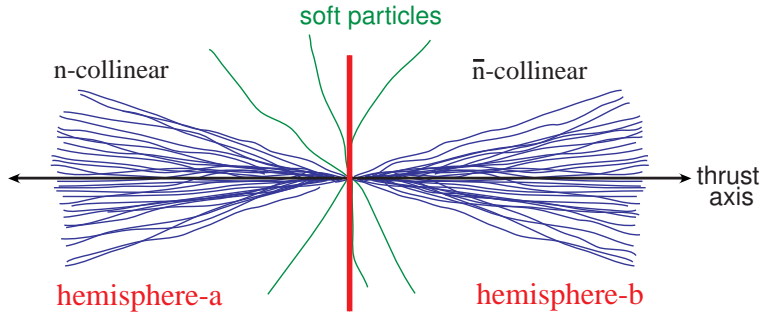}  
  } 
\caption{Final state jets in SCET for stable top-quarks with invariant mass $\sim m^2$. The 
invariant mass is restricted and the top-decay products become explicit by matching onto HQET. }
\label{fig:topjet}
\end{figure}

An example of an external operator that connects different collinear
sectors is the jet production current, which couples to the $\gamma^*$
or $Z^*$. In QCD the production matrix element is $\langle X | {\cal
J}_{a,v}^\mu | 0 \rangle$ where $\langle X|$ is the final state. The
required vector and axial currents are given by
\begin{align}
\label{QCDcurrents}
{\cal J}^\mu_v(x) &=  \bar{\psi}(x) \gamma^\mu \psi(x) \,,
 & {\cal J}^\mu_a(x) & =  \bar{\psi}(x) \gamma^\mu \gamma_5 \psi(x) \,,
\end{align}
and for convenience we will adopt the short-hand notation ${\cal J}^\mu_i
=\bar\psi(x) \Gamma_i^\mu \psi(x)$. The matching relation of these QCD
currents to SCET currents 
is given by the convolution formula~\cite{Bauer:2000yr}
\begin{eqnarray}
\label{currentmatch}
 {\cal J}^\mu_i(0) = \int\!\! d\omega\, d\bar\omega\, C(\omega,\bar\omega,\mu) 
  J^{(0)\mu}_i(\omega,\bar \omega,\mu) \,,
\end{eqnarray}
where $C$ contains short-distance dynamics at the scale $Q$, while $
J_i^{(0)\mu}$ describes fluctuations at all longer distance scales. In the
presence of multiple collinear fields, as well as modes scaling like our
mass-modes and soft-modes, the construction of currents in SCET has been
discussed in great detail in Ref.~\cite{Bauer:2002nz}. Interactions between the
mass-modes and the collinear-modes produce offshell particles, which when
integrated out leave residual interactions through Wilson lines in the SCET
current.  The SCET production current at leading order in $\lambda$ is given by
\begin{eqnarray}
\label{currentscet}
 J^{(0)\mu}_i(\omega,\bar \omega,\mu)
   = \bar \chi_{n,\omega}(0) S_n^\dagger \Gamma^\mu_i S_\bn \chi_{\bn,\bar\omega}(0) \,,
\end{eqnarray}
where $\chi_{n,\omega}(0) = \delta(\omega- \bn\mcdot \cP) (W_n^\dagger
\xi_n)(0)$ and $\chi_{\bn,\bar\omega}(0) = \delta(\bar\omega- n\mcdot \cP)
(W_\bn^\dagger \xi_\bn)(0)$. The mass-mode Wilson lines $S_n^\dagger$ and
$S_\bn$ will be described below. Here the $(0)$ indicates that the fields are at
coordinate $x^\mu=0$, and we recall that this $x^\mu$ dependence carries
information about the residual momenta at the scale $Q\lambda^2=m^2/Q$. The
dependence on larger momenta is encoded in labels on the collinear
fields~\cite{Bauer:2001ct}, and, for example, $\delta(\omega-\bn\cdot P)$ forces
the total minus-label-momentum of $(W^\dagger_n \xi_n)$ to be $\omega$. We also
use the notation $\chi_{n} = (W_n^\dagger \xi_n)$ and $\chi_{\bn} =
(W_\bn^\dagger \xi_\bn)$.

One can decouple the soft and collinear modes in ${\cal L}_{qn}^{(0)}$ by
performing a field redefinition on collinear fields~\cite{Bauer:2001yt}
\begin{eqnarray}  \label{fd}
  \xi _{n} \to Y_n \xi _{n} \,, \qquad
  A_{n}^\mu \to Y_n\, A_{n}^{\mu}\, Y_n^\dagger \,,
\end{eqnarray}
where $Y_n$ is a soft Wilson line
\begin{align} \label{Yn}
 Y_n(x) &=  \overline {\rm P} \:
   \exp\Big(-i g\! \int_{0}^\infty \!\!\!ds\, n\mcdot A_{s}(ns\!+\! x) \Big) 
    \,.
\end{align}
This gives
\begin{align}
 Y_n^\dagger(x) &=   {\rm P} \, 
   \exp\Big(i g\! \int_{0}^\infty \!\!\!ds\, n\mcdot A_{s}(ns\!+\!x) \Big) \,,
\end{align}
which satisfies $Y_n^\dagger Y_n=1$.  For two-jet production the factorization
is most transparent~\cite{Bauer:2002ie} with the reference point $s_0=\infty$
shown in Eq.~(\ref{Yn}). The gluon fields are either antipath-ordered (for
$\overline{\rm P}$) or path-ordered (for ${\rm P}$).  We use the same Wilson
line for both the quark and antiquark parts of $\xi_n$.  Another possibility is
to make different field redefinitions on the particle and antiparticle parts of
the fields~\cite{Chay:2004zn}.  In fact, all results are independent of the
choice of reference point in the field redefinition; the path is determined
entirely by changes the field redefinition induces on the operators and the
interpolating fields for the states~\cite{Arnesen:2005nk}.  

The mass-mode Wilson line $S_n(x)$ is defined in an identical manner to
Eq.~(\ref{Yn}), but with $n\cdot A_s\to n\cdot A_m$. In order to avoid double
counting with the effects contained in the soft Wilson lines the mass-mode
$A_m^\mu$ fields are defined with zero-bin subtractions for the
soft-region~\cite{Manohar:2006nz}, and we have mass-mode top-quarks $\psi_m$
with a mass $m$.  Any graphs with mass-mode gluons that do not involve a
top-bubble from $\psi_m$ fields are exactly canceled by these zero-bin
subtractions. Thus the mass-modes only contribute in these vacuum polarization
graphs. The soft-gluons can also couple to the $\psi_m$ fields, however they do
so with a multipole expansion, and therefore do not inject momentum into the
closed $\psi_m$ loop.

After the change of variable in Eq.~(\ref{fd}) the leading order SCET collinear
quark Lagrangian and current become
\begin{align}
{\cal L}^{(0)}_{nq}  &= \bar{\xi}_n\Big [ i n\cdot \partial + g n \cdot
A_n + (i \Dslash _c^\perp \!-\! m) W_n\frac{1}{\bar{{\cal P}}}W_n^\dagger
(i\Dslash _c^\perp \!+\! m) \Big ] \frac{\bnslash}{2} \xi_n  
 \,,\nn\\
 J^{(0)\mu}_i &=  \overline \chi_{n,\omega} Y_n^\dagger 
  S_n^\dagger \Gamma_i^\mu S_\bn
  Y_\bn \chi_{\bn,\bar\omega}(0)\,,
\end{align}
where we used the property $Y^\dagger_n \> n\cdot D_s \> Y_n = n\cdot
\partial$.  The only coupling of the soft gluon to the collinear quark
was through $i n\cdot D_s$ which is no longer present (and a similar
property occurs in the collinear gluon action).  These soft couplings
reappear as Wilson lines in the current as shown above.  Hence we have
achieved soft-collinear decoupling in the Lagrangian and the current.

In the two-jet process we wish to factorize we must also consider the
transformation property of the state $|X\rangle$ under
Eq.~(\ref{fd}). For manipulations in the factorization theorem for
two-jet production we decompose the state into collinear and soft
pieces,
\begin{align} \label{state}
  \langle X\ | =  \langle X_n X_\bn X_s | \,.
\end{align}
Note that this decomposition is only valid for the states we are interested in
for describing the dijet region, not for a general state in QCD.  Since there is
always at least one $n$-collinear and one $\bn$-collinear particle, we do not
consider any mass-modes, $X_m$, in these states either.  The presence of a mass mode
would induce an invariant mass $p_X^2 = (p_n+p_m)^2 \simeq p_n^- p_m^+ \simeq Q
m \gg m^2$, which would make it impossible to satisfy the invariant mass
condition required to study the peak region.  Therefore the mass-modes will only
appear as virtual contributions.  The collinear states $\langle X_n|$ and
$\langle X_\bn |$ are a color triplet and color antitriplet, just like a quark
and antiquark state.  Therefore, we must consider how these collinear states
transform under the change in the action induced by Eq.~(\ref{fd}).  However,
because these color triplet states can be derived from the out states at large
time, $t\to \infty$, they are not affected by the field redefinition with reference
point at $\infty$~\cite{Arnesen:2005nk}.  With the current at $x$, we therefore
have
\begin{align} \label{melt}
  \langle X | {\rm T}\big\{ \overline \chi_{n,\omega} S_n^\dagger \Gamma S_\bn
   \chi_{\bn,\bar\omega} \big\} | 0 \rangle \to 
    \langle X | {\rm T} \big\{ \overline \chi_{n,\omega} Y_n^\dagger\, S_n^\dagger\, 
     \Gamma\, S_\bn\, Y_\bn \chi_{\bn,\bar\omega} \big\} | 0 \rangle \,.
\end{align}
Here the ${\rm T}$ reminds us to keep the proper time-ordering of the
$A^a(x)$ gluon fields in the $Y$'s. There is no ordering issue between
fields in $Y_\bn$ with 
those in $Y_n^\dagger$, since they are space-like separated and
commute~\cite{Bauer:2002ie}. We also need the complex conjugate of
Eq.~(\ref{melt}) for the matrix element, which is
\begin{align} \label{cmelt}
  \langle 0 | {\rm T}\big\{ \overline \chi_{\bn,\bar\omega} S_\bn^\dagger \Gamma
  S_n \chi_{n,\omega} \big\} | X \rangle 
   \to 
  \langle 0 | \overline {\rm T}\big\{ \overline \chi_{\bn,\bar\omega}
  Y_\bn^\dagger\, S_\bn^\dagger\, \overline
    \Gamma\, S_n\, Y_n \chi_{n,\omega} \big\} | X \rangle
  \,,
\end{align}  
where $\overline {\rm T}$ is anti-time-ordering.
Note that
\begin{align} \label{TY}
 {\rm T}\  (Y_\bn)^T 
  &= (Y_\bn^\dagger)^* 
   = \overline {Y_\bn}^\dagger
   =   {\rm P} \: \exp\Big( i g\! \int_{0}^{\infty} \!\!\!ds\, 
      \bn\mcdot \overline {A}_{s}(\bn s\!+\! x) \Big) 
  \,, \nn\\
  {\overline {\rm T}}\ (Y_\bn^\dagger)^T 
  &= (Y_\bn)^* 
   = \overline {Y_\bn}  
   =   \overline {\rm P}\: \exp\Big(-i g\! \int_{0}^{\infty} \!\!\!ds\, 
      \bn\mcdot \overline {A}_{s}(\bn s\!+\!x) \Big) 
   \,,
\end{align}
where $\overline {A}_{s} = A^A_{s} \overline {T}^A$, with $\overline {T}^A = -
(T^A)^T$ the generator for the $\overline 3$ representation, and the superscript
$T$ is the transpose with respect to the color indices of the fundamental
representation.  If we switch to these barred Wilson lines then the
time-ordering and anti-time-ordering becomes redundant. Eq.~(\ref{TY}) applies equally
well for the $S$ Wilson lines. Considering the squared matrix element for the
cross-section we find
\begin{align} \label{melt2}
 & \langle 0 | \overline {\rm T}\big\{ \overline \chi_{\bn,\bar\omega'}
    Y_\bn^\dagger\, S_\bn^\dagger\, \overline
    \Gamma\, S_n\, Y_n \chi_{n,\omega'} \big\} | X \rangle 
   \langle X | {\rm T} \big\{ \overline \chi_{n,\omega} Y_n^\dagger\, S_n^\dagger\,
     \Gamma\, S_\bn\, Y_\bn \chi_{\bn,\bar\omega} \big\} | 0 \rangle\nn\\[4pt]
  &=  \langle 0 |  \overline \chi_{\bn,\bar\omega'}^a  (\overline {Y}_\bn)^{ba}
  \, (\overline {S}_\bn)^{b'b}
    (\overline\Gamma\, S_n Y_n \chi_{n,\omega'})^{b'}  | X \rangle 
   \langle X | (\overline \chi_{n,\omega} Y_n^\dagger\, S_n^\dagger 
     \Gamma)^{c'}\, (\overline {S}^\dagger_\bn)^{cc'}
    (\overline {Y}^\dagger_\bn)^{dc} \chi_{\bn,\bar\omega}^d  | 0
     \rangle \nn\\
   &=  {\cal M}(m,\mu) \ 
   \langle 0 |  \overline \chi_{\bn,\bar\omega'}^a  (\overline {Y}_\bn)^{ba}
  \,  (\overline\Gamma\, Y_n \chi_{n,\omega'})^{b'}  | X \rangle 
   \langle X | (\overline \chi_{n,\omega} Y_n^\dagger\, 
     \Gamma)^{c'}\, 
    (\overline {Y}^\dagger_\bn)^{dc} \chi_{\bn,\bar\omega}^d  | 0
     \rangle
  \,,
\end{align}
where $a,b,c,d,b',c'$ are color indices. The decoupling of soft gluons in
Eq.~(\ref{melt2}) is identical to that in massless two-jet production, and
ignoring the mass-mode Wilson lines the discussion above agrees with the SCET
derivation in Ref.~\cite{Bauer:2002ie}, as well as the original derivation in
Refs.~\cite{Korchemsky:1994is,Korchemsky:2000kp}. To obtain the last line in
Eq.~(\ref{melt2}) we note that the Dirac structures $\Gamma$ and
$\overline\Gamma$ are color singlets, and that the mass-mode Wilson lines can be
separated into vacuum matrix elements since there are no mass-modes in the
states. Furthermore
\begin{align}
  \langle 0 | (\overline {S}_\bn)^{b'b} (S_n)^{b'a} | 0 \rangle &= 
   \frac{\delta^{ba}}{N_c}  \langle 0 | (\overline {S}_\bn)^{b'a'} (S_n)^{b'a'} | 0 \rangle\,,
\end{align}
with an analogous result for $\langle 0 | (S_n^\dagger)^{ac'} (\overline
{S}^\dagger_\bn)^{cc'} | 0\rangle$, so this contracts the color indices on
either side of the product of soft Wilson-line factors. Thus defining 
\begin{align}
 {\cal M}(m,\mu) &\equiv \frac{1}{N_c^2}  \big| \langle 0 | \overline
 {S}_\bn^{ab} S_n^{ab} | 0 \rangle \big|^2 \,,
\end{align} 
we are left with the matrix element shown on the last line of Eq.~(\ref{melt2}).
Here ${\cal M}(m,\mu)=1+{\cal O}(\alpha_s^2)$.

The soft-collinear decoupling property is crucial to organizing the physics of
the massive two-jet problem.  As we will show in Sec.~\ref{section3}, there
is a factorization theorem in SCET that decouples the soft and collinear modes
at leading order which allows us to study the physics of each jet
independently.  The cross-talk is confined to a simple top mass
independent vacuum matrix element involving the
$Y$ Wilson lines,
\begin{align}\label{Yme}
 \langle 0 | (\overline {Y}_\bn)^{ba} \, 
    (Y_n)^{bc}  | X_s \rangle 
   \langle X_s | (Y_n^\dagger)^{cd}\, (\overline {Y}^\dagger_\bn)^{ad}   | 0
   \rangle
   \,,
\end{align}
which agrees with the corresponding soft matrix element for massless
quark production~\cite{Korchemsky:2000kp,Lee:2006nr,Bauer:2003di} and
which will eventually determine the soft function $S_{\rm
hemi}(\ell^+,\ell^-)$ to be used in Eq.~(\ref{FactThm}). As we also
show in Sec.~\ref{section3}, the precise definition of the soft
function $S$ depends on the prescription that is used how the momenta
of soft particles enter the top and antitop invariant masses $\hat
s_t$ and $\hat s_{\bar t}$, respectively. In the next subsection we
describe how the matrix element in Eq.~(\ref{Yme}) is modified when we
integrate out the top-quark mass.

Finally, in SCET because the top-quark mass $m$, and mass of the $W$-boson,
$m_W$, are still low energy scales, the decay of an $n$-collinear top-quark is
simply described by the full electroweak interaction,
\begin{eqnarray} \label{Lnew}
  {\cal L}_{ew} = \frac{g_2}{\sqrt{2}}\, \bar b W_\mu^- \gamma^\mu P_L t 
  + \frac{g_2}{\sqrt{2}}\, \bar t W_\mu^+ \gamma^\mu P_L b \,,
\end{eqnarray}
where $G_F=\sqrt{2} g_2^2/(8m_W^2)$ is the Fermi constant. This treatment is
consistent since we can treat the top decay as fully inclusive up to ${\cal
  O}(m^2/Q^2)$.

The collinear $A_n$ and $A_\bn $ gluons in SCET can induce
fluctuations $\hat s_t, \hat s_{\bar t}\sim m$. Once we restrict
ourselves to events with $\Gamma\sim  \hat s_t, \hat s_{\bar t}  \ll
m$, i.e.\,we force the top quark and antiquark to remain close to
their mass shell, the situation looks very much like two distinct
copies of HQET in boosted frames. There was nothing special in the
dynamics that sets the scale $m^2/Q$ for the soft interactions, and so
we call $\Delta$ the scale that controls the soft-cross talk. In the
field theory $\Delta$ will be defined as the scale where we model or
fit the primordial soft function. Generally we will take $m\gg
\Delta \sim \Gamma$, although any value $\Delta > \Lambda_{\rm QCD}$ can be
considered. So we must switch from SCET onto these HQET theories, and also
consider what happens to the decay interaction in Eq.~(\ref{Lnew}). We describe
the boosted HQET theories in detail in the next section, and we also discuss how
the soft cross-talk interactions remains active when the fluctuations at the
top mass scale $m$ are integrated out.

Since the above Lagrangians and currents are LO in $\lambda$, it is
natural to ask about the role of power corrections.  As it turns out,
higher order Lagrangians and currents give corrections to our analysis
at ${\cal O}(\alpha_s m/Q)$, ${\cal O}(\Delta/Q)$, ${\cal
O}(m^2/Q^2)$, or ${\cal O}(\Gamma/m)$. The absence of ${\cal O}(m/Q)$
implies that the $m/Q$ expansion does not significantly modify the
top-mass determination.  The leading action contains all $m/Q$
corrections that do not involve an additional perturbative gluon, so
the corrections are ${\cal O}(\alpha_s m/Q)$. We have also verified that at
tree level the $m/Q$ corrections to the SCET
current~\cite{Rothstein:2003wh} vanish when contracted with the
leptonic tensor. Furthermore, many of the higher order $m/Q$ corrections
have the form of normalization corrections, and thus do not change the
shape of the invariant mass distribution.  Subleading soft
interactions are ${\cal O}(\Delta/Q)$.  The interplay of our
hemisphere invariant mass variable with the top decay can induce
${\cal O}(m^2/Q^2)$ corrections, as we discuss later on. Finally there
will be power corrections of ${\cal O}(\Gamma/m)$ in bHQET.


\subsection{Boosted HQET with Unstable Particles and Soft Cross-Talk}
\label{buHQET}


{\it Boosted Heavy Quarks.}
HQET \cite{Georgi:1990um, Eichten:1989zv, Grinstein:1990mj, Isgur:1989vq,
  Isgur:1989ed} is an effective theory describing the interactions of a heavy
quark with soft degrees of freedom, and also plays a crucial role for jets
initiated by massive unstable particles in the peak regions close to the heavy
particles mass shell.  The momentum of a heavy quark interacting with soft
degrees of freedom can be written as
\begin{eqnarray}
\label{HQETmomdecomp}
p ^\mu = m v^\mu + k^\mu , 
\end{eqnarray} 
where $k^\mu$ denotes momentum fluctuations due to interactions with
the soft degrees of freedom and is much smaller than the heavy quark
mass $|k^\mu | \ll m$.  Also typically $v^\mu\sim 1$ so that we are
parametrically close to the top quark quark rest-frame, $v^\mu=(1,\vec
0)$.  

In the top-quark rest frame we have $\Gamma \lesssim k^\mu \ll m$, where $k^\mu$
refers to momentum fluctuations of the top due to interactions with gluons
collinear to its direction, which preserve the invariant mass conditions $\Gamma
\sim \hat s_t,\hat s_{\bar t} \ll m$. For our top-quark analysis, the center of
mass frame is the most convenient to setup the degrees of freedom.  In this
frame the gluons collinear to the top-quark which preserve the invariant mass
condition will be called {\it ultra-collinear (ucollinear)} in the $n$
direction. A different set of $\bn$-ucollinear gluons interact with the antitop
quark which moves in the $\bn$ direction.  The leading order Lagrangian of the
EFT describing the evolution and decay of the top or antitop close to it's mass
shell is given by
\begin{align} 
\label{LbHQET}
 {\cal L}_{+} &=
\bar{h}_{v_+} \big( i v_+ \cdot D_+ - \delta m + \frac{i}{2} \Gamma \big) h_{v_+ } , 
 &{\cal L}_{-} &=
\bar{h}_{v_-} \big( i v_- \cdot D_- -\delta m+  \frac{i}{2} \Gamma \big) h_{v_-} , 
\end{align}
where the $+$ and $-$ subscripts refer to the top and antitop sectors
respectively, and $iD_\pm^\mu = i\partial^\mu+ g A_\pm^\mu$.  These HQETs
represent an expansion in $\Gamma/m$.  The HQET field $h_{v_+}$ annihilates top
quarks, while $h_{v_-}$ creates antitop quarks.  In the c.m. frame the
components of $k^\mu$ are no longer homogeneous in size, and $v_\pm^\mu \ 
\slash\!\!\!\!\!\!\sim 1$.  Instead for the $(+,-,\perp)$ components we have
\begin{eqnarray}
\label{BHQETres}
v^\mu_+ &=& \bigg( \frac{m}{Q} , \frac{Q}{m}, \mathbf{0}_\perp \bigg),
\qquad\quad
k^\mu_+ \sim  \Gamma\,\bigg(\frac{m}{Q}, \frac{Q}{m}, 1 \bigg),
\\
v^\mu_- &= & \bigg( \frac{Q}{m}, \frac{m}{Q} , \mathbf{0}_\perp \bigg), 
\qquad\quad
k^\mu_- \sim \Gamma\bigg(\frac{Q}{m}, \frac{m}{Q}, 1\bigg) 
\nonumber .
\end{eqnarray}
Note that the $\Gamma$ in Eq.~(\ref{BHQETres}) can be replaced by a
larger scale, of order $\hat s$, as long as this scale is much less
than $m$.  Eq.~(\ref{BHQETres}) is easily
obtained by boosting from the rest frame of the top and antitop
respectively with a boost factor of $Q/m$. In this naming scheme we
will continue to call the gluons that govern the cross-talk between
top and antitop jets {\it soft}. We emphasize that they are not
included in ${\cal L}_\pm$, since they have nothing to do with the
gluons in standard HQET. Soft gluon interactions will be added
below. To avoid double counting between the soft gluons, the
ultracollinear gluons are defined with zero-bin
subtractions~\cite{Manohar:2006nz}, so that for example $\bn\mcdot
k_+\ne 0$ and $n\mcdot k_- \ne 0$. Finally, since HQET is applied for $\mu<m$
there are no analogs of the SCET mass-modes in this theory. All effects
associated with virtual top-quark loops are integrated out at the scale $m$.

The leading order Lagrangians ${\cal L}_\pm$ contain a residual mass
term $\delta m$ which has to be chosen according to the desired top
quark mass scheme. For a given top mass scheme $m$, the residual mass
term is determined by its relation to the pole mass $m_{\rm pole} \, =
\, m + \delta m$.  Anticipating that we have to switch to a properly
defined short-distance mass
definition~\cite{Hoang:1998nz,Beneke:1998rk,Uraltsev:1998bk,Hoang:1998ng}
when higher order QCD corrections are included, we note that only
short-distance mass definitions are allowed which do not violate the
power counting of the bHQET theories, $\delta m\sim \Gamma$. This
excludes for example the use of the well known $\overline{\rm MS}$
mass, since in this scheme $\delta m\sim \alpha_s m \gg
\Gamma$. In practice, this means that using the $\overline{\rm MS}$ mass leads
to an inconsistent perturbative expansion as explained in
Sec.~\ref{sec:sdmass}. This is the 
reason why the $\overline{\rm MS}$ mass can not be measured directly from
reconstruction.

The leading order Lagrangians ${\cal L}_\pm$ also contain top-width terms
$i\Gamma/2$.  An effective field theory treatment of the evolution and decay of
a massive unstable particle close to its mass shell was developed
in~\cite{Fadin:1991zw,Beenakker:1999hi, Beneke:2003xh,
  Beneke:2004km,Hoang:2006pd,Hoang:2004tg}. The examples treated in these
references were the resonant production of a single unstable scalar particle,
and the leading and subleading width corrections to threshold $t\bar t$
production.  In our case, we deal with the energetic pair production of massive
unstable fermions, and we arrive at two copies of this unstable HQET
corresponding to the top and antitop sectors.  In these two HQET theories we
treat the top and antitop decays as totally inclusive, since we do not require
detailed differential information on the decay products.  So the total top width
$\Gamma$ appears as an imaginary mass term in ${\cal L}_\pm$, which is obtained
by simply matching the imaginary part of the top and antitop self-energy graphs
from SCET onto bHQET. As we show in Sec.~\ref{section3}, this inclusive
treatment of the top decay is consistent with the hemisphere invariant mass
definition we employ in this work up to power corrections of order $(m/Q)^2$. We
will come back to the role of higher order power corrections in the treatment of
the finite top lifetime at the end of this section.

{\it Soft Interactions.}
Lets consider how the soft gluons interact with our heavy quarks in each bHQET.
For a heavy quark in the boosted frame we consider interactions with soft gluons
of momentum
\begin{align}
  \ell^\mu \lesssim (\Delta,\Delta,\Delta) \,.
\end{align}
Our main interest is in the case $\Delta\lesssim \Gamma$, but it is useful to keep a
more general $\Delta \ge \Lambda_{\rm QCD}$ for the moment. We wish to
demonstrate that these gluons are still entirely described by the cross-talk
matrix element in Eq.~(\ref{Yme}), and that this is true without needing to
expand in the ratio of $\Delta$ to $\Gamma$.  Or in other words, that the simple
eikonal propagators for the soft-gluon attachments to the energetic tops remains
valid even below the mass of the quarks and even in the presence of the
top-width. Our demonstration assumes the reader is quite familar with
Ref.~\cite{Bauer:2001yt}. To prove this we go back to the original SCET Lagrangian in
Eq.~(\ref{Lscet}) prior to the field redefinition, and match the soft
interactions onto the HQET theory. This gives the same Lagrangian as in
Eq.~(\ref{LbHQET}) but with replacements
\begin{align} \label{Drepl}
  i D_{+}^\mu \to i {\cal D}_+^\mu &= i \tilde \partial_+^\mu + g A_{+}^\mu
   +  \frac{\bn^\mu}{2} g n\mcdot A_s
  \,,\nn\\
  i D_{-}^\mu \to i {\cal D}_-^\mu &= i \tilde \partial_-^\mu + g A_{-}^\mu
   + \frac{n^\mu}{2} g \bn\mcdot A_s \,.
\end{align}
The new covariant derivatives ${\cal D}_\pm$ also appear in the pure gluon
action responsible for the ultracollinear gluon kinetic term.  The nature of the
expansion for different momenta in $i\tilde \partial^\mu$ will depend on the
size of the soft scale $\Delta$ relative to the smallest ultracollinear
components $m\Gamma/Q$ displayed in Eq.~(\ref{BHQETres}).  Note in this
comparison that the width is suppressed by a factor of $m/Q$.  Physically this
factor is easy to understand, it is simply the time-dilation of the width of the
energetic top-quark from the point of view of the soft gluons.  The boost factor
is encoded in $v_+\cdot \bn = v_-\cdot n = Q/m$.

For our analysis we also need the effective current in the bHQET theories that
corresponds to the SCET current in Eq.~(\ref{currentscet}). It is
\begin{align} 
\label{JbHQET}
  J^\mu_{\rm bHQET} = (\bar h_{v_+} W_{n}) \Gamma_i^\mu (W_{\bn}^{\dagger}
  h_{v_-}) \,,
\end{align}
where the Wilson lines are the same as $W_n$ and $W_\bn^\dagger$ in SCET, except
here we have gluons $\bn\cdot A_+$ with path along $\bn^\mu$ for $W_{n}$,
and $n\cdot A_-$ with path along $n^\mu$ for $W_{\bn}^\dagger$.  The simplest
way to derive this result is to note that the two collinear sectors in the SCET
current in Eq.~(\ref{currentscet}) do not directly interact, and neither do the
two sectors of the two bHQETs.  In the rest frame of the top-quark for example
the matching is simply $(W^\dagger_n \psi_s) \to (W_n^{\dagger} h_{v_+})$, where
$\psi_s$ is a field for the top-quark near its rest-frame.  Boosting this result gives
the matching for the top-quark field in Eq.~(\ref{JbHQET}), and the result for
the antitop quark is analogous.  The dynamics of the $B_+$ and $B_-$ jet
functions will be defined by the two interpolating field operators
$(W_{n}^\dagger h_{v_+} )$ and $(W_{\bn}^{\dagger} h_{v_-})$, and is governed by
the Lagrangians ${\cal L}_{+} $ and ${\cal L}_{-}$ respectively.

Let us now come back to the derivation of Eq.~(\ref{Yme}) from bHQET. For
convenience we start by taking both scales the same size, $m\Gamma/Q \sim
\Delta$. (Below we will show that the same result is obtained for the case where
$m\Gamma/Q \ll \Delta$, which includes the situation $\Delta\sim\Gamma$.)
For $m\Gamma \sim Q\Delta$ we can formulate the multipole expansion for the
coupling of soft gluons to the heavy quarks by splitting the momenta into a
large label components of size $Q\Gamma/m$ and $\Gamma$, and residual momentum
components of size $m\Gamma/Q$. Thus\footnote{This formulation of the multipole
  expansion is the same as for the coupling of ultrasoft particles to collinear
  particles in SCET~\cite{Bauer:2001yt} where the two types of derivatives are
  formally separated by introducing label operators, and leaving residual
  momenta to be picked out by $i\partial^\mu$.}
\begin{align} \label{ipartial}
  i \tilde \partial^\mu_+ &= \frac{n^\mu}{2}\bn\mcdot {\cal P}_c +
  {\cal P}_{c\perp}^\mu
    + \frac{\bn^\mu}{2} n\mcdot i\partial
  \,,
  &i \tilde \partial^\mu_- &= \frac{\bn^\mu}{2} n\mcdot {\cal P}_c  +
  {\cal P}_{c\perp}^\mu
    + \frac{n^\mu}{2} \bn\mcdot i\partial \,.
\end{align}
The notation indicates that soft momenta only appear in the components
$i\partial^\mu$. On the other hand ultracollinear momenta appear in all four
components, and are picked out by the label operators ${\cal P}_c^\mu$ or the
$i\partial$. Next we make the same field redefinition on bHQET fields that we
made on the SCET fields in Eq.~(\ref{fd})
\begin{align}\label{fd22}
 &  h_{v_+} \to Y_n h_{v_+} \,, 
 &  A_{+}^\mu & \to Y_n A_{+}^\mu Y_n^\dagger \,, 
 &  h_{v_-} & \to Y_\bn h_{v_-} \,,
 &  A_{-}^\mu & \to Y_n A_{-}^\mu Y_n^\dagger \,,
\end{align}
where the fields in $Y_n$ and $Y_\bn$ are soft gluons. Since
\begin{align}
  & (v_+\mcdot \bn) (in\mcdot \partial \plus g n\mcdot A_s) Y_n = 0 \,,
  & & (v_-\mcdot n) (i\bn\mcdot \partial \plus g \bn\mcdot A_s) Y_\bn = 0\,,
\end{align}
this field redefinition gives back exactly Eq.~(\ref{LbHQET}) for the bHQET
Lagrangian and also gives a leading ucollinear gluon action that has no
couplings to soft gluons. In addition when making the field redefinition in the
bHQET currents, Eq.~(\ref{JbHQET}), we get exactly the same soft cross-talk
matrix element for the two-jet production
\begin{align}\label{bYme}
 \langle 0 | (\overline {Y}_\bn)^{ba} \, 
    (Y_n)^{bc}  | X_s \rangle 
   \langle X_s | (Y_n^\dagger)^{cd}\, (\overline {Y}^\dagger_\bn)^{ad}   | 0
   \rangle
   \,.
\end{align}
The only difference between the SCET matrix element in Eq.~(\ref{Yme}) and the
HQET matrix element in Eq.~(\ref{bYme}) is that in the former the soft-gluons
couple to the massive $\psi_m$ fields, while there are no such couplings in the
latter. In matching renormalized soft matrix elements at a scale $\mu\simeq m$
the only effect of these couplings to $\psi_m$ fields is to induce an overall
Wilson coefficient, so that $S^{\rm SCET} = T_0(m,\mu) S^{\rm bHQET}$.  Thus
the main dynamics of the soft gluons is not modified in a substantial way by
passing from SCET to the boosted HQET Lagrangian, nor by the presence of the
width term for unstable quarks.

For completeness, let us now consider the case $\Delta \gg m\Gamma/Q$ and show
that the same result is obtained.  In this case a soft gluon of momentum
$\ell^\mu$, coupling to an $h_{v_+}$ with residual momentum $k_+$, has $\bn\cdot
\ell \gg \bn\cdot k_+$, while for $h_{v_-}$ we have $n\cdot \ell \gg n\cdot
k_-$.  Thus these soft gluons knock the heavy quarks away from their mass shell,
and their interactions can not be formulated in a local manner in the same
theory as the ucollinear gluons.  This is similar to how soft and collinear
gluons interact in the theory \SCETb as discussed in Ref.~\cite{Bauer:2001yt}.
To derive the form of the soft gluon interactions for this situation we can
construct an auxiliary intermediate theory where the ucollinear gluons and heavy
quarks are further from their mass shell and the soft interactions are local.
The form of this theory is identical to
Eqs.~(\ref{LbHQET},\ref{Drepl},\ref{ipartial}), but with $\Gamma$ in
Eq.~(\ref{BHQETres}) replaced by $Q\Delta/m$, and we can make the field
redefinition of Eq.~(\ref{fd22}) in this theory.  Then we lower the offshellness
of the ucollinear particles and match onto the bHQET with scaling exactly as in
Eq.~(\ref{BHQETres}). (This is identical to the procedure used to construct
\SCETb operators from \SCETa which was devised in
Refs.~\cite{Bauer:2002aj,Bauer:2003mg}.) The result of this procedure is exactly
Eqs.~(\ref{LbHQET}) and (\ref{bYme}). Thus, the result for $\Delta \gg
m\Gamma/Q$ is the same as for $\Delta\sim m\Gamma/Q$.

We conclude that at leading order the interaction of the bHQET heavy quarks with
soft gluons are described by Eq.~(\ref{bYme}).  This matrix element can be used
to define a soft function $S$, that describes the cross-talk between massive
top-quarks which have fluctuations below the mass scale $m$, and we can use
Eq.~(\ref{LbHQET}) for the remaining dynamics at LO. Thus, the dynamics
separates in the manner shown in Fig.~\ref{fig:efts}, into two decoupled HQET's
and a decoupled soft-sector. In Sec.~\ref{subsectionfactorizationtheorem} below
we will derive the same result in an alternative manner, starting from the
factorization theorem for the cross-section in SCET. In this approach the
definition of the jet functions and soft-cross talk matrix elements are first
defined in SCET, and then matched onto bHQET. In this case the soft couplings
are fully formulated by the matrix element in Eq.~(\ref{bYme}), and there is no
need to consider soft couplings to fields in the bHQET Lagrangian.

{\it Decay Product Interactions.}
It is conspicuous that in the leading order bHQET setup, gluon exchange
involving top and antitop decay products is not present. We now show that this
treatment is correct and discuss the size of possible power corrections. Since
we are interested in top/antitop invariant masses in the peak region at large
$Q$, we only have to consider ucollinear and soft gluons. Concerning ucollinear
gluons it is convenient to switch for each bHQET into the respective heavy quark
rest frame where $v_\pm^\mu=(1,0,0,0)$ and the ucollinear gluons have momenta
$k^\mu\sim\Gamma\ll m$. For the hemisphere invariant masses we can treat the top
decay as fully inclusive at leading order (see Sec.~\ref{section3}), so we can
address the issue by analyzing possible cuts from the top/antitop final states
in electroweak diagrams contributing to the bHQET matching
conditions~\cite{Hoang:2004tg}.
\begin{figure}[t!]
  \centerline{ 
   \includegraphics[width=14cm]{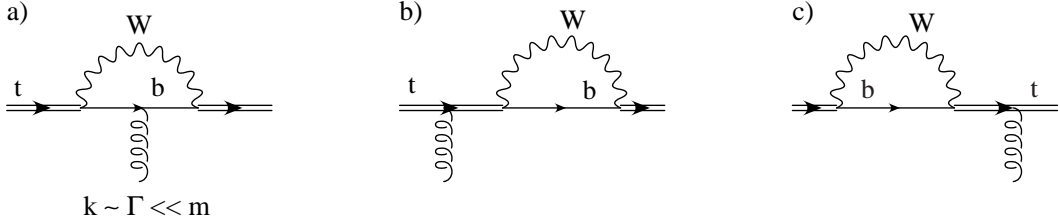}  
  } 
\caption{Example of the cancellation of soft gluon attachments to the decay products.}
\label{fig:ward}
\end{figure}
At leading order in the expansion in $\Gamma/m$ there are cuts from the
top/antitop self energy which lead to the width terms in ${\cal L}_\pm$.
Subleading finite lifetime corrections to the heavy quark bilinear terms are
suppressed by $\Gamma/m$ and physically related to the lifetime-dilations coming
from residual momentum fluctuations of the heavy quark.  Furthermore, due to
gauge invariance finite lifetime matching contributions can not arise for the
$v_\pm\cdot A_\pm$ couplings in the covariant derivatives of ${\cal L}_\pm$.
Diagrammatically this involves a cancellation between the graphs in
Fig.~\ref{fig:ward} including all possible cuts. Diagram a) is a vertex
correction, while diagram b), c) is a wave-function-type contribution. Since
momenta in the cut graphs are of order $m$, at leading order we can take the
ucollinear gluons to have momentum $k^\mu=0$. In this situation the diagrams
cancel due to gauge invariance. Thus, at leading order there are no finite
lifetime effects involving ucollinear gluon exchange. Effects from the sum of
the diagrams in Fig.~\ref{fig:ward} that do not cancel are suppressed by at
least a factor $\alpha_s\Gamma/m$ relative to the leading order factorization
theorem.

Finally we consider soft gluon interactions. Using the proof above for
the universality of the soft cross-talk matrix element in
Eq.~(\ref{bYme}) and repeating the arguments made for the ucollinear
gluon interactions we find that the dominant soft gluon interactions
involving top/antitop decay products are described by possible cuts of
electroweak matching contributions of the $n\cdot A_s$ and $\bar
n\cdot A_s$ couplings in Eq.~(\ref{Drepl}). In this case the same
cancellation as for the ucollinear gluons takes place since the
average soft gluon energy in the top/antitop rest frame is still
$\Delta$ and thus much smaller than $m$. Thus interactions involving
top/antitop decays products and soft gluons are suppressed by at least
a factor $\Delta/m$. Numerical studies in Ref.~\cite{Sjostrand:1999ki}
have estimated QCD interconnection effects based on nonperturbative
models (see also Ref.~\cite{Khoze:1992rq}).

Having defined the EFT's we now turn to the derivation of the
factorization theorem.

\section{Factorized Cross-Section and Invariant Mass Definitions}\label{section3}

\subsection{The QCD Cross-Section}

We start with the general expression of the cross-section for top-antitop quark
production, $\>e^+e^-\rightarrow \gamma^*,Z^*\rightarrow t\bar t+X$. The final
state we are interested in is observed as the top and antitop jets plus soft
radiation $J(t) J(\bar t)X_s$.  We remind the reader that we refer to all the
jets coming from the top and antitop quark decay collectively as top and antitop
jets, respectively. But we stress that despite the language, our analysis is
still perfectly consistent with the fact that the different jets from each the
top and antitop decay can be resolved in the experimental analysis.

The full cross-section is
\begin{eqnarray}\label{qcdcrosssection}
\sigma &=& \sum_X^{res.} (2\pi)^4 \, \delta^4(q-p_X) \sum_{i=a,v}
L_{\mu\nu}^{i}\ \langle 0| {\cal J}^{\nu\dagger}_i(0)  |X\rangle
\langle X | {\cal J}^\mu_i(0) |0\rangle 
 \, ,
\end{eqnarray}
where the initial state total leptonic momentum is $q=p_{e^-}+p_{e^+}$,
$Q^2=q^2$, and the QCD currents ${\cal J}_{v,a}^\mu$ are given in
Eqs.~(\ref{QCDcurrents}).  The superscript $res.$ on the summation symbol
denotes a restriction on the sum over final states $X$, to give $J(t) J(\bar
t)X_s$. These final states contain top and antitop jets with invariant masses
close to the top quark mass. The explicit form of these restrictions depends on
the specific jet and invariant mass definitions used.  For the hemisphere
invariant mass prescription these restrictions will be implemented explicitly in
Sec.~\ref{section_hemi} below, while other methods are discussed in
Sec.~\ref{sectionotheralgo}.

In Eq.~(\ref{qcdcrosssection}) we include photon and $Z$ boson exchange, and
imply an angular average of the leptonic tensor, to obtain the parity
conserving $L_{\mu\nu}^{i}$ with a sum over vector and axial-vector parts,
$i=v, a$. For convenience we also include the charges and boson
propagators, and the cross-section prefactor $1/(2Q^2)$, so that
\begin{align}
 L_{\mu\nu}^{(v)} &= -\frac{8\pi^2 \alpha^2}{3 Q^4} \Big(g_{\mu\nu}-\frac{q_\mu
   q_\nu}{Q^2}\Big)
  \bigg[\, e_t^2 - 
   \frac{2 Q^2\, v_e v_t e_t}{Q^2-m_Z^2} + 
   \frac{Q^4 (v_e^2+a_e^2)v_t^2}{(Q^2-m_Z^2)^2}\, \bigg] \,,\nn\\
 L_{\mu\nu}^{(a)} &=  -\frac{8\pi^2 \alpha^2}{3 Q^4} \Big(g_{\mu\nu}-\frac{q_\mu
   q_\nu}{Q^2}\Big)
   \bigg[\, \frac{Q^4\, (v_e^2+a_e^2)a_t^2}{ (Q^2-m_Z^2)^2 } \bigg] \,.
\end{align}
Here $e_t$ is the top-quark charge, and
\begin{eqnarray}
  v_f = \frac{T_3^f-2 Q_f \sin^2\theta_W}{2\sin\theta_W \cos\theta_W}\,,
  \qquad\qquad
  a_f = \frac{T_3^f}{2\sin\theta_W \cos\theta_W} \,,
\end{eqnarray}
where $T_3^f$ is the third component of weak isospin, and $\theta_W$ is the
weak mixing angle.

\subsection{The SCET Cross-Section}

We now proceed by using the fact that the states are restricted to be
dijet-like through the constraint that the top and antitop jet invariant
masses are close to the top quark mass, as illustrated in
Fig.~\ref{fig:topjet}. 
In this section we reformulate the cross section by using the more specific
SCET currents of Eq.~(\ref{currentscet}) that are 
suitable for this kinematic situation.  We integrate out the hard production 
energy scale $Q$ by matching the SCET currents onto the QCD currents giving us 
via the matching relation~(\ref{currentmatch})
a new expression for the cross-section defined with matrix elements in SCET. 

The SCET currents in Eq.~(\ref{currentscet}) correctly reproduce the long
distance physics of the QCD current, and the difference in the short distance
physics is contained in the Wilson coefficient $C(\omega,\bar\omega,\mu)$. We
will see momentarily that momentum conservation dictates that the final form of
the cross-section depends only on $C(Q,-Q,\mu)\equiv C(Q,\mu)$. In
Ref.~\cite{Manohar:2003vb} the Wilson coefficient at one loop was computed. It
is independent of the Dirac structure $\Gamma _i$ and also of whether or not the
collinear quarks are massive (the latter fact is demonstrated in
Ref.~\cite{FHMS2} where the matching computation for the corresponding vertex
diagrams is carried out explicitly for finite heavy quark mass). The result is 
\begin{eqnarray}
\label{cQmu}
C(Q,\mu) = 1+\frac{\alpha _s C_F}{4\pi}\Big [ 3\log \frac{-Q^2\!-\!i0}{\mu ^2} -\log
^2\frac{-Q^2\minus i0}{\mu ^2} 
-8 + \frac{\pi ^2}{6}\Big ].
\end{eqnarray}
At the matching scale $\mu=Q$ this Wilson coefficient does not contain any large
logarithms.  The product of the Wilson coefficient $C(Q,\mu)$ and the SCET
matrix element is independent of the scale $\mu$, and renormalization group (RG)
evolution determines the Wilson coefficient at a lower scale $\mu$. This RG
evolution of the hard Wilson coefficient sums logarithms of $\mu/Q$ with $\mu\gtrsim
m$. The Wilson coefficient contains an imaginary part that arises from real
QCD intermediates states in the QCD vertex diagram that are not accounted for
in the corresponding SCET diagrams when the collinear action only contains the
two sectors for the $n$ and $\bar n$ directions (see
Sec.~\ref{mass-scet}). However, only $|C(Q,\mu)|^2$ will appear in the
final factorization theorem since we will sum over $\vec n$.

Using Eqs.~(\ref{currentmatch}) and (\ref{currentscet}) in
Eq.(\ref{qcdcrosssection}), the cross-section in SCET takes the form
\begin{eqnarray}
\label{scetcross-section}
\sigma &=&  \sum_{\vec n} \sum_{X_n X_\bn X_s}^{res.} (2\pi)^4 \,
\delta^4(q \!-\! P_{X_n} \!-\! P_{X_\bn}\!-\! P_{X_s})  \sum_{i} L_{\mu\nu}^{(i)} 
 \int\!\! d\omega\,d\bar\omega\,d\omega'\,d\bar\omega'\: 
   \\
 &&\hspace{-0.5cm}
 \times C(\omega,\bar\omega) C^*(\omega',\bar\omega')
  \langle 0|  {\rm T}\{ \bar\chi_{\bn,\bar\omega'} S_\bn^\dagger \bar\Gamma_j^{\nu}
   S_n \chi_{n,\omega'}\} |X_n X_\bn X_s\rangle
 \langle X_n X_\bn X_s| {\rm T}\{ \overline\chi_{n,\omega} S_n^\dagger \Gamma_i^\mu 
   S_\bn \chi_{\bn,\bar\omega} \} |0\rangle
  \,. \nn
\end{eqnarray}
Here we have pulled out the explicit sum over the top jet label directions
$\vec n$ and keep only two collinear sectors ${\cal L}^{(0)}_n$ and ${\cal
L}^{(0)}_{\bar n}$ for the SCET description of top and antitop jets. This
allows us to explicitly carry out the integral over the top jet directions
$\vec n$ in Sec.~\ref{subsectionmomdecomp} in parallel to implementing
factorization. 

In Eq.~(\ref{scetcross-section})
we have decomposed the final states $|X\rangle$ into a soft sector
$|X_s\rangle$ and collinear sectors $|X_n\rangle,|X_\bn \rangle$ in the
$\vec{n}$ and $\vec{\bn}$ directions respectively
\begin{eqnarray} \label{X1}
  | X \rangle =| X_n X_\bn X_s \rangle   \,.
\end{eqnarray}
Since the hard production scale is integrated out by the matching procedure,
these states now form a complete set of final states that can be 
produced by the SCET currents ${\cal J}^\mu_i$. This already implements
part of the restrictions, ``res'', in the sum over states in
Eq.~(\ref{scetcross-section}).  The 
momentum $P_X$ of the final state $|X\rangle$ is also decomposed into the
momentum of the collinear and soft sectors:
\begin{eqnarray}
  P_X=P_{X_n} + P_{X_\bn} + P_{X_s}.
\end{eqnarray} 
Recall that there are no particles with $p_m^\mu\sim (m,m,m)$ scaling that can
cross the final state cut, without taking the invariant mass far from the peak
region, so there are no mass-modes in this decomposition.  Because the set of
hadrons observed in the detector has a well defined set of momenta, it is
possible to impose criteria on the hadrons in the final state to associate them
with one of $X_n$, $X_\bn$, or $X_s$. Thus, the hadronic two-jet state
factorizes as a direct product
\begin{eqnarray} 
\label{XXX}
  | X\rangle = | X_n \rangle | X_\bn \rangle | X_s \rangle \,.
\end{eqnarray}
This factorization is also a manifest property of the hadronic states in SCET.
 
For quark and gluon states in SCET the difference from the purely
hadronic case in Eq.~(\ref{XXX}) is that the states can carry global
color quantum numbers.  After having made the soft-collinear
decoupling field redefinition, the individual Lagrangians for these
sectors are decoupled, and they only organize themselves into color
singlets in the matrix elements which appear in the observable
cross-section.  We can take this as a manifestation of quark-hadron
duality.  Using the soft-collinear decoupling property from
Sec.~\ref{mass-scet} we can write the matrix elements in
Eq.~(\ref{scetcross-section}) as ${\cal M}(m,\mu)$ times
\begin{align} \label{spincolorindices}
 &\big\langle 0 \big|  \overline \chi_{\bn,\bar\omega'}^a  (\overline {Y}_\bn)^{ba} \, 
    (\overline\Gamma\, Y_n \chi_{n,\omega'})^b  \big| X_n X_\bn X_s \big\rangle 
   \big\langle X_n X_\bn X_s \big| (\overline \chi_{n,\omega} Y_n^\dagger\, 
     \Gamma)^c\, (\overline {Y}^\dagger_\bn)^{dc} \chi_{\bn,\bar\omega}^d  \big| 0 \big\rangle 
  \\[5pt]
 &= \big\langle 0 \big|  \overline \chi_{\bn,\bar\omega'}^a \big| X_\bn \big\rangle 
    \big\langle X_\bn \big| \chi_{\bn,\bar\omega}^{a'} \big| 0 \big\rangle
    \big\langle 0 \big|  \chi_{n,\omega'}^b \big| X_n \big\rangle 
    \big\langle X_n \big|  \overline \chi_{n,\omega}^{b'} \big| 0 \big\rangle \nn\\
 &\quad\times   \big \langle 0 \big| (\overline {Y}_\bn)^{ca} 
    (\overline\Gamma Y_n)^{cb}  \big| X_s \big\rangle 
    \big\langle X_s \big| ( Y_n^\dagger
     \Gamma)^{b'c'} (\overline {Y}^\dagger_\bn)^{a'c'}  \big| 0 \big\rangle 
  \,,\nn
\end{align}
where here roman indices are for color and spin and $|X_n\rangle$ and
$|X_\bn\rangle$ are color triplets.  Next we rearrange the color and spinor
indices so that they are fully contracted within each of the $n$-collinear,
$\bn$-collinear, and soft product of matrix elements. This makes explicit the
fact that in SCET each of these contributions to the cross-section must
separately be a spin and color singlet.  Although it is not absolutely necessary
to make this arrangement of indices manifest at this point, it does allow us to
avoid carrying around unnecessary indices (a similar manipulation was used for
$B\to X_s\gamma$ in Ref.~\cite{Lee:2004ja}). For color, our
$|X_\bn\rangle\langle X_\bn|$ forces the indices on $\overline \chi_\bn^a$ and
$\chi_\bn^{a'}$ to be the same, so $\big\langle 0 \big| \overline \chi_\bn^a
\big| X_\bn \big\rangle \big\langle X_\bn \big| \chi_\bn^{a'} \big| 0
\big\rangle = (\delta^{aa'}/N_c) \, \big\langle 0 \big| \overline \chi_\bn^b
\big| X_\bn \big\rangle \big\langle X_\bn \big| \chi_\bn^{b} \big| 0
\big\rangle$.  A similar result holds for the $n$-collinear matrix elements.
For spin we can use the SCET Fierz formula
\begin{align} \label{Fierz}
 1 \otimes 1 
 &= \frac{1}{2} \Big[ 
  \big(\frac{\bnslash}{2}\big)\!\otimes\! \big(\frac{\nslash}{2} \big)
 + \big(\frac{-\bnslash\gamma_5}{2}\big)\!\otimes\!
   \big(\frac{\nslash\gamma_5}{2}\big)
 + \big(\frac{-\bnslash\gamma_\perp^\alpha}{2}\big) \!\otimes\! 
     \big(\frac{\nslash\gamma^\perp_\alpha}{2}\big) 
  \Big] \,,
\end{align}
which is valid when the identity matrices are inserted so that the $\nslash$
terms on the RHS appear between $\overline\chi_{\bn} \cdots \chi_\bn$ without
additional $\bnslash$ factors next to these fields (or the analogous statement
with $n\leftrightarrow \bn$). Combining the color and spin index
rearrangement, the matrix element in Eq.~(\ref{spincolorindices}) becomes
\begin{align}  \label{factor-m-elt}
 & {\rm tr}\Big[ \frac{\nslash}{2} \Gamma_i^\mu \frac{\bnslash}{2}
 \bar \Gamma_j^\nu \Big] 
  \Big[ \big\langle 0 \big|  \overline \chi_{\bn,\bar\omega'}^a \big| X_\bn \big\rangle 
    \big\langle X_\bn \big| \Big(\frac{\nslash}{4N_c} \chi_{\bn,\bar\omega}\Big)^{a} 
       \big| 0 \big\rangle \Big]\
    \Big[ \big\langle 0 \big|  \Big(\frac{\bnslash}{4N_c} \chi_{n,\omega'} \Big)^b 
       \big| X_n \big\rangle 
    \big\langle X_n \big|  \overline \chi_{n,\omega}^{b} \big| 0 \big\rangle \Big]
   \nn\\
   &\qquad\times  \Big[ \big \langle 0 \big| (\overline {Y}_\bn)^{ca'} 
    (Y_n)^{cb'}  \big| X_s \big\rangle 
    \big\langle X_s \big| ( Y_n^\dagger)^{b'c'} (\overline {Y}^\dagger_\bn)^{a'c'} 
      \big| 0 \big\rangle 
   \Big] \nn\\
 & \equiv {\rm tr}\Big[ \frac{\nslash}{2} \Gamma_i^\mu \frac{\bnslash}{2}
 \bar \Gamma_j^\nu \Big] 
  {\rm tr} \Big( \big\langle 0 \big|  \overline \chi_{\bn,\bar\omega'} \big| X_\bn \big\rangle 
    \big\langle X_\bn \big| \slash\!\!\!\hat n \chi_{\bn,\bar\omega} 
       \big| 0 \big\rangle \Big)
    {\rm tr} \Big(\big\langle 0 \big|  \slash\!\!\!\hat\bn \chi_{n,\omega'}
       \big| X_n \big\rangle 
    \big\langle X_n \big|  \overline \chi_{n,\omega} \big| 0 \big\rangle\Big) 
   \nn\\
   &\qquad\times  {\rm tr} \Big( \big \langle 0 \big| \overline {Y}_\bn 
     Y_n  \big| X_s \big\rangle 
    \big\langle X_s \big| Y_n^\dagger \overline {Y}^\dagger_\bn
      \big| 0 \big\rangle \Big) \,,
\end{align}
where for convenience we defined 
\begin{equation}
\slash\!\!\!\hat n \equiv \nslash/(4N_c)\,,\qquad
\slash\!\!\!\hat \bn \equiv \bnslash/(4N_c)\,. 
\end{equation}
Note that only the first term on the RHS of Eq.~(\ref{Fierz})
contributes because the collinear states give at least one matrix
element which is zero when we have a $\gamma_5$ or
$\gamma_\perp^\alpha$.  This factorizes the SCET cross-section into a
product of three singlets under spin and color.  For convenience we
will in the following suppress writing these explicit traces on the
matrix elements.

Using Eq.~(\ref{factor-m-elt}) in Eq.~(\ref{scetcross-section}), the
factorized SCET cross section takes the form
\begin{eqnarray}
\label{factorizedcross-section}
\sigma &=& \!\!  K_0\, {\cal M}
\sum_{\vec n} \!\sum_{X_n X_\bn X_s}^{res.} \!\!
(2\pi)^4 \, \delta^4(q\!-\!P_{X_n}\!-\!P_{X_\bn}\!-\!P_{X_s}) 
 \langle 0| \overline {Y}_\bn\,  {Y}_n |X_s \rangle
\langle X_s| {Y}^\dagger_n\,  \overline {Y}_\bn^\dagger |0\rangle 
 \\
&& \!\! \!\! \!\!\times \!\!
 \int\!\! d\omega\,d\bar\omega\,d\omega'\,d\bar\omega'\: 
   C(\omega,\bar\omega) C^\dagger(\omega',\bar\omega')
\langle 0| \slash\!\!\!\hat \bn \CH n {\omega'} |X_n \rangle 
\langle X_n |\bCH n \omega |0 \rangle 
\langle 0 |\bCH \bn {\bar\omega'} | X_\bn\rangle 
\langle X_\bn| \slash\!\!\!\hat n \CH \bn {\bar\omega} | 0 \rangle 
  \,, \nonumber
\end{eqnarray}
where ${\cal M}={\cal M}(m,\mu)$ and we defined the normalization factor
\begin{eqnarray}
 K_0 &=&  \sum_{i=v,a}
   L_{\mu\nu}^{(i)}  \textrm{Tr}
\Big[\frac{\nslash}{2} {\Gamma}^\mu_i
     \frac{\bnslash}{2} \overline {\Gamma}^{\,\nu}_i \Big]  
   =  -2 g_\perp^{\mu\nu}\sum_{i=v,a}
   L_{\mu\nu}^{(i)} \nn\\
 &=& \frac{32\pi^2 \alpha^2}{3 Q^4} 
  \bigg[\, e_t^2 - 
   \frac{2 Q^2\, v_e v_t e_t}{Q^2-m_Z^2} + 
   \frac{Q^4 (v_e^2+a_e^2)(v_t^2+a_t^2)}{(Q^2-m_Z^2)^2} \bigg]\,. 
\end{eqnarray}
We can further simplify the form of the factorized cross-section. First we
use the identities
\begin{eqnarray} \label{Qconserved}
  \langle X_n| \bCH {n} {\omega'} |0 \rangle
   &=& \langle X_ n|\overline\chi_n\delta_{\omega',\bn\cdot \cP^\dagger} |0 \rangle 
   = \delta_{\omega',p^-_{X_n}}\, \langle X_n | \overline \chi_n | 0\rangle \,,
   \nn\\
   \langle X_{\bn} | \bCH {\bn} {\bar\omega'} |0 \rangle
   &=& \langle X_\bn|\overline\chi_{\bn} \delta_{\bar\omega',n\cdot {\cal P}^\dagger} |0 \rangle 
   = \delta_{-\bar \omega',p^+_{X_\bn}}\, \langle X_\bn | \overline \chi_\bn | 0\rangle \,,
\end{eqnarray}
with similar relations for the other two collinear matrix elements in
Eq.(\ref{factorizedcross-section}).  Combining this with the relation
$\delta_{\omega',p_{X_n}^-} \delta_{\omega,p_{X_n}^-}
=\delta_{\omega',\omega} \delta_{\omega,p_{X_n}^-} $, and analog for
$p_{X_\bn}^+$, we can write the product of collinear matrix elements
in Eq.(\ref{factorizedcross-section}) as
\begin{eqnarray}
 &&  \langle 0 | \slash\!\!\!\hat \bn \CH n {\omega'} |X_n \rangle 
\langle X_n |\bCH n \omega | 0 \rangle 
\langle 0 |\bCH \bn {\bar\omega'} |X_\bn \rangle
\langle X_\bn |\slash\!\!\!\hat n \CH \bn {\bar\omega} | 0 \rangle 
 \nn\\
 &&=
  \delta_{\bar\omega',\bar\omega}\, \delta_{\omega',\omega}\, 
 \langle 0 | \slash\!\!\!\hat \bn \chi_n | X_n \rangle 
 \langle X_n |\bCH n \omega | 0 \rangle 
 \langle 0 |\overline\chi_{\bn} | X_\bn \rangle
 \langle X_\bn | \slash\!\!\!\hat n \chi_{\bn,\bar\omega}  | 0 \rangle  \,.
\end{eqnarray}
Next we do the sums over $\omega' , \bar\omega'$ to arrive at the form
\begin{eqnarray}
\label{factorizedcross-section2}
\sigma &=& K_0\, {\cal M} 
\sum_{\vec n}\! \sum_{X_n X_\bn X_s}^{res.}
(2\pi)^4 \, \delta^4(q\!-\!P_{X_n}\!-\!P_{X_\bn}\!-\!P_{X_s}) 
 \langle 0| \overline {Y}_\bn\,  {Y}_n |X_s \rangle
\langle X_s| {Y}^\dagger_n\,  \overline {Y}_\bn^\dagger |0\rangle 
 \nn\\
&& \!\! \!\! \!\!\times \!\!
 \int\!\! d\omega\,d\bar\omega\:
   |C(\omega,\bar\omega)|^2
\big\langle 0 \big| \slash\!\!\!\hat \bn \chi_n  \big| X_n \big\rangle 
\big\langle X_n \big|\bCH n \omega \big| 0 \big\rangle 
\big\langle 0 \big| \overline\chi_\bn  \big| X_\bn \big\rangle
\big\langle X_\bn \big| \slash\!\!\!\hat n \chi_{\bn,\bar\omega} \big| 0 \big\rangle  
  \,. 
\end{eqnarray}

Before proceeding, we pause to define the thrust axis which is needed to
properly define the invariant mass of jets and to state its relation to the
direction of the energetic collinear degrees of freedom. Then in order to make
the power counting manifest we decompose the final state momenta into label and
residual parts and perform some general manipulations of the phase space
integrals to setup a formula for the cross-section to be used for the
remaining calculation.

\subsection{Thrust or Jet Axis }

The thrust $T$ of any event
is defined to be
\begin{equation}
 \label{thrust-1}
 T = \mathop{\textrm{max}}_{\hat{{\bf t}}} \frac{\sum_i | \hat{{\bf t}} \cdot
  {\bf p}_i |}{Q} \,,
\end{equation}
where the sum is over the momenta ${\bf p}_i$ of all the final state particles
produced. The thrust axis ${\bf \hat{t}}$ is chosen so that is maximizes the sum
of particle momenta projected along ${\bf \hat{t}}$. Intuitively, for a
dijet-like event the thrust axis corresponds to the axis along which most of
the momentum is deposited. Conversely, the thrust $T$ is close to its
maximum for a dijet-like event.
We choose $\vec n$ to point along ${\bf \hat{t}}$. For an event with exactly
two massive stable particles $T= \sqrt{Q^2 - 4m^2}/Q = 1 - 2 m^2 / Q^2 + {\cal
  O}(m^4/Q^4)$, is the maximum allowed thrust.  Since we are interested in
thrusts in the dijet region for the top and antitop jets it is convenient to
define a shifted thrust parameter,
\begin{align} \label{tau}
  \tau &= \sqrt{1- \frac{4m^2}{Q^2} } - T 
   =1 - \frac{2m^2}{Q^2} - T +{\cal O}\Big(\frac{m^4}{Q^4}\Big)\,.
\end{align}
For stable top-antitop production additional jets always result in $\tau>0$.
For unstable top-quarks the values of $\tau<0$ also become allowed. Note that
for massless jet production the thrust ($T$) distribution is peaked close to
$T=1$ while for events containing a heavy quark pair it is peaked close to 
$T=\sqrt{Q^2 - 4m^2}/Q$. Thus a cut on thrust can in principle be used to
discriminate between massive and massless quark
production~\cite{Chekanov:2003cp}.

\subsection{Differential Cross-Section with  Momentum Decomposition}
\label{subsectionmomdecomp}

To insert the invariant mass constraints into our cross-section in
Eq.~(\ref{factorizedcross-section}) we use the identity operator:
\begin{eqnarray}
\label{identity-1a}
1= \int\! d^4p_n \> d^4p_\bn\> d^4p_s\> \delta ^4(p_n- P_{X_n}) \> \delta ^4(p_\bn-
P_{X_\bn}) \> \delta ^4(p_s- P_{X_s})\,, 
\end{eqnarray}
which sets the total collinear and soft momenta of the states $P_{X_n}, P_{X_\bn},
P_{X_s}$ to $p_n,p_\bn,p_s$ respectively.  In Sec.~\ref{section_hemi} we will
use an additional insertion of an identity operator to define the hemisphere
invariant masses, $M_t$ and $M_{\bar t}$.  In this section we carry out
manipulations that are common to any definition of the invariant masses. For now
we ensure that the invariant mass of each hemisphere is close to the top mass by
including in the restrictions, ``res'', on the states the fact that $M_t,
M_{\bar t}$ are in the region
\begin{eqnarray}
\label{sn-range}
| s_{t,\bar t} |  =  | M^2_{t,\bar t}-m^2 |  \ll m^2 .
\end{eqnarray}
From here on we assume that in the sense of power counting $\Delta
\sim \Gamma$.  We now decompose the collinear and soft momenta into
label and residual parts
\begin{align}
\label{labelresidual}
p_n &= \tilde p_n + k_n, 
& p_{\bar n} & = \tilde p_{\bar n} + k_{\bar n} ,
& P_{X_n}^\perp & = K_{X_n}^\perp ,
 \\
P_{X_\bn}^\perp & = K_{X_\bn}^\perp , 
& P_{X_n}^- & = \tilde P_{X_n}^- + K_{X_n}^- , 
& P_{X_\bn }^+ & = \tilde P_{X_\bn}^+  + K_{X_\bn}^+ 
\,, \nn \\
P_{X_n}^+ &= K_{X_n}^+ \,,
& P_{X_\bn}^- &= K_{X_\bn}^- \,, 
& P_{X_s}^\mu & = K_{X_s}^\mu \,,
& p_s^\mu &= k_s^\mu \,. \nn
\end{align}
Note that our choice of $\vec n$ along the thrust axis together with
the restrictions on the states ensures that the perpendicular momentum
of the jets relative to this axis, $P_{X_n}^\perp$ and
$P_{X_\bn}^\perp$, are purely residual.  The last result in
Eq.~(\ref{labelresidual}) indicates that the soft state also has a
momentum that is purely residual.  The integrals in
Eq.(\ref{identity-1a}) can be decomposed into a sum over labels and
integrals over residual momenta as \begin{eqnarray}
\label{momdecomp-1}
\int d^4 p_n \,  \int d^4 p_\bn  &=&
\frac{1}{2} \sum_{\tilde{p}_n} \int dk_n^+ dk_n^- d^2k_n^\perp\,
\frac{1}{2} \sum_{\tilde{p}_\bn} \int dk_\bn^+ dk_\bn^- d^2k_\bn^\perp\, .
\end{eqnarray}

In the total cross-section in Eq.~(\ref{factorizedcross-section}) we
sum over the directions $\vec n$ of the thrust axis. To turn this sum
into an integral over the full solid angle, we need to combine it with
a residual solid angle integration for each $\vec n$. Therefore, we
decompose the residual measure as
\begin{align}
  d^2k_n^\perp = |\vec p_n|^2\:  d\phi\: d\cos(\theta_r)
    = \Big(\frac{Q^2}{4} - p_{n}^2\Big) d\phi\: d\cos(\theta_r) ,
\end{align}
where $\theta_r$ is the small angle of $p_3$ relative to $\vec p$. In the first
equality we used the fact that $\cos(\theta_r)\simeq 1$.
Since we are in the peak region we can approximate $p_n^2= m^2$ up to small
$\Gamma/m$ corrections.  Combining this with the sum over $\vec n$ gives
\begin{eqnarray} \label{SA-relation}
  \sum_{\vec n} d^2k_n^\perp = \Big(\frac{Q^2}{4} - m^2\Big)
     \: d\phi\: d\cos(\theta) = \frac{Q^2}{4} \,
      d\Omega \,,
\end{eqnarray}
where in the last equality we work to leading order in $m^2/Q^2$.
Since the angular averaged two-jet production is independent of the
thrust direction we are free to carry out the remaining integrations
in a frame where $k_n^\perp=0$, and also replace $\int d\Omega =
4\pi$. The differential cross-section now reads
\begin{align}
\label{factorizedcross-section-new}
 \sigma &=\!
 \frac{\pi Q^2 K_0}{4}\, {\cal M} \!\!\!\sum_{X_n X_\bn X_s}^{res.} \!\!\!
(2\pi)^4 \, \delta^4(q\!-\!P_{X_n}\!-\!P_{X_\bn}\!-\!P_{X_s}) 
  \sum_{\tilde{p}_n, \tilde{p}_\bn } \!\int \!\! dk_n^+ dk_n^- \!
  \int \!\! dk_\bn^+ dk_\bn^- d^2k_\bn^\perp d^4k_s
 \nn\\
 &\times \: \delta ^4(p_n- P_{X_n}) \> \delta ^4(p_\bn- P_{X_\bn})
  \> \delta ^4(k_s- P_{X_s})\  
  \langle 0| \overline {Y}_\bn\,  {Y}_n |X_s \rangle
\langle X_s| {Y}^\dagger_n\,  \overline {Y}_\bn^\dagger |0\rangle  \nn\\
& \times 
 \int\!\! d\omega\,d\bar\omega\:
   |C(\omega,\bar\omega)|^2
\big\langle 0 \big| \slash\!\!\!\hat \bn \chi_n(0)  \big| X_n \big\rangle 
\big\langle X_n \big|\bCH n \omega (0) \big| 0 \big\rangle 
\big\langle 0 \big|\bar \chi_\bn  (0) \big| X_\bn \big\rangle
\big\langle X_\bn \big| \slash\!\!\!\hat n \chi_{\bn,\bar\omega} (0) \big| 0 \big\rangle 
 \,. 
\end{align}
In the remainder of this section we will simplify this formula as much as
possible prior to specifying the exact constraints on the restricted sum of
states.  First we decompose the delta functions into label and residual parts as
\begin{align} \label{decomposedelta-1}
\delta^4 (p_n \!-\! P_{X_n}) &=  \delta_{\tilde{p}_n, \tilde{P}_{X_n}} 
   \> \delta^4 (k_n \!-\! K_{X_n}) 
  = \delta_{\tilde{p}_n^-, \omega} 
   \delta_{\tilde{p}_n^\perp, 0 }\!  \int\!\! \frac{d^4x}{(2\pi)^4} \>
   e^{ i\left[  (k_n^+ \!-\! K_{X_n}^+)\frac{x^-}2 
   +  (k_n^- \!-\! K_{X_n}^-)\frac{x^+}2 
   - K_{X_n}^\perp\cdot \, x^\perp \right] }
  , \nn \\
\delta^4 (p_\bn \!-\! P_{X_\bn}) &= \delta_{\tilde{p}_\bn, \tilde{P}_{X_\bn}} \> 
    \delta^4 (k_\bn - K_{X_\bn}) 
  = \delta_{\tilde{p}_\bn^+, -\bar\omega} 
   \delta_{\tilde{p}_\bn^\perp, 0 }
   \> \int\!\! \frac{d^4y}{(2\pi)^4} \> \:
 e^{i\>(k_\bn - K_{X_\bn})\>\cdot \>y}  , \nn \\
\delta^4 (p_s \!-\! P_{X_s}) & = \delta^4 (k_s \!-\! K_{X_s}) 
  = 
   \> \int\!\! \frac{d^4z}{(2\pi)^4} \> \:
 e^{i\>(k_s - K_{X_s})\>\cdot \>z}  
  \,, 
\end{align}
where there is no $k_n^\perp$ in the first line (or below) because we
fixed $k_n^\perp=0$. In the second equality on lines 1 and 2 we
replaced $\tilde{P}_{X_n}^- , \tilde{P}_{X_\bn}^+$ with the labels
$\omega ,
\bar{\omega}$ respectively using the momentum conservation delta-functions
discussed below Eq.~(\ref{Qconserved}). We also decompose
\begin{align} 
\label{decomposedelta-2}
 \delta ^4(q\!-\!P_{X_n}\!-\!P_{X_\bn}\!-\! K_{X_s}) 
 &=  \delta _{Q,\tilde{p}_n^-} \,
  \delta_{Q,\tilde{p}_\bn^+}  \,
  \delta^4(k_n\!+\! k_\bn \!+\! k_{s} )
  \,,
\end{align} 
where we have replaced $P_{X_n}, P_{X_\bn}$ with $p_n,p_\bn$ by
using the delta functions in Eq.~(\ref{decomposedelta-1}).

Next we use Eqs.~(\ref{decomposedelta-1}) and (\ref{decomposedelta-2}) in
Eq.~(\ref{factorizedcross-section-new}) and with the exponential factors of 
$e^{-i K_{X_n}\cdot x}, e^{-iK_{X_\bn}\cdot y}$, and $e^{-i K_{X_s}\cdot z}$ in
Eq.~(\ref{decomposedelta-1}) we translate the collinear and soft fields to the
positions $x$, $y$, and $z$, respectively. This gives
\begin{align}
\label{factorizedcross-section-new-2}
\sigma &=
\frac{\pi}{(2\pi)^8} \frac{Q^2 K_0}{4}\, {\cal M} \!\!
  \sum^{res.}_{X_n X_\bn X_s} \!\! \int\!\! dk_n^+ dk_n^- dk_\bn^+ dk_\bn^-
  d^2k_\bn^\perp d^4 k_s
  \!\! \int\!\! d^4 x \, d^4 y \,  d^4 z  \> \delta^4(k_n\plus k_\bn \plus k_s)
\nn\\
&\times  \big |C(Q,\mu)\big |^2\:
   \text{Exp} \Big [\frac{i}{2} k_n^+ x^- \plus \frac{i}{2} k_n^- x^+ \plus
   ik_\bn\mcdot y \plus i k_s \mcdot z \Big ] \ 
 \big\langle 0 \big| (\overline {Y}_\bn\,  {Y}_n) (z) \big|X_s \big\rangle
  \big\langle X_s\big| ({Y}^\dagger_n\,  \overline {Y}_\bn^\dagger) (0) 
 \big|0\big \rangle 
 \nn \\
&\times
\big\langle 0 \big| \slash\!\!\!\hat \bn \chi_n(x)  \big| X_n \big\rangle 
\big\langle X_n \big|\bCH n Q (0) \big| 0 \big\rangle 
\big\langle 0 \big|\overline \chi_\bn (y) \big| X_\bn \big\rangle
\big\langle X_\bn \big| \slash\!\!\!\hat n \chi_{\bn,-Q} (0) \big| 0 \big\rangle 
  \,,
\end{align}
where here the large label momenta in the jets are fixed to be $Q$,
$\overline\chi_{n,Q}= \overline\chi_n\delta_{Q,\bnP^\dagger}$ and
$\chi_{\bn,-Q}=\delta_{-Q,\cP}\chi_\bn$.  Next we can use the fact that the
$n$-collinear graphs are independent of the $k_n^-$ and $k_n^\perp$, so that the
above $n$-collinear matrix element is proportional to
$\delta(x^+)\delta(x_\perp)$~\cite{Bauer:2001yt}.  Similarly the $\bn$-collinear
matrix element is $\propto \delta(y^-)\delta(y_\perp)$. It is not crucial to use
these $\delta$-functions at this stage, but they do allow us to simplify the
formula by dropping $x^+$, $x_\perp$, $y^-$, and $y_\perp$ dependence in the
exponentials.  Performing a few integrals we arrive at a fairly simple form for
the cross-section
\begin{align}
\label{factorizedcross-section-pre-res}
\sigma\ = & \
  \sigma_0\:
  \big |C(Q,\mu)\big |^2 {\cal M} \!
   \int\!\! dk_n^+ \, dk_\bn^-\, dk_s^+ \, dk_s^-
 \\
& \times \sum^{res.}_{X_n} \frac{1}{2\pi} \int\!\! d^4 x \: e^{ik_n^+ x^-/2} \
  \big\langle 0 \big| \slash\!\!\!\hat \bn \chi_n(x)  \big| X_n \big\rangle 
\big\langle X_n \big|\bCH n Q (0) \big| 0 \big\rangle 
\nn\\
&\times \sum^{res.}_{X_\bn} \frac{1}{2\pi} \int\!\! d^4 y \: e^{ik_\bn^- y^+/2} \
  \big\langle 0 \big|\overline\chi_\bn (y) \big| X_\bn \big\rangle
\big\langle X_\bn \big| \slash\!\!\!\hat n \chi_{\bn,-Q} (0) \big| 0 \big\rangle 
\nn\\
&\times  \sum^{res.}_{X_s} \!\! \:\frac{1}{4N_c (2\pi)^2} \int\!\! dz^+ dz^-  \:
    e^{\frac{i}{2} (k_s^+z^- + k_s^-z^+) }\ 
 \langle 0| \overline {Y}_\bn\,  {Y}_n (z^-,z^+) |X_s \rangle
  \langle X_s| {Y}^\dagger_n\,  \overline {Y}_\bn^\dagger (0) |0\rangle 
  \,. \nn
\end{align}
The result in Eq.~(\ref{factorizedcross-section-pre-res}) is a
factorized product of Fourier transforms over $n$-collinear,
$\bn$-collinear, and soft matrix elements. We introduced a $1/N_c$ in
front of the soft-matrix element in
Eq.~(\ref{factorizedcross-section-pre-res}), and include a
compensating factor $N_c$ in $\sigma_0$. This equation provides a good
starting point for the derivation of any differential cross-section
(for massive or massless dijet events).  The new normalization factor
$\sigma_0$ is just the total Born cross-section
\begin{eqnarray}
 \sigma_0 &=& N_c \frac{Q^2}{8\pi} \, K_0
     = N_c\, \frac{4\pi \alpha^2}{3 Q^2} 
   \bigg[\, e_t^2 - 
   \frac{2 Q^2\, v_e v_t e_t}{Q^2-m_Z^2} + 
   \frac{Q^4 (v_e^2+a_e^2)(v_t^2+a_t^2)}{(Q^2-m_Z^2)^2} \bigg] \,.
\end{eqnarray}
For massive quarks
$\sigma_0$ depends on $\beta_m=(1-4m^2/Q^2)^{1/2}$ through an extra
multiplicative factor of $\beta_m (3-\beta_m^2)/2 = 1 - 6
m^4/Q^4+\ldots$. This is only a 1\% correction to $\sigma_0$ for
$Q/m\sim 5$.

To proceed further we now need to make the prescription how the $n$- and $\bar
n$-collinear and the soft particles enter the invariant masses $s_t$ and
$s_{\bar t}$ explicit. This removes the implicit restrictions on the sums 
over states indicated in Eq.~(\ref{factorizedcross-section-pre-res}). In the
next subsection we implement the prescriptions 
for the hemisphere jet invariant masses. In Sec.~\ref{sectionotheralgo} we
briefly discuss how the implementation changes for other invariant mass
prescriptions.

\subsection{Factorization for Hemisphere invariant masses in SCET}
\label{section_hemi}

In the hemisphere mass case all the final state particles are assigned
to belong to one of two hemispheres defined with respect to the thrust
axis. The boundary between the two hemispheres is perpendicular to the
thrust axis and centered at the $e^+e^-$ collision point, see
Fig.~\ref{fig:topjet}. Thus the top and antitop jets we consider
correspond to all the particles in the respective two hemispheres and
the invariant mass of each jet is defined to be the total invariant
mass of all the final state particles in each hemisphere. As we show
explicitly below, the requirement that these jet invariant masses are
both close to the top mass, automatically restricts the final state to
be dijet-like, and eliminates the need to introduce any additional
event-shape constraint. We stress that some mechanism to control the
soft particles is absolutely crucial for establishing the
factorization theorem and the unique definition of the soft function
$S$. Here this is accomplished by the fact that all soft particles
enter the invariant mass variables $M^2_{t,\bar t}$.

The invariant mass of each hemisphere includes contributions from both
soft and collinear particles. The total momentum of the collinear
particles in the $n$-hemisphere is $P_{X_n}$ and in the
$\bn$-hemisphere is $P_{X_\bn}$.  The total final state soft momentum
$K_{X_{s}}$ is split between the two hemispheres and can be divided
as:
\begin{eqnarray}
K_{X_s} = k^a_s + k^b_s
\end{eqnarray}
where $k^a_s$ and $k_s^b$ correspond to the total momenta of all the soft
partons in the $n$ and $\bn$ hemispheres, respectively. It is useful to think of these
hemisphere momenta as the result of hemisphere projection operators
$\hat{P}_a,\hat{P}_b$:
\begin{eqnarray} \label{PaPb}
\hat{P}_a\> | X_s\rangle = k^a_s \>| X_s\rangle,
  \>\>\>\> \hat{P}_b \>|X_s\rangle = k_s^b \>| X_s\rangle.
\end{eqnarray}
In other words, these projection operators act on each state
$|X_s\rangle$, pick out the soft partons in the respective hemisphere
and add up their total momentum. Note that the eigenvalues are
dependent on the state $X_s$, so $k_s^a=k_s^a[X_s]$ and
$k_s^b=k_s^b[X_s]$.  We can now define the invariant mass of each jet
as $(P_{X_n} + k^a_s)^2$ and $(P_{X_\bn} + k^b_s)^2$ for the $n$ and
$\bn$ hemispheres respectively.  The delta functions $\delta ^4 (p_n -
P_{X_n})\, \delta ^4 (p_\bn - P_{X_\bn})$ in the second line of
Eq. (\ref{factorizedcross-section-new}) allow us to define the jet
invariant masses in terms of $p_n, p_\bn$ as $(p_{n} + k^a_s)^2$ and
$(p_{\bn} + k^b_s)^2$ for the $n$ and $\bn$ hemispheres respectively.
 
Note that this implements a very simple form of a jet algorithm. For a different
jet algorithm we would change the definitions of the operators $\hat P_a$ and
$\hat P_b$.  Running a jet algorithm in inclusive $e^+e^-$
mode~\cite{Catani:1991hj,Butterworth:2002xg} each soft parton is still accounted for, having a
certain probability of being assigned to either the top or the antitop invariant
mass. We discuss other algorithms in Sec.~\ref{sectionotheralgo}.

If the top quark were a stable particle, these invariant mass definitions would
be obvious because $n$- and $\bar n$-collinear particles would be fully radiated
into the $n$- and $\bar n$-hemispheres, respectively.  Due to the finite
lifetime of the top quark, however, we need to convince ourselves that this
invariant mass definition still works if the $n$- and $\bar n$-collinear momenta
of the top and antitop quarks, respectively, are distributed among their decay
products. So let us consider the top quark in the $n$-hemisphere.  Since the top
rest frame is boosted with respect to the $e^+e^-$ c.m.\,frame with a boost
factor $Q/m$, top decay events can have final state particles appearing in the
$\bar n$-hemisphere of the antitop quark only if these final state particles
have an angle (defined in the top rest frame) smaller than $m/Q$ with respect to
the antiparticle direction. On the other hand, the top spin is only about $20\%$
polarized (for unpolarized $e^+e^-$ beams and upon averaging over the directions
of the thrust axis)~\cite{Parke:1996pr}, and thus the top decay products in the
top rest frame are distributed isotropically to a rather good approximation. The
fraction of events in this kinematical situation is therefore suppressed by
$(m/Q)^2$ and can be neglected at leading order in the power counting. Of course
the analogous conclusion also applies to the antitop quark in the $\bar n$
hemisphere.  So at leading order in the power counting it is consistent to
employ the invariant mass definition of the previous paragraph.

The jet invariant mass definitions can be implemented into the cross-section of
Eq.~(\ref{factorizedcross-section-pre-res}) by inserting underneath the
$\sum_{X_s}$ the identity relation
\begin{eqnarray}
\label{identity-2}
1&=& \int \!\! dM_t^2 \> \delta \big((p_n + k_s^{a})^2-M_t^2\big) 
 \int d M^2_{\bar t}\> \delta \big((p_\bn +k_s^{b} )^2 -M_{\bar t}^2 \big)
 \nn\\
 &=& \int \!\! dM_t^2 \> \delta \big((p_n + k_s^{a})^2-m^2-s_t\big) 
 \int d M^2_{\bar t}\> \delta \big((p_\bn +k_s^{b} )^2 -m^2 - s_{\bar t} \big),
\end{eqnarray}
where $s_t=s_t(M_t)$ and $s_{\bar t}=s_{\bar t}(M_{\bar t})$ from
Eq.~(\ref{massshell}), i.e. it should be understood that $s_{t,\bar
t}$ are functions of $M_{t,\bar t}^2$.  In the second line $m$ is
defined as the pole mass. It is straightforward to switch the final
result to a suitable short distance mass definition, as we explain in
Sec.~\ref{sec:sdmass}.  Decomposing the $\delta$-functions at leading
order gives
\begin{eqnarray} \label{decomposedelta-3}
\delta ((p_n+k_s^a)^2-m^2-s_t) &=& \frac{1}{Q} \> \delta \Big(k_n^+ +k_s^{+a} -
\frac{m^2 + s_t}{Q}\Big) \,, \nn \\
\delta ((p_\bn+k_s^b)^2-m^2-s_{\bar t}) &= &  \frac{1}{Q} \> \delta \Big(k_\bn^-
+k_s^{-b}- \frac{m^2 + s_{\bar t}}{Q} \Big) \,,
\end{eqnarray} 
where we set $\tilde{p}_n^-=\tilde{p}_\bn^+=Q$ due to
$\delta$-functions from Eq.~(\ref{decomposedelta-2}).  Carrying out
the integration over $k_s^+$ and $k_s^-$ in
Eq.~(\ref{factorizedcross-section-pre-res}) sets the arguments of the
soft function to $z^\pm=0$.  Inserting the identity relation
\begin{eqnarray}
  1 = \int\!\!  d\ell^+ d\ell^-  \delta(\ell^+ - k_s^{+a}) \delta(\ell^- - k_s^{-b})
\end{eqnarray}
the differential cross-section then reads
\begin{align}
\label{factorizedcross-section-new-1}
\frac{d^2\sigma}{dM_t^2 dM_{\bar t}^2} &=
 \frac{\sigma_0}{Q^2} \:
  \big |C(Q,\mu)\big |^2 {\cal M} \!
   \int\!\! dk_n^+ \, dk_\bn^-\, d\ell^+ \, d\ell^-  
  \delta \Big(k_n^+ \!+\ell^+\! -\! \frac{m^2 \!+\! s_t}{Q}\Big)
  \delta \Big(k_\bn^- \! +\ell^-\! -\! \frac{m^2 \!+\! s_{\bar t}}{Q} \Big) 
\nn\\
& \times \sum_{X_n} \frac{1}{2\pi} \int\!\! d^4 x \: e^{ik_n^+ x^-/2} \
  {\rm tr} \big\langle 0 \big| \slash\!\!\!\hat \bn \chi_n(x)  \big| X_n \big\rangle 
\big\langle X_n \big|\bCH n Q (0) \big| 0 \big\rangle 
\nn\\
&\times \sum_{X_\bn} \frac{1}{2\pi} \int\!\! d^4 y \: e^{ik_\bn^- y^+/2} \
  {\rm tr}  \big\langle 0 \big|\overline\chi_\bn (y) \big| X_\bn \big\rangle
\big\langle X_\bn \big| \slash\!\!\!\hat n \chi_{\bn,-Q} (0) \big| 0 \big\rangle 
\nn\\
&\times  \sum_{X_s} \!\! \:\frac{1}{ N_c} 
  \delta(\ell^+ - k_s^{+a}) \delta(\ell^- - k_s^{-b})
 {\rm tr}  \langle 0| \overline {Y}_\bn\,  {Y}_n (0) |X_s \rangle
  \langle X_s| {Y}^\dagger_n\,  \overline {Y}_\bn^\dagger (0) |0\rangle 
   \,, 
\end{align}
where we have dropped the ``res.'' label on the sums, because all restrictions
are now explicitly implemented.

To see that the hemisphere definition of $s_t$ and $s_{\bar t}$ can be used to
select dijet-like events, we can check that Eq.~(\ref{sn-range}) plus the
$\delta$-functions in Eq.~(\ref{factorizedcross-section-new-1}) are sufficient
to constrain the thrust to the dijet region. At leading order the total thrust
of an event is given by 
\begin{align}
  Q T = | p_n^z | + | p_\bn^z | + |k_s^{a\,z}| + |k_s^{b\,z}| \,,
\end{align}
where $2|p_n^z| = Q + k_n^- - k_n^+$ and $2|p_\bn^z| = Q + k_\bn^+ - k_\bn^-$.
So the shifted thrust defined in Eq.~(\ref{tau}) is
\begin{align} \label{tauss}
  \tau &= -\frac{2m^2}{Q^2} + \frac{1}{2Q}\Big[ k_n^+ - k_n^- + k_\bn^- - k_\bn^+
  + k_s^{a\,+} - k_s^{a\,-} + k_s^{b\,-} - k_s^{b\,+} \Big]  \nn\\
   &= -\frac{2m^2}{Q^2} + \frac{1}{Q}\Big[ k_n^+  + k_s^{a\,+}
   + k_\bn^- +  k_s^{b\,-}    \Big] \nn\\
   &= \frac{s_t + s_{\bar t} }{Q^2} \,.
\end{align}
To obtain the second line we used the separate conservation of the 
$+$ and $-$ momentum components to eliminate
$k_n^- + k_s^{b\,-}$ and $k_\bn^+ + k_s^{a\,+}$. For the last line we used
the $\delta$-functions in Eq.~(\ref{decomposedelta-3}) to get $s_t$ and $s_{\bar
  t}$. Thus, the restriction to small hemisphere invariant masses $s_{t,\bar
t}$ automatically gives small $\tau$ and restricts the events to the dijet
region. The presence of a third hard jet takes us away from the
dijet region and directly shows up by a substantial positive shift of $s_t +
s_{\bar t}$ away from the peak region. 

Next we simplify the form of the cross-section by defining the massive
collinear jet functions $J_n, J_\bn$ as
\begin{align}
\label{jetfunc-1}
 \sum_{X_n} {\rm tr}\,
\big\langle 0 \big| \slash\!\!\!\hat \bn \chi_n(x)  \big| X_n \big\rangle 
\big\langle X_n \big|\bCH n Q (0) \big| 0 \big\rangle 
&=
 Q\int \frac{d^4r_n}{(2\pi)^3} e^{-i r_n\cdot x} 
J_{n}(Qr_n^+ -m^2,m) 
 \\
&
 = Q\, \delta(x^+) \delta ^2(x_\perp ) \int \!\!{dr^+_n}\:
  e^{-\frac{i}{2} r_n^+ x^-}\:  J_{n}(Qr_n^+ -m^2,m)\,,
\nn\\
\sum_{X_\bn} {\rm tr}\,
\big\langle 0 \big|\overline\chi_\bn (y) \big| X_\bn \big\rangle
\big\langle X_\bn \big| \slash\!\!\!\hat n \chi_{\bn,-Q} (0) \big| 0 \big\rangle 
&=
  Q \int \frac{d^4r_\bn}{(2\pi)^3} e^{- i r_\bn\cdot y} 
J_{\bn}(Qr_\bn^- -m^2,m) 
 \nn \\
&
 = Q\, \delta(y^-) \delta ^2(y_\perp ) \int \!\! {dr^-_\bn}\:
  e^{-\frac{i}{2} r_\bn^-  y^+}\:  J_{\bn}(Qr_\bn^--m^2,m)
 \,. \nn
\end{align}
Here $m$ is the pole mass just as in Eq.~(\ref{identity-2}) and we do
not display the $\mu$ dependence.  Note that the subscript $Q$ on the
LHS does not change the mass-dimension of a $\chi$-field away from
$3/2$, since $\delta_{Q,\bnP}$ is dimensionless.  We remind the reader
that $\slash\!\!\!\hat n =\nslash/(4N_c)$, $\hat \bn\!\!\! \slash =
\bnslash/(4N_c)$ and that tr is a trace over both color and spin
indices.  The arguments of the jet functions, $J_{n}$ and $J_{\bn}$,
in Eq.~(\ref{jetfunc-1}) are just the off-shellness of the jets,
$p_n^2-m^2$ and $p_\bn^2-m^2$, respectively, but given in expanded
form.  Here the labels $Q$ on the $\overline\chi_n$ and $\chi_{\bn}$
fields ensures that there is only a contribution from the required
``quark'' and ``antiquark'' cut since $Q>0$. To see this recall that
the sign of the label $p$ on $\xi_{n,p}$ picks out the quark
annihilation, or antiquark production part of the
field~\cite{Bauer:2001ct}. We note that the sums over collinear states
in the collinear jet functions are unrestricted since the restrictions
are now implemented automatically through the amount the jet invariant
mass differs from $m^2$. Thus, the jet functions can be written as the
discontinuity of a forward scattering amplitude after summing over the
collinear states:
\begin{eqnarray}
\label{jetfunc2}
J_{n}(Qr_n^+ - m^2,m) 
&=& 
\frac{-1}{2\pi Q}\, \textrm{Disc}\! \int\!\! d^4 x \: e^{i r_n\cdot x} \,
\langle 0|\text{T}\{ \bCH n Q (0)\slash\!\!\!\hat \bn  \chi_n(x)\}|0 \rangle \, ,
\nn
\\
 J_{\bn}(Qr_\bn^- - m^2,m) 
&=& 
 \frac{1}{2\pi Q }\, \textrm{Disc}\! \int\!\! d^4 x \: e^{i r_\bn\cdot x} \,
\langle 0|\text{T}\{ \bar \chi_\bn  (x) \slash\!\!\!\hat n \chi_{\bn,-Q}(0)\} |0 \rangle \, .
\end{eqnarray}
The collinear fields in the SCET jet functions $J_n$ and $J_\bn$ are defined
with zero-bin subtractions~\cite{Manohar:2006nz}, which avoids double counting
with the soft-function.  Using Eq.(\ref{jetfunc-1}) and performing all the
remaining integrals in the cross-section of
Eq.(\ref{factorizedcross-section-new-1}) we arrive at the SCET result for double
differential hemisphere invariant mass cross-section
\begin{align}
\label{SCETcross-hem}
 \frac{d^2\sigma }{dM^2_t\>dM^2_{\bar t}} &= 
    \sigma_0
     \> H_Q(Q,\mu)\: {\cal M}(m,\mu) \\
  &\ \times \int_{-\infty}^{\infty}\!\!\!d\ell^+ d\ell^-
  \> J_n(s_t - Q\ell^{+},m,\mu)  J_\bn(s_{\bar t} - Q\ell^{-},m,\mu) 
  S_{\rm hemi}(\ell^+,\ell^-,\mu,m)
  \,, \nn
\end{align}
where the hard function $H_Q(Q,\mu) = | C(Q,\mu)|^2$. Here the hemisphere soft
function in SCET is
defined by 
\begin{align} \label{SSS}
 S_{\rm hemi}(\ell^+,\ell^-,\mu,m) &=  \frac{1}{N_c} \sum _{X_s} 
 \delta(\ell^+ - k_s^{+a}) \delta(\ell^- - k_s^{-b})
  \langle 0| \overline {Y}_\bn\,  {Y}_n (0) |X_s \rangle
\langle X_s| {Y}^\dagger_n\,  \overline {Y}_\bn^\dagger (0) |0\rangle   \,.
\end{align}
At tree level for stable top quarks $H_Q=1$, $J_{n}(s_t) = \delta(s_t)$,
$J_{\bn}(s_{\bar t}) = \delta(s_{\bar t})$, and $S_{\rm hemi}(\ell^+,\ell^-) =
\delta(\ell^+)\delta(\ell^-)$, and integrating Eq.~(\ref{SCETcross-hem}) over
$s_t$ and $s_{\bar t}$ gives the total tree-level Born cross-section $\sigma_0$.
This provides a check for the normalization of Eq.~(\ref{SCETcross-hem}).  The
argument $m$ on the soft-function in Eq.~(\ref{SSS}) and ${\cal M}(m,\mu)$ in
Eq.~(\ref{SCETcross-hem}) account for massive top-quark bubbles that are
perturbative and start at ${\cal
  O}(\alpha_s^2(m))$~\cite{Kniehl:1989kz,Burgers:1985qg,Hoang:1995ex}. Note that
Eq.~(\ref{SCETcross-hem}) extends the SCET computation of the massless dijet
cross-section in Ref.~\cite{Bauer:2002ie,Bauer:2003di} to all orders in
perturbation theory for the jet-functions.

In the factorization theorem in Eq.~(\ref{SCETcross-hem}) the jet-functions
$J_n$ and $J_\bn$ describe the dynamics of the top and antitop jets.  In the
next section we will see that these jet functions can be computed in
perturbation theory and at the tree level are just Breit-Wigner distributions.
The soft matrix elements $ \langle 0| {Y}^\dagger_n {Y}_\bn (0)|X_s \rangle
\langle X_s| \tilde{Y}^\dagger_\bn \tilde{Y}_n (0)|0\rangle$, on the other hand,
depends on the scale $\Lambda _{QCD}$, and thus the soft function $S_{\rm
  hemi}(\ell^+,\ell^-)$ is governed by non-perturbative QCD effects. The
momentum variables $\ell^\pm$ represent the light cone momentum of the soft
particles in each of the two hemispheres, and $S_{\rm hemi}(\ell^+,\ell^- )$
describes the distribution of soft final state radiation.
Eq.~(\ref{SCETcross-hem}) already demonstrates that the invariant mass spectrum
for unstable top quarks is not a Breit-Wigner function even at tree level
because the convolution with the soft function $S_{\rm hemi}$ modifies the
observed distribution. The effects of the convolution on the observable
invariant mass distribution are discussed in Sec.~\ref{section4}.

To sum large logs in Eq.~(\ref{SCETcross-hem}) the SCET production current can
be run from $\mu=Q$ down to $\mu= m$, which then characterizes the typical
virtuality of the collinear degrees of freedom in massive SCET. In the process,
large logarithms of $Q/m$ are summed into the hard function $H_Q(Q,\mu)$.  In
the next section we integrate out the scale $m$ and match these SCET jet
functions onto bHQET jet functions.

\subsection{Factorization of Jet mass effects in HQET}
\label{subsectionfactorizationtheorem}

The main result of the last subsection is the factorization of the
scales $Q$ and $m$ in the differential cross section of
Eq.~(\ref{SCETcross-hem}).  In this section we further factorize the
scale $m$ from the low energy scales $\Gamma$, $\hat s$, and
$\Delta$. This will allow us to sum large logs of $\Gamma/m$ and $\hat
s_{t,\bar t}/m$ in the jet functions, and lower the scale of the soft
functions to $\Delta$.  This step is also important for treating the
width effects. As explained earlier, one can formulate width effects
in a gauge invariant way with a natural power counting in HQET,
whereas doing so in a relativistic theory such as SCET is notoriously
difficult.

To perform the scale separation and sum the logarithms requires us to match and
run below the scale $\mu=m$. This can be done in a standard way, by matching and
running of the bHQET current in Eq.~(\ref{JbHQET}), as we described in
Sec.~\ref{buHQET}.  However, due to the factorization properties of SCET
which leads to a decoupling of the $n$-collinear, $\bn$-collinear, and soft
sectors, the matching and running below the scale $\mu=m$ can also be done
independently for $J_n$, $J_\bn$, and $S$. In the following we explain this
second method.

\begin{figure}
  \centerline{ 
   \includegraphics[width=15cm]{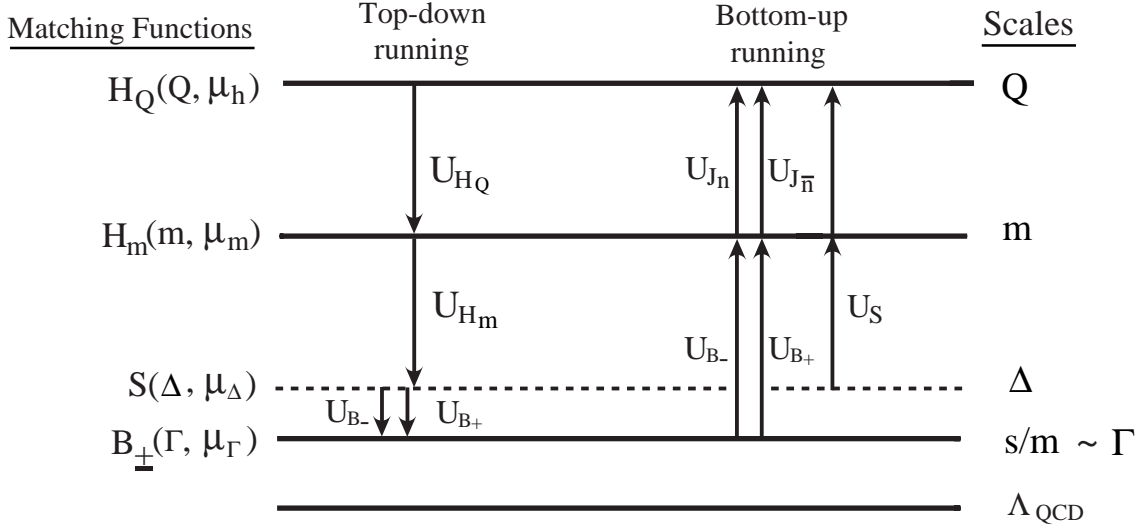}  
  } 
\caption{Scales and functions appearing in the formula for the
  invariant mass distribution. The result is determined by matching at the
  physical scales and running to sum large logs as shown. We show both the
  top-down and bottom-up approach to the running. The evolution for $U_H$ and
  $U_C$ is local, while all other evolution functions involve convolutions. Note
  that the evolution functions obey $U_H=U_{J_-}\otimes U_{J_+}\otimes U_S$ and
  $U_C=U_{B_-}\otimes U_{B_+}\otimes U_S$ where $\otimes$ indicates
  convolutions.  }
\label{fig:theory}
\end{figure}

As discussed in Sec.~\ref{buHQET} the soft function above and below the scale
$m$ is identical, except for certain vacuum polarization effects from graphs
with top-quark bubbles that only exist in SCET. For the soft-function in bHQET
we have
\begin{align}
 S_{\rm hemi}(\ell^+,\ell^-,\mu) &=  \frac{1}{N_c} \sum _{X_s} 
 \delta(\ell^+ - k_s^{+a}) \delta(\ell^- - k_s^{-b})
  \langle 0| \overline {Y}_\bn\,  {Y}_n (0) |X_s \rangle
\langle X_s| {Y}^\dagger_n\,  \overline {Y}_\bn^\dagger (0) |0\rangle   \,,
\end{align}
where there is no $m$ dependence. The matching condition between the soft
functions in the two theories is
\begin{align}
 S_{\rm hemi}(\ell^+,\ell^-,\mu,m ) = T_0(m,\mu) S_{\rm hemi}(\ell^+,\ell^-,\mu)
 \,,
\end{align}
where $T_0(m,\mu)$ is a Wilson coefficient.  Large logarithms in the soft
function can be summed by computing the anomalous dimension of the soft function
and using RG evolution to run between $\Delta$ and $m$ and between $m$ and $Q$
as illustrated by the line labeled $U_S$ in Fig.~\ref{fig:theory}.

For the SCET collinear degrees of freedom the power counting for the
virtuality is $p_c^2 \sim m^2$. Thus, $J_n$ and $J_\bn$ describe the
physics of jets with an invariant mass up to $M^2\sim \mu^2 \sim
m^2$. However, the restriction of being in the peak region means that
$M^2-m^2\ll m^2$. This disparity gives rise to the large
logarithms in the collinear jet functions.  Intuitively, this can also
be understood by noting that if one starts out with a top quark that
is close to its mass shell, a typical collinear SCET gluon will knock
the top far offshell so that $p_{c}^2- m^2 \sim m^2 $.  By restricting
the jet functions to $p_{c}^2-m^2 \ll m^2$ we forbid such real
radiation contributions, but not virtual contributions. The latter must be
integrated out explicitly by switching to the description of the jet
functions in the boosted unstable HQET theories discussed in
Sec.~\ref{buHQET}. In these HQETs the only fluctuations are due to low
energy ultracollinear gluons that preserve the condition $M^2 -m^2
\ll m^2$.

To determine the definitions of the bHQET jet functions we follow the
same procedure as for the bHQET current in Eq.~(\ref{JbHQET}), namely
boost the SCET jet function in Eq.~(\ref{jetfunc2}) to the heavy quark
rest frame, giving $\bar \psi(x) W(x) W(0)\psi(0)$, then match onto
HQET $\psi(x)\to h_v(x)$. We then boost back to the moving frame where
$v\to v_\pm$. The spin structure can also be simplified to give
\begin{align} 
 \frac{1}{Q} \overline\chi_{n,Q} \hat\bnslash \chi_n \to 
 \frac{1}{Q}\, \bar h_{v_+} \hat\bnslash\, h_{v_+} = 
\frac{v_+\cdot\bn}{4N_c Q}\, \bar  h_{v_+} h_{v_+}
  =   \frac{1}{4N_c m}\, \bar h_{v_+} h_{v_+} \,.
\end{align}
Thus the bHQET jet functions are defined as
\begin{eqnarray}
\label{hqetjet}
B_+(2v_+\mcdot k) &=&
\frac{-1}{8\pi N_c m} \, \textrm{Disc}\! \int\!\! d^4 x \: e^{i k \cdot x} \, 
 \, 
\langle 0|{\rm T}\{\bar{h}_{v_+}(0) W_n(0) W_n^{\dagger}(x) h_{v_+} (x)\}|0 \rangle\, ,
\nn
\\
 B_{-}(2v_-\mcdot k) &=& 
\frac{1}{8\pi N_c m} \, \textrm{Disc}\! \int\!\! d^4 x \: e^{i k\cdot x}  \, 
\langle 0|{\rm T}\{\bar{h}_{v_-}(x) W_\bn(x) W_\bn^{\dagger}(0) h_{v_-} (0)\}|0 \rangle .
\end{eqnarray}
These bHQET jet functions can be calculated using the usual Feynman rules of
HQET except that the gluons have ucollinear scaling as in Eq.~(\ref{BHQETres}).
The $W$-Wilson lines in $B_\pm$ also contain these boosted gluons. Since $p_n^2
-m^2=2m v_+ \cdot k$ and $p_\bn^2-m^2 = 2 m v_-\cdot k$, we can identify
the arguments of the bHQET jet functions as
\begin{eqnarray}
  2 v_+\cdot k = \frac{s_t}{m} = \hat s_t \,,\qquad\qquad
  2 v_-\cdot k = \frac{s_{\bar t}}{m} = \hat s_{\bar t} \,.
\end{eqnarray} 
In the factorization theorem these arguments are shifted by the soft gluon
momenta as shown in Eq.~(\ref{bHQETcross-hem}) below. Recall that the fields
$h_{v_+}$ and $h_{v_-}$ are defined with zero-bin subtractions on their
ultracollinear momenta. For Eq.~(\ref{hqetjet}) these subtractions can be
thought of as being inherited from the SCET fields in the matching. They remove
the light-cone singularities $n\cdot k \to 0$ and $\bn\cdot k\to 0$ in $B_+$ and
$B_-$ respectively, and are important to ensure that the width $\Gamma$ is
sufficient to make $B_{\pm}$ infrared finite.

In general the matching of the jet functions in SCET onto those in bHQET could
take the form
\begin{eqnarray}\label{matchscetbhqet} 
J_{n,\bn}(m\hat s,m,\Gamma,\mu ) 
   = \int_{-\infty}^\infty \!\! d\shat' \: \>T_{\pm} (\shat,\shat',m,\mu )\>
B_\pm(\shat',\Gamma,\mu ),
\end{eqnarray} 
where the convolution takes into account the fact that depending on
the definition, the observable $\hat s$ could be sensitive to scales
of ${\cal O}(m)$ and ${\cal O}(\Gamma)$. In such a case, since $\hat
s'$ does not know about the scale $m$, it can not be identical to
$\hat s$. The convolution with $T_{\pm}(\hat s,\hat s',m,\mu)$ then
compensates for this difference. In our case (and most reasonable
cases) the definition of the invariant mass is not sensitive to $m$,
so we have $T_\pm(\hat s,\hat s',m,\mu) =
\delta(\hat s-\hat s') T_\pm(m,\mu)$ and the matching equations are simply
\begin{align} \label{matchscetbhqet2}
  J_n(m\hat s,m,\Gamma,\mu_m) &= T_+(m,\mu_m)\: B_+(\shat,\Gamma,\mu_m) \,,\nn\\
  J_\bn(m\hat s,m,\Gamma,\mu_m) &= T_-(m,\mu_m)\: B_-(\shat,\Gamma,\mu_m) \,.
\end{align}
Since there are no mass-modes in bHQET the function ${\cal M}$ also appears as
part of the Wilson coefficient.  From this we define a hard-coefficient that
contains the mass corrections as\footnote{In explicit computations scheme
  dependence may effect the manner in which the mass-corrections are divided up
  between $T_\pm$, $T_0$, and ${\cal M}$, however this dependence will cancel
  out in the product that gives $H_m$.}
\begin{align} \label{Cm}
 H_m(m,\mu_m) = T_+(m,\mu_m)T_-(m,\mu_m) T_0(m,\mu_m) {\cal M}(m,\mu_m) \,.
\end{align}
By charge conjugation we know that the jet functions for the top and antitop
have the same functional form, and that $T_+ = T_-$. When we sum large logs into
the coefficient $H_m$ it develops an additional dependence on $Q/m$ through its
anomalous dimension which depends on $v_+\cdot \bn = v_-\cdot n =Q/m$. Note that
in principle $H_m(m,\mu)$ and the factors in Eq.~(\ref{Cm}) can also have $Q/m$
dependence at NNLL order. For related discussions see
Refs.~\cite{Becher:2007cu,Chiu:2007yn}.

Since the
functions $T_\pm$ are independent of the top width $\Gamma$, we are
free to set $\Gamma=0$ (i.e.\,use stable top quarks) for the matching
calculations at any order in perturbation theory.  At tree level we
need to compute the
\begin{figure}
  \centerline{ 
   \includegraphics[width=12cm]{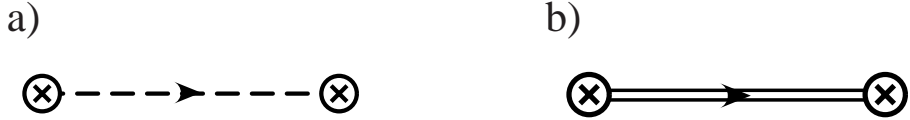}  
  } 
\caption{Tree level top-quark jet functions in a) SCET and b) bHQET.  }
\label{fig:Bjet}
\end{figure}
discontinuity of the graphs in Fig.~\ref{fig:Bjet} which have a trace over spin
and color indices. For $\Gamma=0$ this gives 
\begin{align}
  B_{+}^{\rm tree}(\hat s,{\Gamma=0})
  &= \frac{-1}{8\pi N_c m} (-2N_c)\: {\rm Disc} \Big( \frac{i}{v_+ \cdot k +
    i0} \Big)
    = \frac{1}{4\pi m} {\rm Im} \Big( \frac{-2 }{v_+ \cdot k + i0} \Big) \nn\\
    &= \frac{1}{m}\: \delta(2v_+\cdot k) 
     = \frac{1}{m}\: \delta(\hat s) =  \delta(s) \,,
\end{align}
which is identical to the result for the corresponding SCET jet function, so
at tree level $T_+=T_-=1$.

Plugging Eq.~(\ref{matchscetbhqet2}) into Eq.~(\ref{SCETcross-hem}), and 
incorporating renormalization group evolution, the form for the differential cross
section is
\begin{align}
\label{bHQETcross-hem}
\left( \frac{d^2\sigma }{dM^2_t\>dM^2_{\bar t}} \right)_{\rm hemi} \!\!\! &= 
    \sigma_0
     \> H_Q(Q,\mu_m) H_m\Big(m,\frac{Q}{m},\mu_m,\mu\Big) \\
 &\quad\times \int_{-\infty}^{\infty}\!\!\!d\ell^+ d\ell^-
  \> B_+\Big(\hat s_t- \frac{Q\ell^{+}}{m},\Gamma,\mu\Big) \:
     B_-\Big(\hat s_{\bar t} - \frac{ Q\ell^{-}}{m},\Gamma,\mu\Big)\: 
  S_{\rm hemi}(\ell^+,\ell^-,\mu)
  . \nn
\end{align}
Eq.~(\ref{bHQETcross-hem}) is our final result in terms of the pole mass $m$.
The analogous result for a short distance mass is given in the next section.
Here $H_m(m,Q/m,\mu_m,\mu)$ is the hard coefficient $H_m(m,\mu_m)$ run down from
$\mu_m$ to $\mu$, and we still have $H_Q(Q,\mu_m) = | C(Q,\mu_m)|^2$, and the
soft function with Wilson lines evaluated at $x=0$,
\begin{align}
 S_{\rm hemi}(\ell^+,\ell^-,\mu) &= \frac{1}{N_c} \sum _{X_s} 
 \delta(\ell^+ \minus k_s^{+a}) \delta(\ell^- \minus k_s^{-b})
  \langle 0| (\overline {Y}_{\!\bn})^{ca'} ({Y}_n)^{cb'} |X_s \rangle
\langle X_s| ({Y}^\dagger_n)^{b'c'} (\overline {Y}_{\!\bn}^\dagger)^{a'c'}
|0\rangle  
 \,.
\end{align}
For completeness we wrote out the color indices from Eq.~(\ref{factor-m-elt}).
Its interesting to note that in the result in Eq.~(\ref{bHQETcross-hem}) the 
final matrix elements only involve Wilson lines (since the coupling of gluons to
a heavy quark field $h_{v_+}$ in $B_+$ is the same as to a Wilson line $W_{v_+}$).

To conclude this section we finally repeat the computation of the tree level
bHQET jet functions, but now for the realistic case with $\Gamma\ne 0$ in the
HQET propagators. The computation is done at a scale $\mu\gtrsim \Gamma$, but the
$\mu$ dependence does not show up at tree level.  Fig.~\ref{fig:Bjet}b gives
\begin{align}
\label{Bpmtree}
  B_\pm^{\rm tree} (\hat s,\Gamma) 
  &= \frac{-1}{8\pi N_c m} (-2N_c)\: {\rm Disc} \Big( \frac{i}{v_\pm \cdot k +
    i\Gamma/2} \Big)
    = \frac{1}{4\pi m} {\rm Im} \Big( \frac{-2 }{v_\pm \cdot k + i\Gamma/2} \Big) \nn\\
    &= \frac{1}{\pi m}\: \frac{\Gamma}{\hat s^2 + \Gamma^2} \,.
\end{align}
Thus we see that $B_\pm(\hat s)$ are equal to Breit-Wigners at lowest
order in $\alpha_s$. At higher orders in perturbation theory the width
will cut off the IR divergences that would otherwise occur at $\hat
s=0$. The functions $B_\pm$ at the scale $\mu$ can therefore be
computed perturbatively to any desired order in $\alpha_s$. In general
the perturbative ``matching'' corrections will lead to distortions of
the tree-level Breit-Wigner distributions shown in
Eq.~(\ref{Bpmtree}), as does the potential separate running between
$\mu_\Delta$ and $\mu_\Gamma$ discussed below in Sec.~\ref{sec:RGE}
and shown in Fig.~\ref{fig:theory}.

\subsection{A Short-Distance Top-Mass for Jets} \label{sec:sdmass}

The derivation of the factorization formulae (\ref{bHQETcross-hem}) in
the previous section was given in the pole mass scheme\footnote{In
Eq.~(\ref{bHQETcross-hem}) we used $m$ for the pole mass, but in this
section we write $m_{\rm pole}$, and reserve ``m'' for a generic
mass-scheme.}, $m_{\rm pole}$. It is, however, well known that the
pole mass definition leads to an artificially enhanced sensitivity to
small momenta in Feynman diagrams (see Ref.~\cite{Beneke:1998ui} for a
review) and, as a consequence, to artificially large perturbative
corrections. This behavior is particularly important for observables
that have a strong dependence on the heavy quark
mass~\cite{Hoang:1998nz,Beneke:1998rk,Hoang:1998ng,Uraltsev:1998bk,Hoang:2000yr}.
From a nonperturbative point of view, this feature is related to an
intrinsic ambiguity in the heavy quark pole mass parameter of order
the hadronization scale $\Lambda_{\rm QCD}$, and is sometimes referred
to as the ${\cal O}(\Lambda_{\rm QCD})$-renormalon problem of the pole
mass. Heavy quark mass definitions that do not have such an ${\cal
O}(\Lambda_{\rm QCD})$ ambiguity are called short-distance mass
schemes.\footnote{In practice, determining the pole mass from the
analysis of experimental data leads to values that depend strongly on
the order of perturbation theory that has been employed for the
theoretical predictions. This makes the treatment of theoretical
errors difficult.}  In the factorization formulae in
Eq.~(\ref{bHQETcross-hem}), the top-mass appears in the hard function
$H_m$ and in the two jet functions $B_+(\hat s_t)$ and $B_-(\hat
s_{\bar t})$.  The most important sensitivity to the top-mass scheme
is in $\hat s_t= (M_t^2-m^2)/m$ and $\hat s_{\bar t}= (M_{\bar
t}^2-m^2)/m$, where $M_{t}^2$ and $M_{\bar t}^2$ are scheme
independent observables.

A specific short-distance top quark mass scheme ``$m$'' can be defined by a
finite residual mass term $\delta m\neq 0$, as
\begin{eqnarray} \label{chgscheme}
  m_{\rm pole} \, = \, m + \delta m \,,
\end{eqnarray}
where $\delta m$ starts at ${\cal O}(\alpha_s)$ or higher, and must be
strictly expanded perturbatively to the same order as other ${\cal
O}(\alpha_s)$ corrections.  (This strict expansion does not apply to
powers of $\alpha_s$ times logs that are summed up by renormalization
group improved perturbation theory.) Let $B_+(\hat s,\mu,\delta m)$
denote the jet-function in the short-distance mass scheme specified by
$\delta m$.  We can calculate $B_+(\hat s,\mu,\delta m)$ in two
equivalent ways. i) Use the pole-mass scheme initially by
setting $\delta m=0$ in Eq.~(\ref{LbHQET}). In this case the
mass-dependence appears in $\hat s_{\rm pole}=(M^2-m_{\rm
pole}^2)/m_{\rm pole}$ in $B_+$ and we change the scheme with
Eq.~(\ref{chgscheme}). Alternatively, ii) treat $\delta m\ne 0$ in
Eq.~(\ref{LbHQET}) as a vertex in Feynman diagrams, and take $\hat s$ to
be defined in the short-distance mass scheme right from the start, so
$\hat s= (M^2-m^2)/m$.

As discussed in Sec.~\ref{buHQET}, it is necessary that the residual mass
term is consistent with the bHQET power counting, i.e.~
\begin{eqnarray} \label{massscheme}
 \delta m\sim \hat s_t\sim \hat s_{\bar t}\sim\Gamma  \,.
\end{eqnarray}
Eq.~(\ref{massscheme}) restricts us to a suitable class of
short-distance mass schemes for jets.  In any short-distance mass scheme which
violates Eq.~(\ref{massscheme}) the EFT expansion breaks down, and thus the
notion of a top-quark Breit Wigner distribution becomes invalid. The most
prominent example for an excluded short-distance mass scheme is the
$\overline{\mbox{MS}}$ mass scheme, $\overline m$, for which $m_{\rm
  pole}-\overline m=\delta \overline m$.  Here $\delta \overline m \simeq
8\,{\rm GeV} \gg \Gamma$, or parametrically $\delta \overline m\sim\alpha_s
\overline m\gg\Gamma$. Using Eq.~(\ref{Bpmtree}) and converting to the
$\overline{\mbox{MS}}$ scheme with the ${\cal O}(\alpha_s)$ residual mass term
we have
\begin{eqnarray}
 B_+(\hat s, \mu,\delta  \overline m\,)
  &=& \frac{1}{\pi\overline m}\, \Bigg\{ \frac{\Gamma}{\big[\frac{(M_t^2-\overline
      m^2)^2}{\overline m^2} + \Gamma^2 \big]}
   \, + \, 
    \frac{(4\, \hat s\, \Gamma)\,  \delta \overline m}
     {\big[\frac{(M_t^2-\overline m^2)^2}{\overline m^2} + \Gamma^2\big]^2}
     \Bigg\} \,.
\end{eqnarray}
Here the first term is $\sim 1/(\overline m \Gamma)$ and is swamped by the
second term $\sim \alpha_s/\Gamma^2$, which is supposed to be a perturbative
correction.  This means that it is not the $\overline{\mbox{MS}}$ mass that is
ever directly measured from any reconstruction mass-measurement that uses a top
Breit-Wigner at some level of the analysis.  We stress that this statement
applies to any top mass determination that relies on the reconstruction of the
peak position of an invariant mass distribution.

To define a short distance scheme for jet reconstruction measurements, $m_J$, we
choose the residual mass term $\delta m_J$ such that, order-by-order, the jet
functions $B_\pm$ have their maximum at $\hat s_t=\hat s_{\bar t}=0$, where
$B_+(\hat s)$ is the gauge invariant function defined in Eq.~(\ref{hqetjet}). So
order-by-order in perturbation theory the definition is given by the solution to
\begin{eqnarray}
  \frac{ dB_+(\hat s,\mu,\delta m_J)}{d\hat s} \bigg|_{\hat s=0} = 0 \,.
\end{eqnarray}
We call this mass definition the {\it top quark jet-mass}, $m_J(\mu)=m_{\rm
  pole}-\delta m_J$.  Since the bHQET jet functions have a nonvanishing
anomalous dimension, the top jet-mass depends on the renormalization scale
$\mu$, at which the jet functions are computed perturbatively. Thus the jet-mass
is a running mass, similar to the $\overline{\mbox{MS}}$ mass, and different
choices for $\mu\gtrsim \Gamma$ can in principle be made. 

For simplicity we will use the notation $\tilde B_\pm(\hat s,\mu)$ for
the bHQET jet-functions in the jet-mass scheme. At next-to-leading
order in $\alpha_s$,
\begin{eqnarray} \label{Bshift}
\tilde B_\pm(\hat s,\mu) 
& = &
 B_\pm(\hat s,\mu ) + 
 \frac{1}{\pi m_J} \, \frac{(4\,\hat s\,\Gamma)\, \delta m_J}{(\hat s^2 + \Gamma^2)^2}
\,,
\end{eqnarray}
where $m_J=m_J(\mu)$ and $B_+$ is the pole-mass jet function to ${\cal
O}(\alpha_s)$. Here we dropped all corrections that are power suppressed by
$\Gamma/m$. The one-loop relation between the pole and jet-mass
is~\cite{FHMS2}
\begin{eqnarray} \label{mJmpole}
  m_J(\mu) = m_{\rm pole} - \Gamma \frac{\alpha_s(\mu)}{3} \Big[
  \ln\Big(\frac{\mu}{\Gamma}\Big) + \frac32 \Big] \,.
\end{eqnarray}
For $\mu=\Gamma$ we have $\delta m_J \simeq 0.26\, {\rm GeV}$, so the
jet-mass is quite close to the one-loop pole mass.
Equation~(\ref{mJmpole}) also shows that the jet-mass is substantially
different from the short-distance masses that are employed for $t\bar
t$-threshold analyses~\cite{Hoang:2000yr}, where $\delta m\sim
\alpha_s^2 m\sim 2\,{\rm GeV}$ is of order the binding energy of the
$t\bar t$ quasi-bound state.  Nevertheless, in some of the threshold
mass schemes~\cite{Beneke:1998rk,Uraltsev:1998bk} $\delta m$ is
proportional to a cutoff scale that could in principle be adapted such
that they are numerically close to the jet-mass we are proposing. A
detailed discussion on the impact of switching from the pole to the
jet-mass scheme at the one-loop level and at higher orders will be
given in Refs.~\cite{FHMS2} and \cite{FHMS3}, respectively. We remark
that many other schemes satisfying Eq.~(\ref{massscheme}) can in
principle be defined, but the existence of one such scheme suffices.
However, for any suitable short-distance mass scheme the
renormalization scale in $\alpha_s$ contained in $\delta m$ has to be
equal to the scale $\mu$ used for the computation of the bHQET jet
functions.

The other function that must be modified in the factorization theorem is
$H_m(m,Q/m,\mu_m,\mu_\Delta)$. However this function only depends
logarithmically on $m$, and 
\begin{eqnarray}
  \ln\Big(\frac{m_{\rm pole}}{\mu}\Big) = \ln\Big(\frac{m_J}{\mu}\Big) 
  + {\cal O}\Big(\frac{\alpha_s\Gamma}{m_J}\Big) \,.
\end{eqnarray}
So dropping these perturbatively suppressed power corrections we can simply
replace $m\to m_J$ in $H_m$. We note that any $\mu$ dependence from $m_J(\mu)$
in $H_m$ is also power suppressed.

 Thus our final result for the cross-section in terms of the short-distance
jet-mass is
\begin{align}
\label{final-cross}
\left( \frac{d^2\sigma }{dM^2_t\>dM^2_{\bar t}} \right)_{\rm hemi} \!\!\! &= 
    \sigma_0
     \> H_Q(Q,\mu_m)  H_m\Big(m_J,\frac{Q}{m_J},\mu_m,\mu\Big) \\
 &\times\! \int_{-\infty}^{\infty}\!\!\!d\ell^+ d\ell^-
  \> \tilde B_+\Big(\hat s_t- \frac{Q\ell^{+}}{m_J},\Gamma,\mu\Big) \:
     \tilde B_-\Big(\hat s_{\bar t} - \frac{ Q\ell^{-}}{m_J},\Gamma,\mu\Big)\: 
  S_{\rm hemi}(\ell^+,\ell^-,\mu)
  \,, \nn
\end{align}
where the running jet-mass $m_J=m_J(\mu)$.

\subsection{Renormalization-Group Evolution} \label{sec:RGE}

In order to explain the $\mu$-dependence of the factorization theorem in
Eq.~(\ref{final-cross}) we give a brief discussion of the renormalization group
evolution. A more detailed discussion is given in Ref.~\cite{FHMS2}.
Equation~(\ref{final-cross}) depends on two renormalization scales, $\mu_m$ and
$\mu$.  The matching scale $\mu_m\sim m$ was the endpoint of the evolution of
the hard function $H_Q(Q,\mu_m)$. From the matching at $m$ we get the dependence
on $\mu_m$ in $H_m$, and from running below $m$ we get an additional dependence
on $\mu$ as well as $Q/m$ (which is discussed in more detail in
Ref.~\cite{FHMS2} and signifies the presence of a cusp term in the anomalous
dimension, see Ref.~\cite{Korchemsky:1991zp}). The $\mu$-dependence in $H_m$ cancels
against the $\mu$-dependence in the bHQET jet functions and the soft function.

To sum the remaining large logarithms we have in principle two choices. We can
either run the Wilson coefficient $H_m$, or we can run the individual functions
$\tilde B_\pm$ and $S$. The first option essentially corresponds to running the bHQET
top pair production current of Eq.~(\ref{JbHQET}), and we will call this method
{\it ``top-down''}. The relation
\begin{equation}
H_m\Big(m,\frac{Q}{m},\mu_m,\mu\Big) =  H_m(m,\mu_m) U_{H_m}\Big(\mu_m,\mu,\frac{Q}{m}\Big)
\end{equation}
defines the corresponding evolution factor $U_{H_m}$ that is shown in
Fig.~\ref{fig:theory}. The second option means running the jet functions
$\tilde B_\pm$ and the soft function  $S_{\rm hemi}$ independently with the evolution
factors $U_{B_\pm}(\mu,\mu_m)$ and $U_S(\mu,\mu_m)$ respectively, and is also
illustrated in Fig.~\ref{fig:theory}. This running involves
convolutions, such as
\begin{align} \label{Brun}
 \mu\frac{d}{d\mu} \tilde B_+(\hat s,\mu) &= \int\!\! d\hat s' \: \gamma_{B_+}(\hat
 s-\hat s') \: \tilde B_+(\hat s',\mu) \,,\nn\\
  \tilde B_+(\hat s,\mu_m) 
 & = \int \!\! d\hat s'\: U_{B_+}(\hat s-\hat s',\mu_m,\mu)\: \tilde B_+(\hat s',\mu) 
  \,,
\end{align}
and analogously for $\tilde B_-$ and $S_{\rm hemi}$. Since this method for the running
usually involves taking the functions $B_\pm$ and $S_{\rm hemi}$ as an input at
the low scale (to avoid the appearance of large logs) we will call this option
{\it ``bottom-up''}.  Because the running of $H_m$ is local (i.e.~has no
convolution), this RG evolution only affects the normalization of the cross
section and {\em does not change} the dependence on $s_t$ and $s_{\bar t}$ in a
non-trivial way. This is more difficult to discern from the bottom-up running,
but when the convolutions for $B_\pm$ and $S$ are combined they must become
local. These cancellations are discussed in detail in Ref.~\cite{FHMS2}
where also the full leading log evolution is derived.

Generically, we may wish to run the soft function and jet function to slightly
different low energy scales. Lets examine the case shown in
Fig.~\ref{fig:theory} where we run the soft function to $\mu_\Delta$, but run
the bHQET jet functions to a slightly lower scale $\mu_\Gamma$. (The opposite
case could of course also be realized.) In this case the running
is local up to the scale $\mu_\Delta$, and below this scale we have convolution
running for $B_\pm$. Using Eq.~(\ref{Brun}) the factorization formula for split
low energy renormalization scales is
\begin{align}
\label{bHQETcross-hem2}
 \frac{d^2\sigma }{dM_t^2\>dM^2_{\bar t}}  &= 
    \sigma_0
     \> H_Q(Q,\mu_m) H_m\Big(m_J,\frac{Q}{m_J},\mu_m,\mu_\Delta\Big)\!\! \\
 &\times
   \int_{-\infty}^{\infty}\!\!\!\!\!  d\hat s_t'\: d\hat s_{\bar t}' \:
  \: U_{B_+}(\hat s_t\minus \hat s_t',\mu_\Delta,\mu_\Gamma)
  \: U_{B_-}(\hat s_{\bar t}\minus \hat s_{\bar t}',\mu_\Delta,\mu_\Gamma)
  \nn\\
 &\times
 \int_{-\infty}^{\infty}\!\!\!d\ell^+ d\ell^-  
  S_{\rm hemi}(\ell^+,\ell^-,\mu_\Delta) 
  \> \tilde B_+\Big(\hat s_t' - \frac{Q\ell^{+}}{m_J},\Gamma,\mu_\Gamma\Big) \:
     \tilde B_-\Big(\hat s_{\bar t}' - \frac{Q\ell^{-}}{m_J},\Gamma,\mu_\Gamma\Big)
  \,, \nn
\end{align}
where parametrically $\mu_\Delta \sim \mu_\Gamma$ and here we take
$m_J=m_J(\mu_\Gamma)$.  In this paper we will use common low energy
scales for our numerical analysis, as shown in
Eq.~(\ref{final-cross}), and leave the discussion of the more general
case in Eq.~(\ref{bHQETcross-hem2}) to Ref.~\cite{FHMS2}.

\subsection{Thrust and Other Event Shape Variables} \label{sec:eventshapes}

Starting from the two-dimensional distribution, $d^2\sigma/dM_t^2dM_{\bar t}^2$
in Eq.~(\ref{final-cross}) it is straightforward to derive results for other
event shape variables. For example, for the thrust $T$ defined in
Eq.~(\ref{thrust-1}), we have $1-T=(M_t^2+M_{\bar t}^2)/Q^2$ which follows using
Eq.~(\ref{tauss}) with Eqs.~(\ref{massshell}) and (\ref{tau}). Inserting the 
identity 
\begin{align}
   1 = \int\!\!  dT\  \delta\Big(1-T - \frac{M_t^2+M_{\bar t}^2}{Q^2}\Big) 
\end{align}
into Eq.~(\ref{final-cross}) and integrating over $M_t^2$ and $M_{\bar t}^2$ we find
\begin{align} \label{Tfactorization}
 \frac{d\sigma}{dT} &= 
    \sigma_0^H(\mu)
      \int_{-\infty}^{\infty}\!\!\!ds_t\: ds_{\bar t}\:
  \> \tilde B_+\Big( \frac{s_t}{m_J},\Gamma,\mu\Big) \:
     \tilde B_-\Big( \frac{s_{\bar t}}{m_J},\Gamma,\mu\Big)\: 
  S_{\rm thrust}\Big(1-T-\frac{(2m_J^2 + s_t+s_{\bar t})}{Q^2},\mu\Big)
  \,, 
\end{align}
where $\sigma_0^H(\mu) = \sigma_0 H_Q(Q,\mu_m) H_m(m_J,Q/m_J,\mu_m,\mu)$. 
Here the thrust soft-function is simply a projection of the hemisphere soft function, 
\begin{align} \label{Sthrust}
  S_{\rm thrust}(\tau,\mu) & = \int_0^\infty\!\!\! d\ell^+\: d\ell^-
    \delta\Big(\tau - \frac{(\ell^+ + \ell^-)}{Q}\Big) S_{\rm hemi}(\ell^+,\ell^-,\mu)
  \\
  &= \frac{1}{N_c} \sum_{X_s} \delta\Big(\tau - \frac{k_s^{+a}+k_s^{-b}}{Q}\Big)
   \langle 0| \overline {Y}_\bn\,  {Y}_n (0) |X_s \rangle
\langle X_s| {Y}^\dagger_n\,  \overline {Y}_\bn^\dagger (0) |0\rangle 
  \nn \,.
\end{align}

Another well known distribution, which is also frequently analyzed for massless
jets, is the heavy jet mass. It can be defined by the dimensionless variable
\begin{align} \label{rho}
\rho = \frac{1}{Q^2}\: {\rm Max} \big\{ M_t^2, M_{\bar t}^2 \big\} \,.
\end{align}
Using the same steps as above for $\rho$, the factorization theorem for top
initiated jets is
\begin{align}\label{HJMfactorization}
  \frac{d\sigma}{d\rho} &= \sigma_0^H(\mu) \int_{-\infty}^{\infty}\!\!\!ds_t\:
  ds_{\bar t}\: \> \tilde B_+\Big( \frac{s_t}{m_J},\Gamma,\mu\Big) \: \tilde
  B_-\Big( \frac{s_{\bar t}}{m_J},\Gamma,\mu\Big)\: 
 S_{\rm HJM}(\rho-\frac{m_J^2}{Q^2},s_t,s_{\bar t}) \,,
\end{align}
where the relevant soft-function is
\begin{align}
 S_{\rm HJM}(\rho,s_t,s_{\bar t}) &= \! \int_0^\infty\!\!\!\! d\ell^+\: d\ell^-\:
    \delta\Big(\rho - \frac{1}{Q^2} 
   {\rm Max}\big\{ Q\ell^+ \plus  s_t, Q\ell^- \plus s_{\bar t}\big\} \Big)
  S_{\rm hemi}(\ell^+,\ell^-,\mu) \,.
\end{align}
Factorization theorems for other event shapes that are related to
$d^2\sigma/dM_t^2 dM_{\bar t}^2$ can be derived in an analogous manner. As
should be obvious from the definitions of thrust and the heavy jet mass
distribution in Eqs.~(\ref{Tfactorization}) and (\ref{HJMfactorization}), these
event shape distributions are also characterized by a peak at shape parameter
values that are sensitive to the short-distance top-quark mass. It is therefore
possible to use these event shapes to measure the top-mass with a precision
comparable to the invariant mass distribution discussed in the previous
subsection. A brief numerical analysis of the thrust distribution is given in
Sec.~\ref{section40}.

\section{Analysis of the Invariant Mass Distribution}\label{section4}

\subsection{A Simple Leading Order Analysis}\label{section40}

The main result of this paper is the formula in Eq.~(\ref{final-cross}) for the
double invariant mass distribution with a short distance top-quark mass suitable
for measurements using jets. In this section we discuss the implications of
Eq.~(\ref{final-cross}) for top-mass measurements.  For convenience we rewrite
the cross-section in terms of dimension one invariant mass variables
\begin{align} \label{sigmaMM}
  \frac{ d^2\sigma }{dM_t\, dM_{\bar t}} 
    & = \frac{4 M_t M_{\bar t} \: \sigma_0^H}{ (m_J\Gamma)^2}\
    F(M_t,M_{\bar t},\mu) \,,
\end{align}
where $\sigma_0^H= \sigma_0 H_Q(Q,\mu_m)  H_m(m_J,Q/m_J,\mu_m,\mu)$ is the
cross-section normalization factor with radiative corrections, $Q$ is the
c.m. energy, and we have defined a dimensionless function
\begin{align} \label{F}
  F(M_t,M_{\bar t},\mu) &= (m_J\Gamma)^2\!
   \int_{-\infty}^\infty\!\!\!\! d\ell^+\, d\ell^- 
  \tilde B_+\Big(\hat s_t - \frac{Q\ell^+}{m_J}, \Gamma,\mu \Big)
  \tilde B_-\Big(\hat s_{\bar t} - \frac{Q\ell^-}{m_J}, \Gamma,\mu\Big) S_{\rm
    hemi}(\ell^+,\ell^-,\mu)
   .
\end{align}
In terms of $M_t$ and $M_{\bar t}$ the variables $\hat s_{t,\bar t}$ are
\begin{align} \label{ssM}
  \hat s_t = 2 M_t -2 m_J \,,\qquad\quad 
  \hat s_{\bar t} = 2 M_{\bar t} -2 m_J
  \,,
\end{align}
up to small $\Gamma/m$ power corrections.  In Eqs.~(\ref{sigmaMM}-\ref{ssM}) the
jet hemisphere invariant masses are $M_t$ and $M_{\bar t}$ and the
short-distance top-quark mass that we wish to measure is $m_J$.  In
$d^2\sigma/dM_tdM_{\bar t}$ the function $F$ dominates the spectrum, while $4
M_t M_{\bar t}\, \sigma_0^H/(m_J\Gamma)^2$ acts as a normalization constant
(since $M_t M_{\bar t}$ is essentially constant in the peak region of interest).
A measurement of the normalization is not optimal for determining $m_J$; it only
has logarithmic dependence on the short-distance mass, and has larger
theoretical uncertainties.  On the other hand, the spectrum is very sensitive to
$m_J$, so henceforth we focus on $F(M_t,M_{\bar t},\mu)$.

From Eq.~(\ref{F}) $F$ is given by the convolution of the computable $\tilde
B_\pm$ functions, with a non-perturbative hemisphere soft-function, $S_{\rm
  hemi}$, that describes soft final-state radiation.  The majority of the
important features of Eq.~(\ref{F}) can be explained without discussing
perturbative corrections, so we focus here on the leading order result.  From
Eq.~(\ref{Bpmtree}), $\tilde B_\pm$ are simply Breit-Wigner's at leading order,
\begin{align} \label{Bpmtree2}
  \tilde B_+(\hat s_t) &= \frac{1}{\pi (m_J\Gamma)}\: \frac{1}{(\hat
    s_t/\Gamma)^2+1} \,, 
& 
 \tilde B_-(\hat s_{\bar t}) &=
  \frac{1}{\pi (m_J\Gamma)}\: \frac{1}{(\hat s_{\bar t}/\Gamma)^2+1} \,.
\end{align}
For our numerical analysis we use the two-loop standard model prediction for the
top-width $\Gamma=1.43\,{\rm GeV}$~\cite{Czarnecki:1998qc} and we take the short
distance jet-mass to be fixed at $m_J=172\,{\rm GeV}$.  As demonstrated in
Secs.~\ref{sectionefts} and \ref{section3}, $S_{\rm hemi}$ is the same
function that controls the soft radiation for massless dijets, which was studied
in Refs.~\cite{Korchemsky:1998ev,Korchemsky:1999kt,Korchemsky:2000kp}.  Hence,
it is convenient for our analysis to adopt the model used to fit the massless
dijet data~\cite{Korchemsky:2000kp},
\begin{align} \label{SM1}
  S_{\rm hemi}^{\rm M1}(\ell^+,\ell^-) = \theta(\ell^+)\theta(\ell^-)
   \frac{ {\cal N}(a,b) }{\Lambda^2}
  \Big( \frac{\ell^+\ell^-}{\Lambda^2}\Big)^{a-1} \exp\Big(
  \frac{-(\ell^+)^2-(\ell^-)^2-2 b \ell^+\ell^-}{\Lambda^2} \Big).
\end{align}
Here the normalization constant ${\cal N}(a,b)$ is defined so that $\int d\ell^+
d\ell^- S(\ell^+,\ell^-) = 1$, the parameter $\Lambda \sim \Lambda_{\rm QCD}$
sets the scale for $\ell^\pm$ and hence the soft radiation, and the parameter
$a$ controls how fast the soft-function vanishes at the origin. The
dimensionless parameter $b>-1$ controls the correlation of energy flow into the
two hemispheres. Any $b\ne 0$ implies cross-talk between the two hemispheres. A
fit to the heavy jet mass distribution using $e^+e^-$ dijet data from LEP and
SLD with $Q=m_Z$ gives~\cite{Korchemsky:2000kp}
\begin{align}\label{abL}
 a=2 \,, \qquad\quad
 b=-0.4 \,, \qquad\quad
 \Lambda=0.55\,{\rm GeV} \,,
\end{align}
These values were shown to yield accurate predictions for the heavy jet-mass and
$C$-parameter event shapes for a wide range of energies, $Q=35$--$189\,{\rm
  GeV}$~\cite{Korchemsky:2000kp}, as well as available thrust distributions with
$Q=14$--$161\,{\rm GeV}$~\cite{Korchemsky:1999kt}. We adopt Eq.~(\ref{abL}) as
the central values for our analysis, but will also discuss how our predictions vary
with changes to these model parameters.

\begin{figure}[t!]
  \centerline{ 
   \includegraphics[width=10cm]{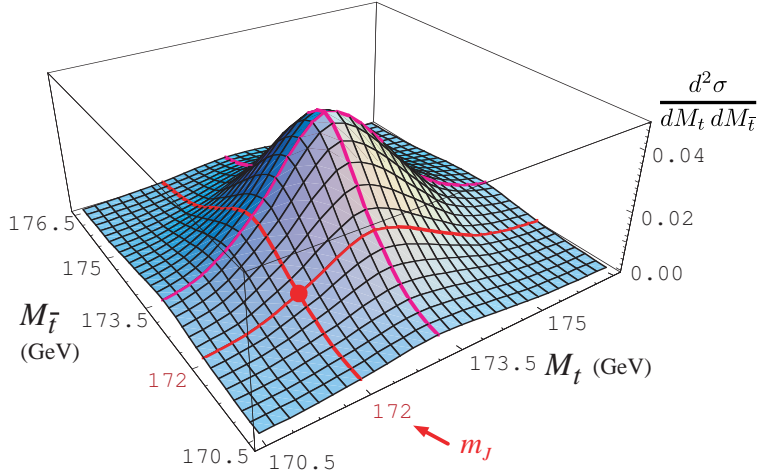}  
  } 
\caption{Plot of $F(M_t,M_{\bar t})$, which is the
  double differential hemisphere invariant mass
  cross-section $d^2\sigma/dM_t dM_{\bar t}$ in units of 
  $4 \sigma_0^H/\Gamma^2$. The observed peak position (intersection of
  the magenta
  lines) is not given by the true top-quark mass, $m=m_J=172\,{\rm GeV}$ (red
  lines).  This peak shift depends on the energy $Q$, the width $\Gamma$, and the
  soft-radiation function.  The result is shown for $Q/m_J=4.33$ and the
  parameters in Eq.~(\ref{abL}).  }
\label{fig:plot3D}
\end{figure}
In Fig.~\ref{fig:plot3D} we plot $F(M_t,M_{\bar t})$ using
Eqs.~(\ref{Bpmtree2}-\ref{abL}) and taking $Q= 745\,{\rm GeV}$.  The key feature
to note is that the observed peak position {\em is not given by} the
short-distance top-quark mass $m_J$, but is instead shifted upward by $\simeq
1.5\,{\rm GeV}$. The positive sign of this shift is a prediction of
Eq.~(\ref{F}) irrespective of the choice of parameters.  The precise value for
this shift depends on $Q/m_J$, $\Gamma$, as well as the parameters of the soft
function. A less obvious feature of Fig.~\ref{fig:plot3D} is that the width of
the observed peak has also increased beyond the width $\Gamma$ of
Eq.~(\ref{Bpmtree2}). Physically, the reason for this behavior is that soft
radiation contributes to the invariant masses, while the Breit-Wigner is {\em
  only} a leading order approximation for the spectrum of the top-quark and
accompanying collinear gluons. Thus the arguments of $\tilde B_\pm$ in
Eq.~(\ref{F}) subtract the dominant soft momentum component from $\hat s_{t,\bar
  t}$. If we approximate $S_{\rm hemi}(\ell^+,\ell^-)$ as a very narrow Gaussian
centered at $\ell^\pm=\ell_0^\pm$, then the observed peak simply occurs at
$M_{t,\bar t} \sim m_J + Q\ell^\pm_0 /(2m_J)$. Although this model is too naive,
we demonstrate in the next section that the linear dependence of the peak shift
on $Q/m_J$ is in fact generic and independent of the soft-function parameters.
The peak width also increases linearly with $Q/m_J$.

The presence of the shift is due to the inclusion of soft radiation in the
definition of the invariant masses $M_t$ and $M_{\bar t}$. Although we adopted a
hemisphere mass definition, the same type of shift will be present for any jet
algorithm that groups all the soft radiation into the jets identified for the
top and anti-decay products, as we discuss in Sec.~\ref{sectionotheralgo}.
The numerical analysis performed in this section applies equally well to these
situations, though the appropriate definition and model for the soft
functions $S$ for such analyses will in general be different than that in
Eq.~(\ref{SM1}) with Eq.~(\ref{abL}). We are not aware of studies
where models for such soft functions were discussed.

It is important to emphasize that the shift of the observed peak
position away from $m_J$ is not an artifact of the
mass-scheme. At the order used to make Fig.~\ref{fig:plot3D} we could set
$m_J=m^{\rm pole}$ since as explained in Sec.~\ref{sec:sdmass} they differ by
${\cal O}(\alpha_s \Gamma)$.\footnote{In general use of $m^{\rm pole}$ is not a
  good idea, since in fits it would induce an unphysical change in the required
  parameters $a,b,\Lambda$ order-by-order in perturbation theory} In a generic
short distance top-quark jet-mass scheme there is a small shift $\sim
\alpha_s\Gamma$ in the peak position due to perturbative corrections in the
matrix element defining $\tilde B_\pm$ (as discussed in detail in
Ref.~\cite{FHMS2}).  In Sec.~\ref{sec:sdmass} we defined $m_J$ using a
jet-mass scheme which keeps the peak of $\tilde B_\pm$ fixed order-by-order in
perturbation theory. In this scheme the shift in the peak location relative to
the short-distance mass is entirely due to the non-perturbative soft radiation.

\begin{figure}[t!]
  \centerline{ 
   \includegraphics[width=10cm]{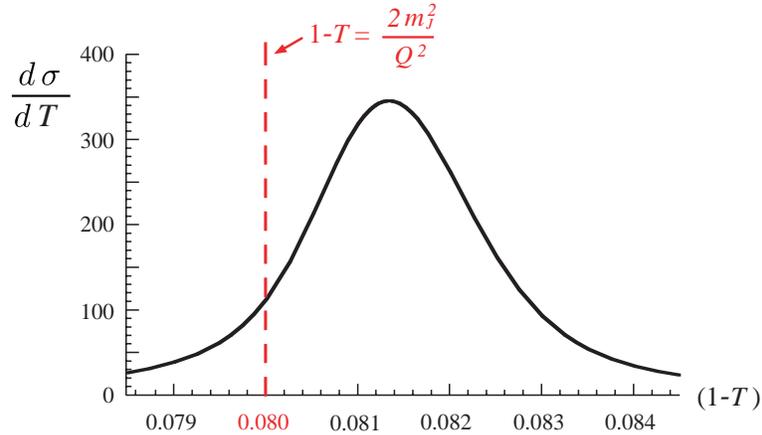}  
  } 
\caption{Plot of the thrust distribution, $d\sigma/dT$ in units of $\sigma_0^H$, for 
top-initiated events in the peak region. We use $Q/m_J=5$, $m_J=172\,{\rm GeV}$ and 
the soft function parameters in Eq.~(\ref{abL}).  }
\label{fig:thrust}
\end{figure}
Although $m_J$ is not determined by the peak-position, the shape of the
cross-section is very sensitive to $m_J$, and hence for precision $\delta m_t
\lesssim 1\,{\rm GeV}$ the top-quark mass should be determined by a fit to $F$
in Eq.~(\ref{sigmaMM}).  In Sec.~\ref{sec:eventshapes} factorization theorems
for related event shape variables were derived, including thrust $d\sigma/dT$,
and the heavy-jet mass $d\sigma/d\rho$. These event shapes also exhibit a peak.
They are sensitive to the top-quark mass parameter $m_J$, and can be used for
top-mass measurements. As an example, in Fig.~\ref{fig:thrust} we plot
$d\sigma/dT$ using $Q/m_J=5$, $m_J=172\,{\rm GeV}$, and the parameters in
Eq.~(\ref{abL}). The expected peak in the thrust distribution is at $1-T\simeq
2m^2/Q^2 = 0.08$, and is shifted to the right by $\Delta(1-T)=1.3\times 10^{-3}$
by the soft-radiation. Again the direction of the shift is a prediction, but the
precise amount of the shift depends on the soft-model parameters in
Eq.~(\ref{abL}) as well as $Q/m_J$. An analysis of any other event shape
distributions that are related to $d^2\sigma/dM_t^2dM_{\bar t}^2$ can be made in a
similar fashion.

In Sec.~\ref{section4a} we explore the functional dependence of the peak
shift for $d^2\sigma/dM_tdM_{\bar t}$ in greater detail. In
Sec.~\ref{section4b} we discuss the implications of our results for fits to
determine the short-distance mass.

\subsection{Analysis of the Peak Shift and Broadening } \label{section4a}

In this section we analyze the parameter dependence of the peak shift and
broadening of the width, and demonstrate that they have a linear dependence on
$Q$. The main analysis is carried out assuming that the soft-function model
parameters have been determined from massless jet observables with small
uncertainties and adopting the parameters in Eq.~(\ref{abL}). It is,
however, also instructive to study the dependence of the invariant
mass distribution on variations of the model parameters, anticipating
that the soft function is different when the definition of the invariant
masses is modified (see the discussion in Sec.~\ref{sectionotheralgo}).
We carry out such an analysis near the end of this section.

\begin{figure}[t!]
  \centerline{ 
    \begin{minipage}{7.2cm}
    \includegraphics[width=7cm]{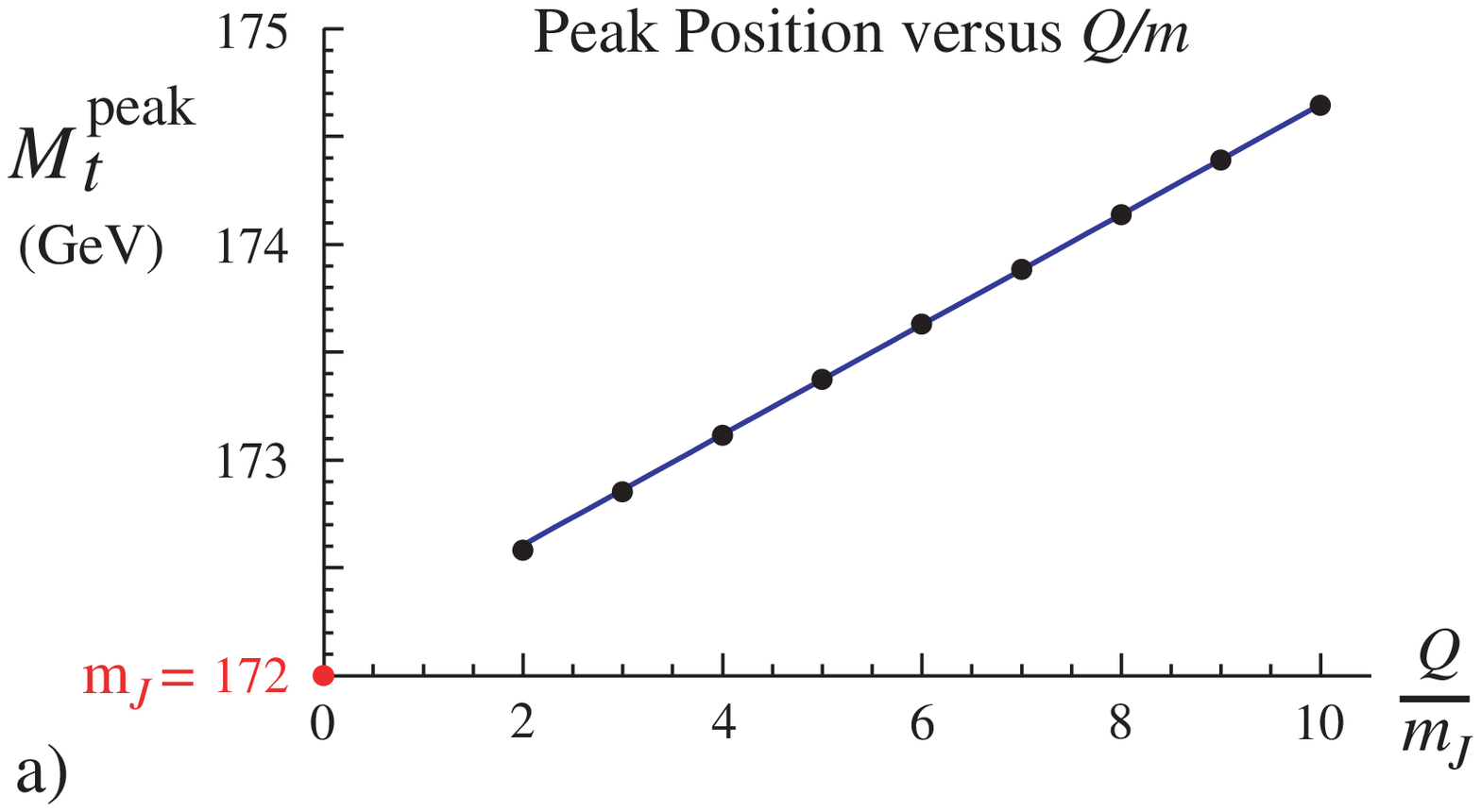}  \\[15pt]
   \hspace{.2cm}\includegraphics[width=7cm]{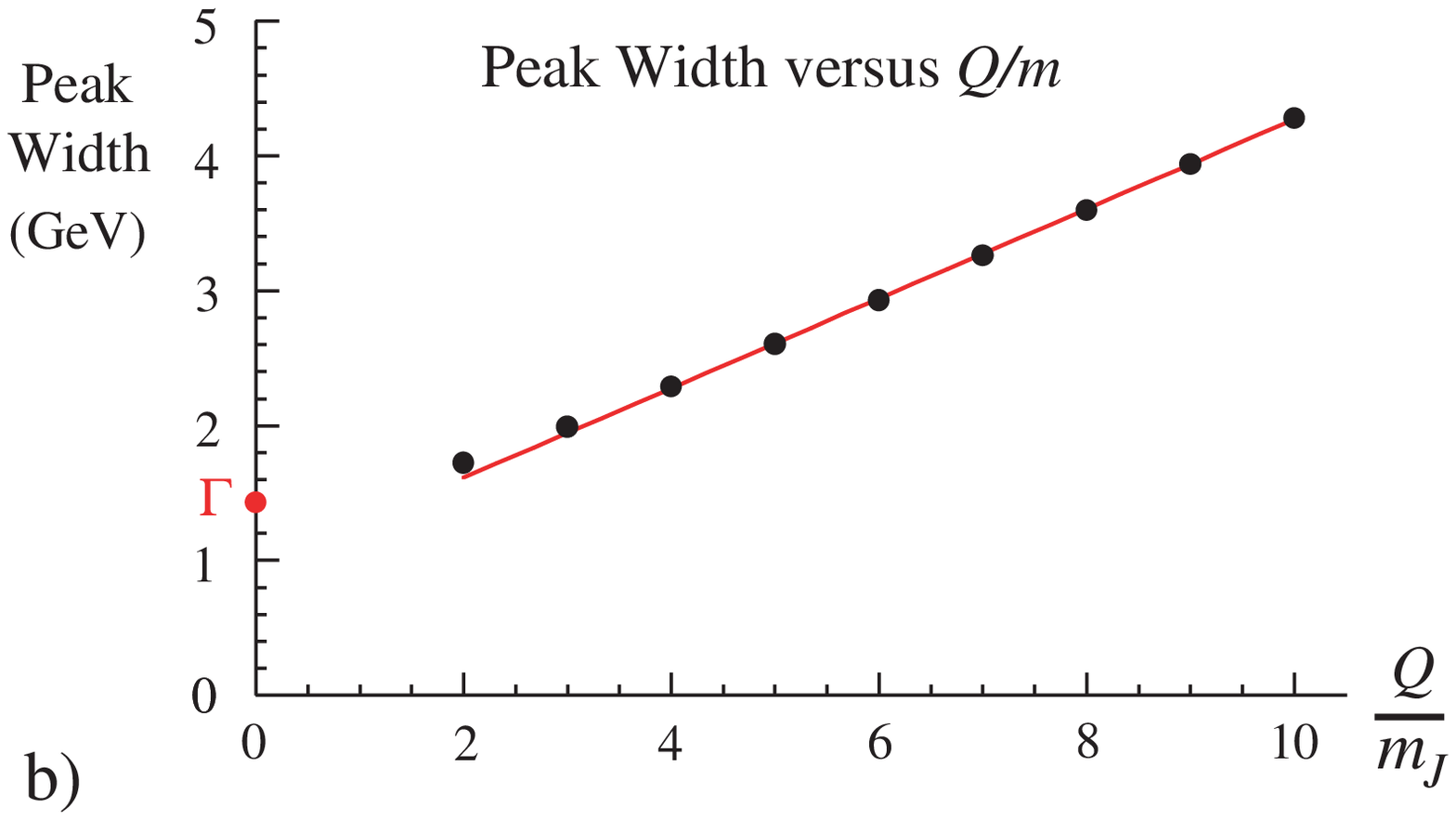} 
   \end{minipage} 
  \hspace{0.2cm}
  \raisebox{-3.5cm}{\includegraphics[width=9cm]{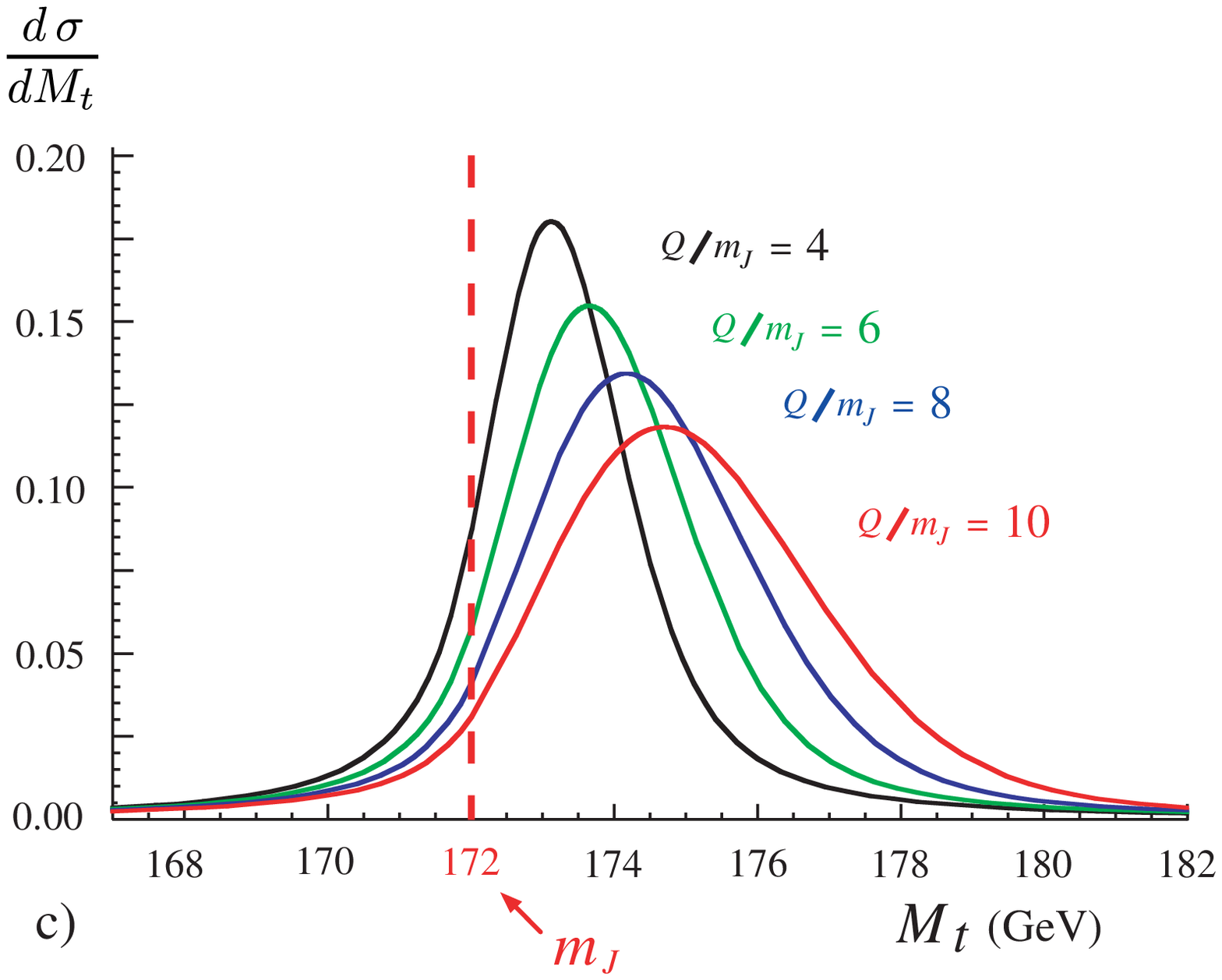}  }
   }
\caption{Effect of a change in $Q$ on the invariant mass distribution. Results on
  the left are generated from $d^2\sigma/dM_tdM_{\bar t}$, a) shows the
  peak position versus $Q/m_J$, and b) gives the full width 
  at half-max versus $Q/m_J$. 
  In c) we show $d\sigma/dM_t$ in units of $2\sigma_0^H/\Gamma$ for
  different values of $Q/m_J$. The curves use $m_J=172\,{\rm GeV}$,
  $\Gamma=1.4\,{\rm GeV}$, and the parameters in Eq.~(\ref{abL}).}
\label{fig:plot1dBW}
\end{figure}

In Fig.~\ref{fig:plot1dBW}a we plot the peak location, $M_t^{\rm peak}$, for
nine values of $Q$.  $M_t^{\rm peak}$ is obtained from the two-dimensional
distribution, and corresponds to the intersection of the magenta lines in
Fig.~\ref{fig:plot3D}. Since $d^2\sigma/dM_tdM_{\bar t}$ is symmetric the value
of $M_{\bar t}^{\rm peak}$ is the same. Note that for $Q\simeq 2m_J$ where the
tops are near $t\bar t$ production threshold, our effective theory expansions do not apply.  The
straight blue line in Fig.~\ref{fig:plot1dBW}a is a linear fit to the points
with $Q/m_J \ge 4$, and clearly shows that the peak location grows linearly with
$Q$. In Fig.~\ref{fig:plot1dBW}b we plot the ``Peak Width'', defined as the
full-width at half-max of $d^2\sigma/dM_tdM_{\bar t}$ in the top-variable $M_t$,
while fixing the antitop $M_{\bar t}=M_{\bar t}^{\rm peak}$. The red solid line
is a linear fit for $Q/m_J \ge 4$.
This figure demonstrates that we also have linear growth with $Q$ for
the width of the measured invariant mass distribution. Note that the
values for the peak position and peak width shown are consistent with
our power counting since $\hat s_{t,\bar t}$ can be order $\Gamma$ as
well as greater than $\Gamma$.

To get a better picture of how the distribution changes with $Q$ we plot the
single invariant mass distribution $d\sigma/dM_t$ in Fig.~\ref{fig:plot1dBW}c.
In particular we plot
\begin{align} \label{F1}
  F_1(M_t) &= \frac{2}{\Gamma} \int_{M_{\rm lower}}^{M_{\rm upper}} dM_{\bar t}\
  \:    F(M_t,M_{\bar t}) ,
\end{align}
which gives $d\sigma/dM_t$ in units of $2\sigma_0^H/\Gamma$. In the numerical
analysis we center the integration interval $[M_{\rm lower},M_{\rm upper}]$ on
$M_{\bar t}^{\rm peak}$ with a size that is twice the measured peak width. Hence
the size of the interval depends on $Q$, but keeps the number of events
collected at each $Q$ constant for the comparison. For different choices of $Q$
we find that the peak position and width of $F_1(M_t)$ behave in an identical
manner to Figs.~\ref{fig:plot1dBW}a,b, including having essentially the same
slopes. In order to keep the area under the curves constant the peak height
drops as $Q$ is increased.  Note that for values $Q/m_J\simeq 8$--$10$ the
observed peak location may be as much as $2.0$--$2.5\,{\rm GeV}$ above the value
of the Lagrangian mass $m_J$ one wants to measure. In our analysis $m_J$ is held
fixed as shown by the dashed line in Fig.~\ref{fig:plot1dBW}c. 

To gain an analytic understanding of this linear behavior we consider the effect of
$Q$ on the mean of the cross-section, which is a good approximation to the peak
location. Taking the first moment with respect to $\hat s_t/2 =(M_t-m_J)$ over
an interval of size $2L\gg Q\Lambda$ and the zeroth moment in $\hat s_{\bar
  t}/2=(M_{\bar t}-m_J)$ gives
\begin{align} \label{mom1}
F^{[1,0]} &\equiv \frac{1}{m_J^2\Gamma^2}  
  \int _{-L }^{L}\!\!\!\! ds_{t}\, \frac{\hat s_t}{2}
 \! \int _{-\infty}^{\infty} \!\!\!\!\! ds_{\bar{t}}\ F(M_t,M_{\bar t}) 
  = \!\! \int_{-\infty}^{\infty}\!\!\!\! d\ell^+\!\!
   \int _{-L }^{L}\!\!\!\! ds_{t}\: \frac{\hat s_t}{2} \:
  \tilde B_+\Big(\hat s_t \minus \frac{Q\ell^+}{m_J}\Big)\!
  \int_{-\infty}^{\infty}\!\!\!\!\! d\ell^- S_{\rm hemi}(\ell^+,\ell^-) 
  \nn\\
 &\simeq \frac12 \int_{-\infty}^{\infty}\!\!\! d\ell^+ 
   \int _{-L }^{L}\!\!\! d s_{t}\ \Big(\hat s_t + \frac{Q\ell^+}{m_J}\Big)\ 
  \tilde B_+(\hat s_t )
  \int_{-\infty}^{\infty}\!\!\! d\ell^- S_{\rm hemi}(\ell^+,\ell^-) 
  \nn\\
 & = \frac{Q}{2m_J}  S_{\rm hemi}^{[1,0]}   \ .
\end{align}
Thus the mean grows linearly with $Q/m_J$ with a slope determined by the
first-moment of the soft function, $S_{\rm hemi}^{[1,0]}=\int d\ell^+d\ell^-\:
\ell^+ S_{\rm hemi}(\ell^+,\ell^-)$. In the first equality of Eq.~(\ref{mom1})
the $\tilde B_-$ function drops out because we integrate over all $\hat s_{\bar
  t}$. The approximation in Eq.~(\ref{mom1}) is that terms of $\sim 1/L$ are
dropped. We can also directly consider the location of the peak in $M_t$, again
integrating over $M_{\bar t}$ for convenience. We use the fact that the
tree-level $\tilde B_+(\hat s_t)$ is symmetric, and solve for $M_t^{\rm
  peak}=m_J+\hat s_t^{\rm peak}/2$ by setting
\begin{align}
  0 &=\frac{1}{m_J^2\Gamma^2} \int_{-\infty}^\infty\!\!\! d\hat s_{\bar t}\
   \frac{dF(M_t,M_{\bar t})}{d\hat s_t} 
  =   \int_{-\infty}^{\infty}\!\!\! d\ell^+ 
  \tilde B_+^\prime\Big(\hat s_t - \frac{Q\ell^+}{m_J}\Big)
  \int_{-\infty}^{\infty}\!\!\! d\ell^- S_{\rm hemi}(\ell^+,\ell^-) 
  \nn\\
  &=  \int_{-\infty}^{\infty}\!\!\! d\ell^+ 
  \bigg[ (\hat s_t \minus \frac{Q\ell^+}{m_J}\Big) \tilde B_+^{\prime\prime}(0)
  \plus \frac{1}{3!} \Big(\hat s_t \minus \frac{Q\ell^+}{m_J}\Big)^3 \tilde B_+^{(4)}(0)
  \plus \ldots \bigg]
  \int_{-\infty}^{\infty}\!\!\! d\ell^- S_{\rm hemi}(\ell^+,\ell^-) 
   \,.
\end{align}
For $Q\Lambda\gg m\Gamma$ we can keep only the first term which yields
\begin{align}  \label{MtpeakLinear}
  M_t^{\rm peak}\simeq m_J + \frac{Q}{2m_J}\, S_{\rm hemi}^{[1,0]}.
\end{align}
Thus we find the same shift as for the moment in Eq.~(\ref{mom1}). Our
default model in Eq.~(\ref{abL}) gives $S_{\rm
hemi}^{[1,0]}/2=0.31\,{\rm GeV}$ for the slope in $Q/m_J$. This can be
compared with the fit to the two-dimensional peak position,
Fig.~\ref{fig:plot1dBW}a, which gives a slope of $0.26\,{\rm
GeV}$. The fit to the peak position of $F_1(M_t)$ in
\ref{fig:plot1dBW}c has a similar slope, $0.25\,{\rm GeV}$. Finally,
the first moments of $F_1(M_t)$ also display linear behavior in
$Q/m_J$ with a slope of $0.28\,{\rm GeV}$. We see that $S_{\rm
hemi}^{[1,0]}/2$ accounts for the largest portion of these slopes,
with the remainder being accounted for by other moments. Note that the
linear behavior in $Q/m_J$ observed in Fig.~\ref{fig:plot1dBW} is much
more accurate than the statement that $S^{[1,0]}_{\rm hemi}/2$
determines the proper slope at lowest order.  We want to point out
that a first measurement of the short distance mass could be made
using a couple of different $Q$ values and a simple extrapolation with
Eq.~(\ref{MtpeakLinear}).

\begin{figure}[t]
  \centerline{ 
  \begin{minipage}{8.4cm}
   \includegraphics[width=8.4cm]{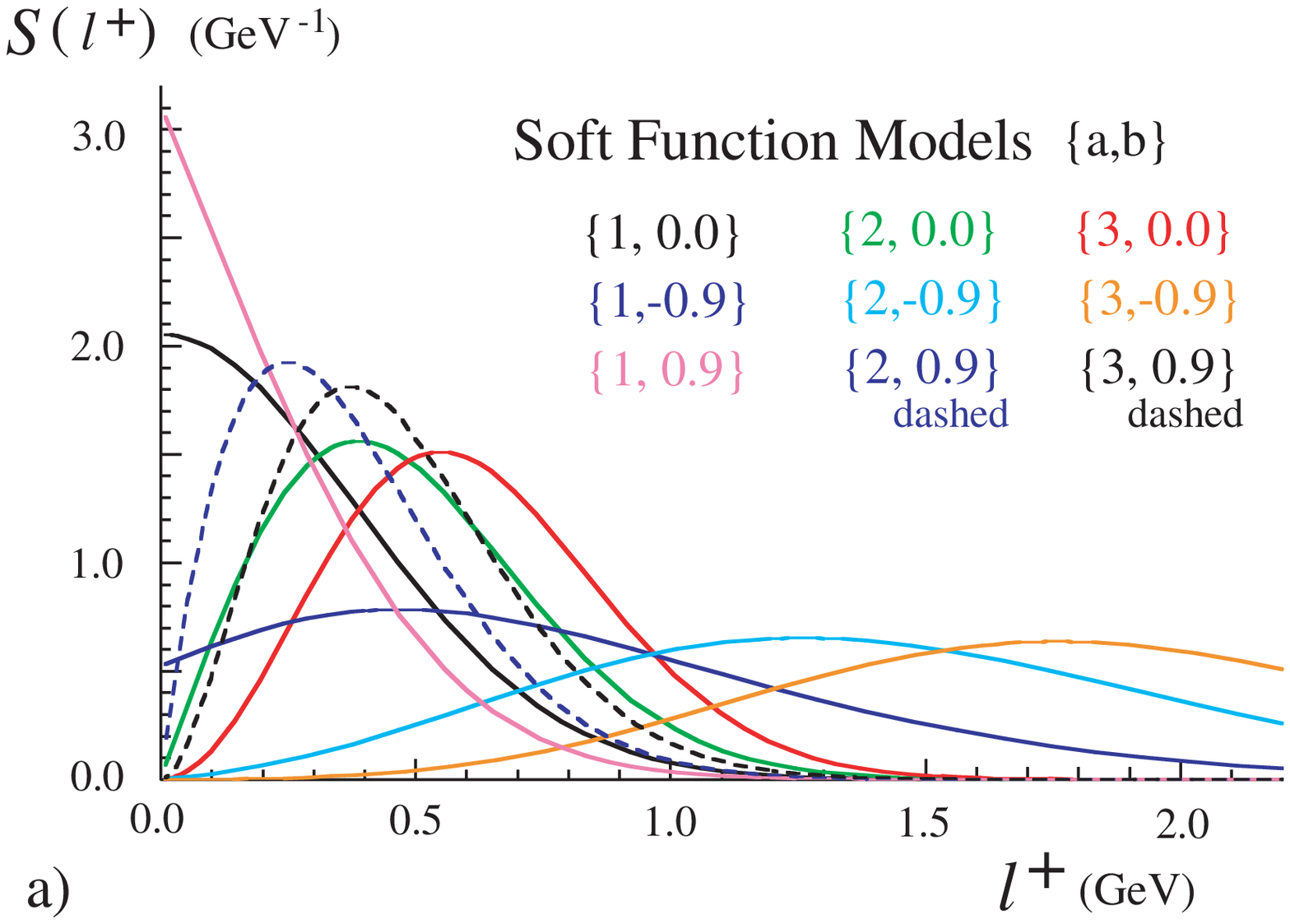} \\[0pt]
   \includegraphics[width=8.4cm]{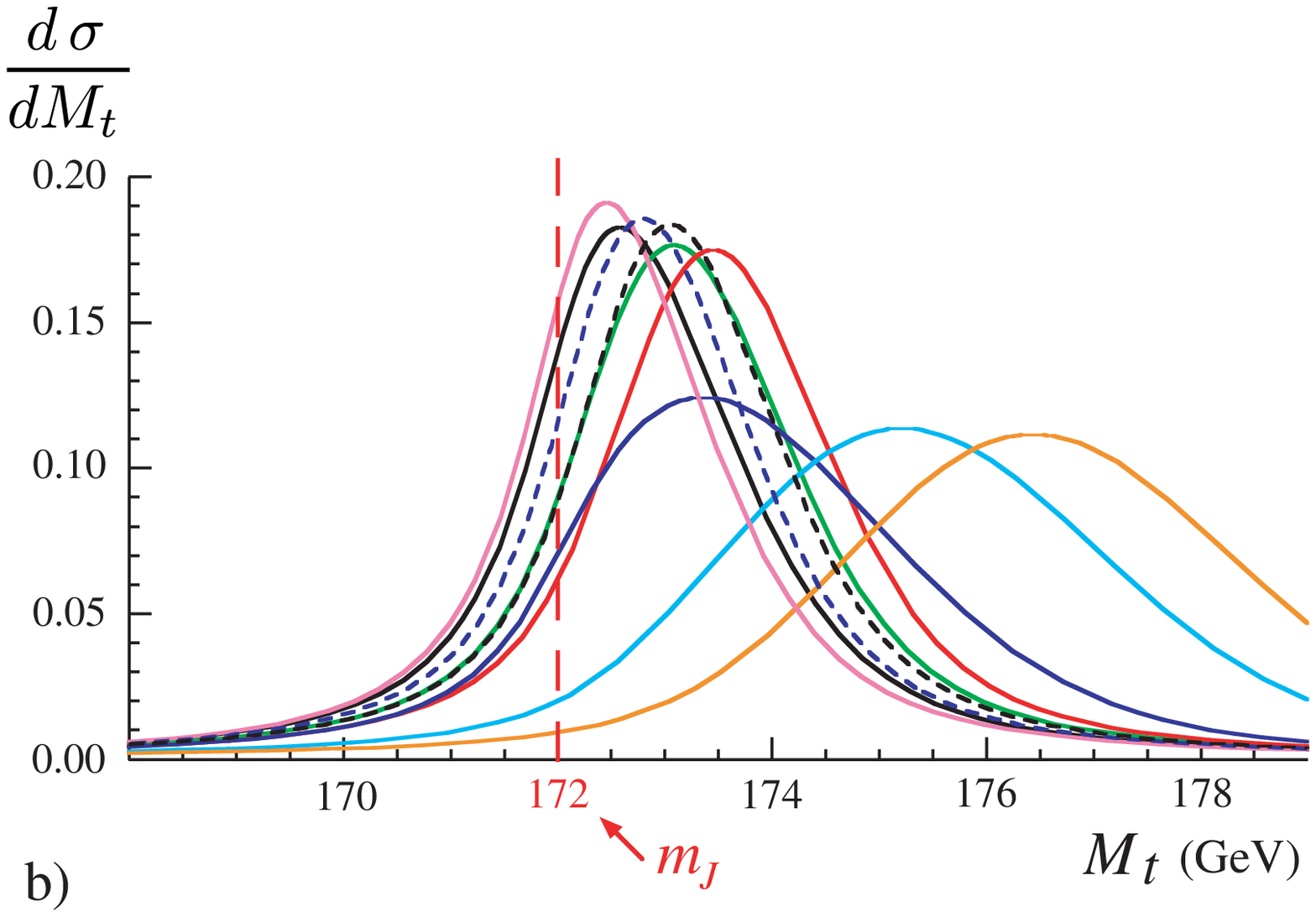}
   \end{minipage} 
  \hspace{0.2cm}
   \begin{minipage}{8cm}
    \includegraphics[width=7.6cm]{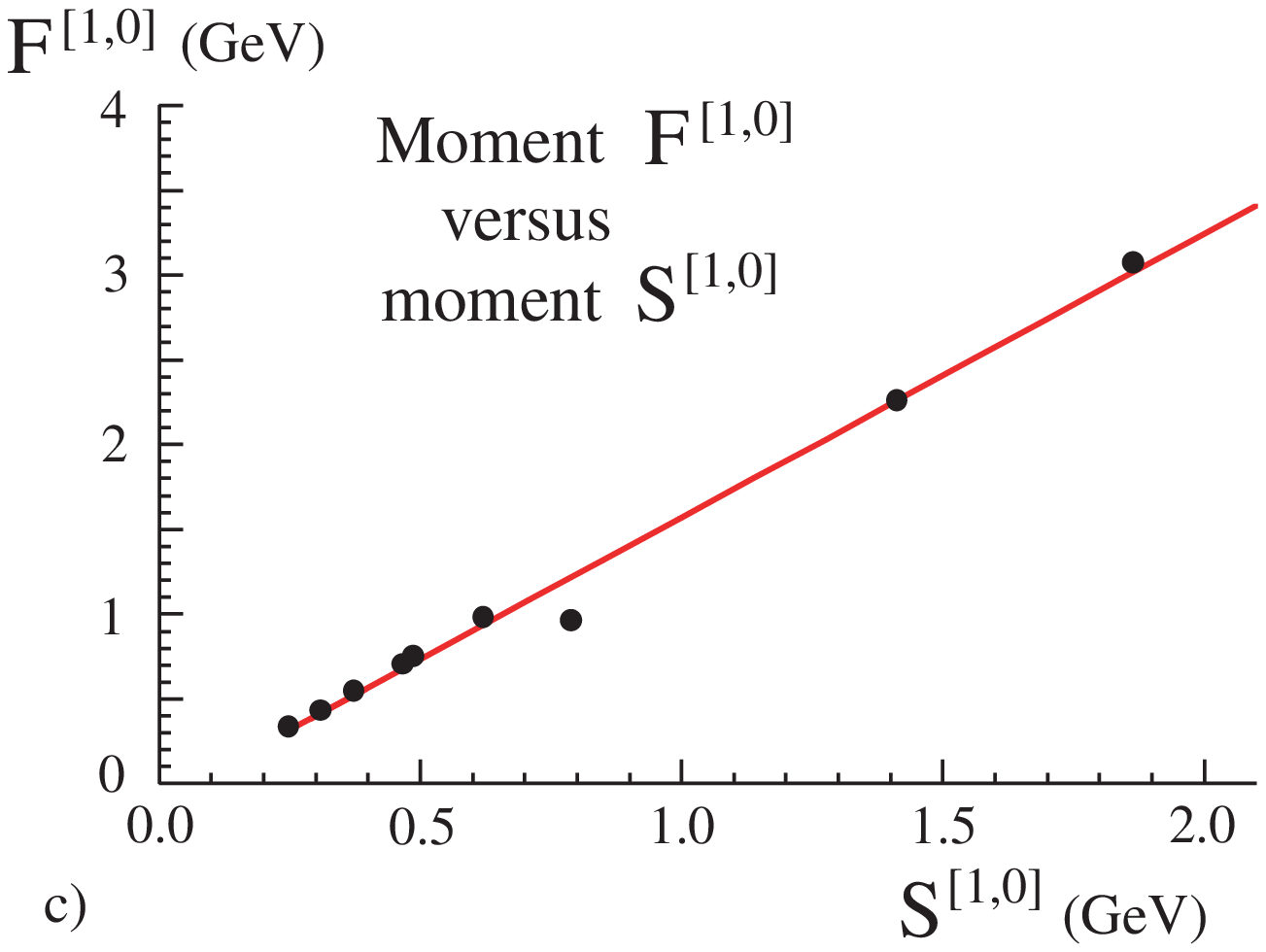}  \\[4pt]
   \includegraphics[width=7.6cm]{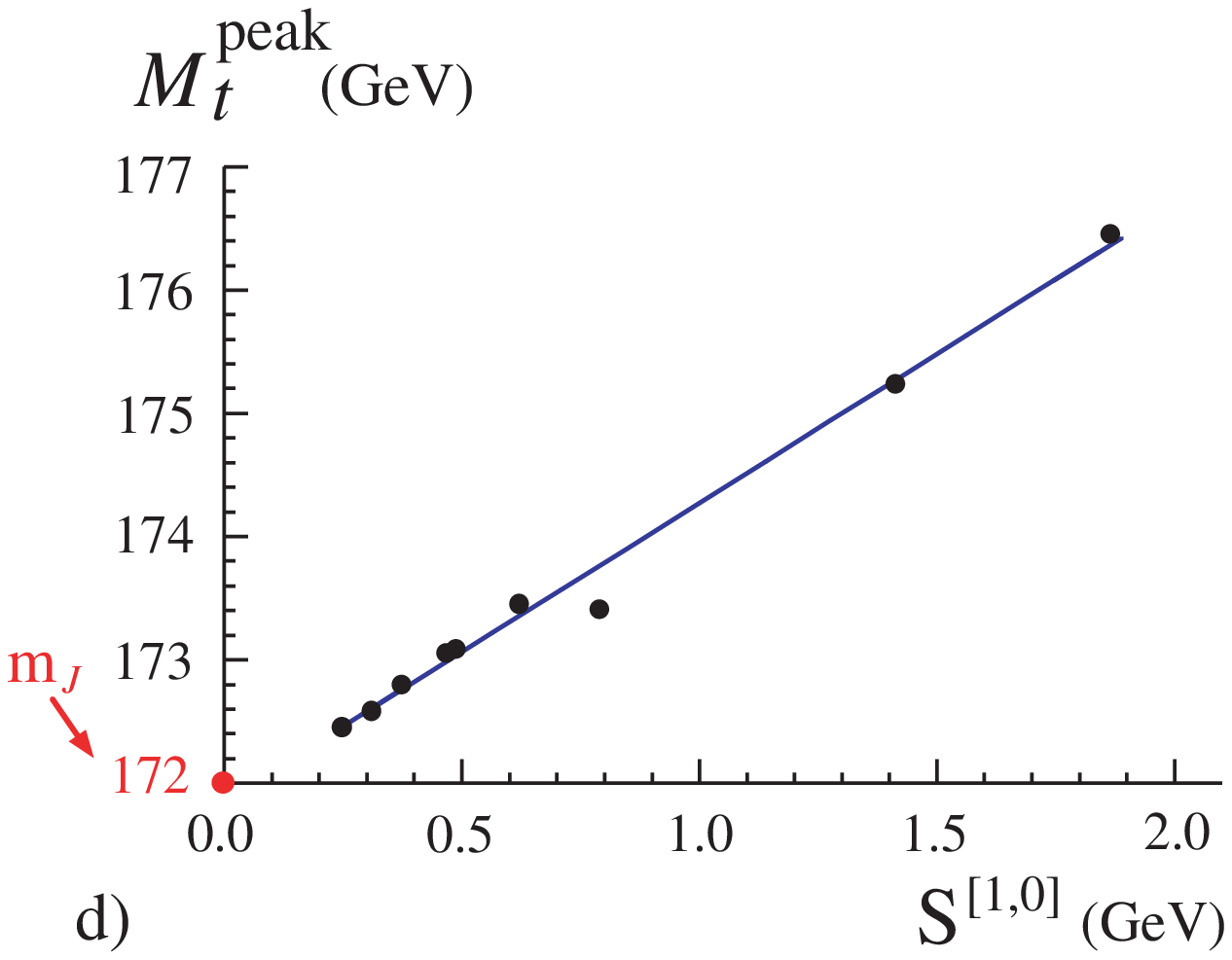}
   \end{minipage} 
    }
\caption{ 
  Dependence of the invariant mass distribution on the soft-function
  model parameters.  In a) we show 9 models with different $a$ and $b$
  parameters, and in b) we show the resulting $d\sigma/dM_t$ in units
  of $2\sigma_0^H/\Gamma$. In c) we plot the first moment of the
  invariant mass distribution, $F^{[1,0]}$, versus the first moment of
  the soft-function $\ell^+$, $S^{[1,0]}$.  In d) we plot the peak
  position of $d\sigma/dM_t$ versus $S^{[1,0]}$.  These plots are made
  with $Q/m_J=5$ and $m_J=172\,{\rm GeV}$.  Note that these scans are
  only relevant experimentally if the universality of $S_{\rm hemi}$
  with massless dijet events is not used to determine the parameters.}
\label{fig:models}
\end{figure}

Finally we consider the effect on the invariant mass shift from a scan over
model parameters. $F(M_t,M_{\bar t})$ depends on the parameters
\begin{align}
  m_J ,\quad  \Gamma,\quad  \beta = \frac{Q\Lambda}{m_J \Gamma},
  \quad a, \quad b \,.
\end{align}
Here the scale $\Lambda$ for the soft-function only shows up along
with $Q/m_J$ in the effective boost parameter $\beta$.  To demonstrate
that it is $\beta$ that appears in $F(M_t,M_{\bar t})$, switch
integration variables to $x=\ell^+/\Lambda$ and $y=\ell^-/\Lambda$,
and let $\hat s_{t,\bar t} =z_{t,\bar t} \Gamma$. This yields a soft
function $\Lambda^2 S_{\rm hemi}(\Lambda x,\Lambda y)$ that is
independent of $\Lambda$, and $\tilde B_+(\hat s_t -Q\ell^+/m_J)=
\tilde B_+(\Gamma(z_t- \beta x))$ which is only a function of
$(z_t-\beta x)$ times $(\Gamma m_J)^{-1}$. Hence $F(M_t,M_{\bar t})$
is only a function of $\beta$, $z_{t,\bar t} = (M_{t,\bar
t}-m_J)/\Gamma$, $(m_J\Gamma)$, and the model parameters $a$ and $b$.
Hence changing $\Lambda$ has the same effect as changing
$Q/m_J$. Below we will only consider variations of the model
parameters $a$ and $b$, while keeping $\Lambda=0.55\,{\rm GeV}$.

We generate 9 soft-function models from the intersection of
$a=\{1,2,3\}$ and $b=\{-0.9,0.0,0.9\}$, and in Fig.~\ref{fig:models}a
give the profile of these models by plotting $S(\ell^+)=\int d\ell^-
S_{\rm hemi}(\ell^+,\ell^-)$.  Increasing $a$ shifts the distribution
to larger average momenta.  For each model the result for the single
invariant mass distribution $F_1(M_t)$ is shown in
Fig.~\ref{fig:models}b with curves of a matching color. We again take
the $M_{\rm \bar t}$ integration interval, $[M_{\rm lower},M_{\rm
uppper}]$, to be centered on the measured peak, with a size that is
twice the measured peak width.  The peak positions for $F_1(M_t)$ are
ordered in the same manner as the peak positions for the $S(\ell^+)$
models. We note that models with $b=0,0.9$ generate smaller peak
shifts than those with $b\sim -0.9$.  To examine the peak shifts more
quantitatively we plot the first moment $F^{[1,0]}$ versus the first
moment $S^{[1,0]}_{\rm hemi}$ in Fig.~\ref{fig:models}c. To compute
$F^{[1,0]}$ we restrict the two integrals in Eq.~(\ref{mom1}) to the
same interval choice $[M_{\rm lower},M_{\rm upper}]$. From the figure
we observe that the mean of the invariant mass distributions for
different models fall close to a straight line. In Fig.~\ref{fig:models}d we
plot the peak position $M_t^{\rm peak}$ for each model versus the
first moment $S^{[1,0]}_{\rm hemi}$, and observe that the behavior is
also quite linear. We conclude that the main effects of $a,b$ on the peak
shift are controlled by the first moment parameter $S^{[1,0]}_{\rm
hemi}$.

\subsection{Implications for top-quark mass measurements} \label{section4b}

In this section we take a step back to consider the more general implications of
our method for top-quark mass measurements.  In a realistic top-quark mass
analysis at a hadron collider with $pp$ or $p\bar p$ collisions, the set of
issues that effect the accuracy for a $m_t$-measurement and that can potentially be
improved by theoretical progress includes: i) the choice of the observable to be
measured, ii) the top mass definition, iii) hadronization effects, iv) color
reconnection, v) final state radiation, vi) initial state radiation, vii)
underlying events, viii) cuts to remove the beam remnant, and ix) parton
distribution functions. In our analysis we treat $e^+e^-$ collisions which
allows us to investigate strong interaction effects in categories i)-v). We
briefly discuss what our result for $d^2\sigma/dM_tdM_{\bar t}$ implies for
these uncertainties. 

The main advantage of the factorization approach is that it keeps careful track
of how changing the observable effects corrections from the other categories.
For example, switching from invariant mass variables to thrust gives a different
function for the non-perturbative soft radiation, but the soft-functions in
these observables are related by Eq.~(\ref{Sthrust}), and one model can be used
to fit both of them. In our analysis, the inclusive nature of the hemisphere
invariant mass observable reduces the uncertainty from hadronization effects. In
particular it yields jet-functions which sum over hadronic states with invariant
mass up to $\sim m\Gamma$, and remain perturbatively computable due to the
low-momentum cutoff provided by the top-width.  Final state gluon radiation from
the decay products contributes to the width, while emission of soft-gluons in
the c.m. frame organize themselves with radiation from the top-quark to give a
single universal soft-radiation function. Thus, non-trivial color reconnection
effects between the decay products are power suppressed. The level of control
provided by the factorization theorem therefore provides a significant reduction
in the associated uncertainties. Of course the nature of this control is
observable dependent, and will undoubtably change in the hadronic collider
environment, in particular, hemisphere masses are not suitable for mass
measurements at the Tevatron or LHC.  Nevertheless, we expect the control
provided by the factorization approach to provide useful applications to this
case as well.  Finally, as discussed in detail in Sec.~\ref{sec:sdmass}, the
inherent theoretical ambiguity in the pole mass, $\delta m^{\rm pole}\sim
\Lambda_{\rm QCD}$, can be avoided by switching to a short-distance jet-mass
$m_J$.  This mass definition is suitable for reconstruction measurements in 
$e^+e^-$ collisions, and in principle also for $p\bar p$, and $pp$
collisions. 

For the case studied in detail here, a measurement of $d^2\sigma/dM_tdM_{\bar
  t}$ from energetic top-jets in $e^+e^-$ collisions, there are at least two
ways the result in Eq.~(\ref{sigmaMM}) can be used to fit for the short-distance
mass $m_J$. In the first one takes the soft-function model parameters $a$, $b$,
and $\Lambda$ from a fit to massless dijet event shapes, and then analyzes
$d^2\sigma/dM_tdM_{\bar t}$ to fit for $m_J$. This method makes use of the
universality of the soft-hemisphere function $S_{\rm hemi}(\ell^+,\ell^-)$
between massive and massless jets.  Alternatively, one can vary $Q$ and do a
simultaneous fit to $m_J$, $a$, $b$, and the effective boost parameter $\beta$,
to determine the soft-parameters from the same data used to determine $m_J$.
This may be advantageous if the energy resolution, jet energy scale, or other
experimental effects have non-trivial interactions with the soft radiation that
are particular to $t\bar t$ decays. In the next section we consider how the
factorization theorem is modified by the use of an inclusive $k_T$ algorithm
rather than using hemisphere masses. 

To conclude this section we note that in our analysis we have not accounted
for QED effects since in this work we are interested in treating the effects
of the strong interactions. For a realistic description of experimental data
obtained at a future Linear Collider QED effects will of course have to be
included. Effects from QED can contribute to the categories ii), v), vi)
as well as to the QED analogues of categories iv) and ix). 
As such the treatment of these QED effects is straightforward and can be
included naturally in our factorization approach. 
To be more concrete, the QED analogue to parton distribution functions entails
to account for initial state radiation, beam strahlung and the beam energy
spread through a luminosity spectral function which has to be convoluted with
the QCD cross section. The luminosity spectrum is obtained from analysing
Babbha scattering~\cite{Boogert:2002jr}.  
The effect of soft photons showing up in the two hemispheres can be
incorporated as additional perturbative contributions in the soft function,
and the effects of collinear photons can be incorporated as additional
perturbative contributions to the jet functions in our factorization theorem.
Finally, the effect of hard photons not alligned with the top and antitop
jets is analogous to the productions of additional hard jets, which leads to
contributions in the hemisphere masses away from the resonance region.
Compared to the QCD effects treated in this work the above QED corrections
lead to changes of the invariant mass distribution that are
suppressed by the small QED coupling.

\section{Factorization for Masses Based on  Jet Algorithms}
\label{sectionotheralgo}

In this work we have defined the top and antitop invariant masses up to now as
the invariant masses of all particles in the two hemispheres defined through the
thrust axis of each event, see Fig.~\ref{fig:topjet}. In past experimental
studies, on the other hand, a $k_T$ algorithm was employed so that each event
results in exactly six jets for the all-hadronic decay mode, $e^+e^-\to t\bar
t+X\to 6~\mbox{jets}$~\cite{Chekanov:2002sa,Chekanov:2003cp}. Of these six jets,
three jets were combined to the top and the other three to the antitop invariant
mass. We remind the reader that jet algorithms for $e^+e^-$ collisions do not
need to remove any 'beam remnants', so every final state particle of an event is
eventually either assigned to the top or the antitop invariant mass. It is this
crucial aspect of jet algorithms for $e^+e^-$ collisions that makes them share a
number of important properties with the hemisphere invariant masses that we have
analyzed so far.  One of these properties is that having both invariant masses
in the peak region close to the top quark mass automatically ensures that that
the event is dijet-like, such that the EFT setup discussed in the previous
sections can be applied in the same way as it was for the hemisphere masses.

In this section we show that using a jet algorithm with the property mentioned
above for the top and antitop invariant mass reconstruction, the double
differential top and antitop invariant mass distribution in the peak region can
be written in the factorized form of Eq.~(\ref{final-cross}), but with a
different soft function $S(\ell^+,\ell^-)$ which depends on the jet algorithm.
All other ingredients, the jet functions $B_{\pm}$ and the matching and
evolution factors are identical. For the proof we assume that the top and
antitop decay jets\footnote{ For this discussion we deal with the case that the
  top quarks decay all-hadronically into jets. However, our arguments also work
  in principle for final states with leptons plus jets.  }  obtained from the
jet algorithm can be assigned unambiguously to the top and the antitop, i.e.~we
neglect the combinatorial background. This simplification is possible at leading
order in $m/Q$ because hard jets from the top decay only have a very small
probability of order $(m/Q)^2$ to show up in the hemisphere of the antitop
quark, as was already pointed out in Sec.~\ref{section_hemi}. The analogous
statement is of course also true for hard jets from the antitop decay. Moreover
we assume that the jet algorithm uses simple addition of four-vectors as its
recombination scheme for merging final state objects.

The proof can be carried out in the EFT setups described in
Sec.~\ref{subsectionfactorizationtheorem} and Fig.~\ref{fig:theory}. The crucial
point that has to be shown is that, at leading order in the power counting,
the total $n$-collinear momentum $P_{X_n}$ enters exclusively the top invariant
mass, while the total $\bar n$-collinear momentum enters exclusively the antitop
invariant mass, just as for the hemisphere mass definitions explained in
Sec.~\ref{subsectionmomdecomp}. Furthermore the prescriptions to determine the
soft function has to be provided for a given jet algorithm. This corresponds to
defining appropriate projection operators $\hat P_a$ and $\hat P_b$ in
Eq.~(\ref{PaPb}) or equivalently the momenta $k_s^a$ and $k_s^b$ for each state
$|X_s\rangle$. Apart from that, the derivation of the factorization theorem goes
along the same lines as for the hemisphere case described in detail in the
previous sections.

Concerning the assignment of $n$- and $\bar n$-collinear momenta it is
easy to see that the top and antitop collinear momenta are attributed
correctly to the top and and antitop invariant masses since we can
neglect combinatorial background for the assignment of top and antitop
decay jets at leading order.  Assuming for example a $k_T$ jet
algorithm similar to Refs.~\cite{Chekanov:2002sa,Chekanov:2003cp}
where all final state particles are combined to exactly six jets, one
can also conclude that at leading order in $m/Q$ the $n$-collinear
gluons are properly assigned to the top invariant mass, since these
gluons are radiated into the $n$-hemisphere and therefore assigned to
one of the three hard jets from the top quark decay. The analogous
conclusion, of course, also applies to the $\bar n$-collinear
gluons. This shows that the top mass reconstruction based on a jet
algorithm treats $n$- and $\bar n$-collinear momenta essentially in
the same way as the hemisphere method. It also means that the double
differential top and antitop invariant mass distribution based on a
jet algorithm can be derived in complete analogy to the hemisphere
case and has the form shown in Eq.~(\ref{final-cross}). The soft
function depends on the jet algorithm that is employed, and in
particular on the distance measure implemented in the algorithm.
Whether a soft gluon of a given energy ends up contributing to the top
or the antitop invariant mass depends on its relative angles with
respect to the hard jets coming from the top and antitop decay. So
upon averaging over the hard jet-configurations, a soft gluon with a
given energy and a given angle with respect to the thrust axis
contributes either to the top or to the antitop invariant mass
governed by a probability function that is determined by the jet
algorithm. This means that a soft gluon in let's say the
$n$-hemisphere has in general a nonvanishing probability to be
eventually assigned to the antitop invariant mass. The equivalence of
the top-down and the bottom-up approaches to the RG evolution in the
EFT's used to derive the factorization theorem further ensures that
the RG running of the soft function for a given jet algorithm agrees
with the running of the hemisphere soft function $S_{\rm hemi}$,
although their scale-independent terms differ. (We assume that the
jet algorithm is symmetric in its treatment of top and antitop final
states.) The explicit one-loop expressions for the RG running and the
scale-independent contributions of the soft function for a general jet
algorithm and for the hemisphere masses will be given in
Ref.~\cite{FHMS2}.



\section{Conclusions} \label{section6}

The reconstruction of top quark invariant mass distributions is one of the major
methods for measuring the top-mass $m$ at present and future collider
experiments.  Using a sequence of effective theories to separate effects at
different mass scales we presented an analytic factorization approach for
the top invariant mass distribution in the peak region. To be definite, we
derived the double differential top/antitop invariant mass distribution
$d^2\sigma/dM_tdM_{\bar t}$ in $e^+e^-$ collisions for c.m.\,energies $Q\gg m$,
where $M_{t,\bar t}$ are defined as the total invariant masses of all particles
in the two hemispheres determined with respect to the event thrust axis. The
factorization formula is given in Eq.~(\ref{final-cross}) and represents the
leading order result in an expansion in $m/Q$ and $\Gamma/m$, where $\Gamma$ is
the top quark total width.

The factorization formula consists of two jet functions for top and antitop
quarks, which depend strongly on the top quark Lagrangian mass, and can be
computed perturbatively order-by-order in $\alpha_s$. It also involves a
nonperturbative soft function that describes the momentum distribution of soft
final state radiation. Using alternative invariant mass prescriptions, for which
the soft particles are assigned differently to $M_t$ and $M_{\bar t}$, the same
factorization formula applies, but with a different soft function. The
observable invariant mass distribution is obtained from a convolution of the
perturbative jet functions with the nonperturbative soft function. Through this
convolution the energy of the maximum and the width of the observed distribution
are dependent on the c.m.~energy $Q$. For a lowest order analysis see
Figs.~\ref{fig:plot3D} and \ref{fig:plot1dBW}, and the accompanying discussion.

A very important outcome of the derivation is that the soft function for the
hemisphere mass prescription also governs event shape distributions for massless
dijet events for which plenty of data has been collected at LEP and previous
$e^+e^-$ experiments. Since the soft function can be determined from these data,
it is possible to predict the top invariant mass distribution based on the
hemisphere prescription as a function of the c.m.~energy $Q$, the strong
coupling $\alpha_s$ and the Lagrangian top mass in different mass schemes
without hadronization uncertainties at leading order in the expansion in
$m/Q$, $\Gamma/m$ and $\Lambda_{\rm QCD}/m$. In principle, this allows to
measure a short-distance top quark mass from reconstruction with a precision
better than $\Lambda_{\rm QCD}$. 

We have proposed a new short-distance mass scheme called {\it top quark jet
  mass} which can be measured with minimized theoretical uncertainties from data
obtained at a future Linear Collider and which can be reliably related to other
known short-distance masses such as the threshold or the $\overline{\mbox{MS}}$
masses. We also expect that, quite generally, the jet-mass scheme will provide an
appropriate mass scheme for jet related observables involving heavy quarks.
  
The factorization approach developed in this work can be applied to
determine the top quark mass from reconstruction of the 
top/antitop quark invariant mass distributions at a future $e^+e^-$ Linear
Collider. The at present most precise method to measure the top quark mass at
a future Linear Colliner is the threshold scan method. It relies on the
determination of the hadronic $R$-ratio for c.m.\,\,energies around twice the
top mass and will provide a short-distance top quark mass measurement with
theoretical and experimental uncertainties at the level of $100$~MeV. Compared
to the measurement of the $R$-ratio for the threshold scan, the reconstruction
of the top/antitop invariant mass distribution is without doubt substantially
more complicated. But it has the advantage that it can be carried out at any
c.m.\,\,energy above threshold and that substantially more luminosity can be
spent of it. Given that our factorization approach allows to control the
perturvative and nonperturbative effects contributing to the invariant mass
distributions we believe that it could eventually become a method that 
competes with the threshold scan.

The factorization ideas proposed in this work can be applied to mass
distributions of other final state particles produced in $e^+e^-$ collisions
in a straightforward manner.  Notable examples include single top quark
production, the production of $W$ bosons or of new heavy colored unstable
particles such as squarks or gluinos in a certain supersymmetric new physics
scenarios. They will also be relevant for predicting invariant mass
distributions at hadron colliders. However, at hadron colliders there are
additional complications that still need to be 
resolved as discussed in Sec.~\ref{section4b}.  These include the initial
state radiation and incorporating parton distribution functions, which lead to a
distribution for $Q$ and require modifications of the concept of event shapes,
the large $p_T$ cuts needed to get clear signals away from the beam remnant, and
the effects of underlying events, which need to be taken into account. Finally,
the algorithm for defining and measuring the invariant mass of jets that contain
the top-decay products is different in the LHC environment. We plan to address
these issues in future work.


\acknowledgments

We would like to thank A.~Juste, S.~Kluth, S.~Menke, and M.~Wise
for helpful discussions. We also thank C.~Bauer and M.~Dorsten for collaboration
in an early stage of this work.  S.F. and S.M. thank the visitor program of the
Max-Planck-Institute for Physics for support. This work was supported in part by
the Offices of Nuclear and Particle Physics of the U.S.\ Department of Energy
under DE-FG02-94ER40818, DE-FG03-92ER40701, and DE-FG02-06ER41449, and in part
by the EU network contract MRTN-CT-2006-035482 (FLAVIAnet).  I.S. and S.F.~were
supported in part by the DOE OJI program, and I.S.~was supported in part by the
Sloan Foundation.

\appendix

\bibliography{topjet}

\end{document}